\documentclass[aps,pra,reprint,amsmath,superscriptaddress,amssymb,showpacs,nofootinbib]{revtex4-2}
\usepackage{amsmath,braket,amssymb}
\usepackage{multirow}
\usepackage[final]{graphicx}
\usepackage{subfigure}
\usepackage[english]{babel}
\usepackage[utf8]{inputenc}
\usepackage{mathtools}
\usepackage{empheq}
\usepackage{tensor}
\usepackage{cancel}
\usepackage{enumitem}
\usepackage{simplewick}
\usepackage{placeins}

\usepackage{enumitem}
\usepackage{slashed}
\usepackage{graphicx}
\usepackage{xcolor}

\colorlet{Green}{black!30!green}

\definecolor{THc}{rgb}{0.9,0.3,0.2}

\usepackage{tikz}
\usetikzlibrary{calc,fadings,decorations.pathreplacing,shapes,shapes.multipart,arrows,shapes.misc,intersections,positioning,patterns}
\tikzset{arrow data/.style 2 args={%
		decoration={%
			markings,
			mark=at position #1 with \arrow{#2}},
		postaction=decorate}
}
\usepackage{bm}
\usepackage[colorlinks=true,citecolor=blue,linkcolor=blue,urlcolor=blue]{hyperref}
\usepackage{dsfont}
\usepackage{changepage}
\usepackage{array}
\usepackage{soul}
\usepackage[capitalise]{cleveref}
\crefname{section}{Sec.}{Secs.}
\Crefname{section}{Sec.}{Secs.}
\usepackage{xifthen}
\usepackage{xargs}
\usepackage{hhline}
\usepackage{pifont}

\usepackage{amsthm}
\theoremstyle{definition}

\theoremstyle{plain}

\usepackage{bm}

\graphicspath{{./}{./figures/}}

\newcommand{\bit}{\begin{itemize}}
	\newcommand{\eit}{\end{itemize}}

\renewcommand{\>}{\right\rangle}
\newcommand{\<}{\left\langle}
\newcommand{\ba}{\begin{align}}
	\newcommand{\ea}{\end{align}}
\newcommand{\be}{\begin{equation}}
	\newcommand{\ee}{\end{equation}}
\newcommand{\bi}{\begin{itemize}}
	\newcommand{\ei}{\end{itemize}}

\newcommand{\id}{\mathbb{I}}

\DeclareMathAlphabet{\mymathbb}{U}{BOONDOX-ds}{m}{n}

\usepackage{physics}

\usepackage{braket}
\usepackage[normalem]{ulem}

\begin{document}
	\date{\today}
	
	\newcommand{\bbra}[1]{\<\< #1 \right|\right.}
	\newcommand{\kket}[1]{\left.\left| #1 \>\>}
	\newcommand{\bbrakket}[1]{\< \Braket{#1} \>}
	\newcommand{\pll}{\parallel}
	\newcommand{\nn}{\nonumber}
	\newcommand{\transp}{\text{transp.}}
	\newcommand{\nor}{z_{J,H}}
	
	\newcommand{\hL}{\hat{L}}
	\newcommand{\hR}{\hat{R}}
	\newcommand{\hQ}{\hat{Q}}

    	\newcommand{\hh}{\widetilde{h}}

	\title{Engineering discrete local dynamics in globally driven dual-species atom arrays}

\begin{abstract}
We introduce a method for engineering discrete local dynamics in globally-driven dual-species neutral atom experiments, allowing us to study emergent digital models through uniform analog controls. Leveraging the new opportunities offered by dual-species systems, such as species-alternated driving, our construction exploits simple Floquet protocols on static atom arrangements, and benefits of generalized blockade regimes (different inter- and intra-species interactions). We focus on discrete dynamical models that are special examples of Quantum Cellular Automata (QCA), and explicitly consider a number of relevant examples, including the kicked-Ising model, the Floquet Kitaev honeycomb model, and the digitization of generic translation-invariant nearest-neighbor Hamiltonians (e.g., for Trotterized evolution). As an application, we study chaotic features of discretized many-body dynamics that can be detected by leveraging only demonstrated capabilities of globally-driven experiments, and benchmark their ability to discriminate chaotic evolution. 
\end{abstract}

\author{Francesco Cesa}
\thanks{These authors contributed equally to this work.}
\affiliation{Institute for Quantum Optics and Quantum Information of the Austrian Academy of Sciences, 6020 Innsbruck, Austria}
\affiliation{Institute for Theoretical Physics, University of Innsbruck, 6020 Innsbruck, Austria}

\author{Andrea Di Fini}
\thanks{These authors contributed equally to this work.}
\affiliation{Dipartimento di Fisica e Astronomia, Universit\`a di Bologna, via Irnerio 46, I-40126 Bologna, Italy}

\author{David Aram Korbany}
\thanks{These authors contributed equally to this work.}
\affiliation{Dipartimento di Fisica e Astronomia, Universit\`a di Bologna, via Irnerio 46, I-40126 Bologna, Italy}
\affiliation{INFN, Sezione di Bologna, via Irnerio 46, I-40126 Bologna, Italy}

\author{Roberto Tricarico}
\affiliation{Institute for Quantum Optics and Quantum Information of the Austrian Academy of Sciences, 6020 Innsbruck, Austria}
\affiliation{Institute for Theoretical Physics, University of Innsbruck, 6020 Innsbruck, Austria}

\author{Hannes Bernien}
\affiliation{Institute for Quantum Optics and Quantum Information of the Austrian Academy of Sciences, 6020 Innsbruck, Austria}
\affiliation{Institute for Experimental Physics, University of Innsbruck, 6020 Innsbruck, Austria}

\author{Hannes Pichler}
\affiliation{Institute for Quantum Optics and Quantum Information of the Austrian Academy of Sciences, 6020 Innsbruck, Austria}
\affiliation{Institute for Theoretical Physics, University of Innsbruck, 6020 Innsbruck, Austria}

\author{Lorenzo Piroli}
\affiliation{Dipartimento di Fisica e Astronomia, Universit\`a di Bologna, via Irnerio 46, I-40126 Bologna, Italy}
\affiliation{INFN, Sezione di Bologna, via Irnerio 46, I-40126 Bologna, Italy}

\maketitle

\section{Introduction}
\label{sec:intro}

\indent In the era of noisy intermediate-scale quantum devices~\cite{preskill2018quantum}, discrete local dynamics have emerged as a powerful tool for our understanding of quantum many-body phenomena. These models, wherein a system evolves in discrete time steps under strictly local update rules, typically contain minimal yet expressive structures, capable of reproducing complex dynamical features of interacting systems~\cite{fisher2023random,bertini2025exactly}. As a result,  they offer a controlled route to investigate hard questions in non-equilibrium physics~\cite{
nahum2018operator,vonKeyserlingk2018operator,zhou2020entanglement, nahum2017quantum, bertini2019entanglement,klobas2021exact, chan2018solution,Giudici2024PXP}. Dynamics of this type, which constitute a class of Quantum Cellular Automata (QCA)~\cite{farrelly2020review, arrighi2019overview}, have been realized, e.g., with superconducting processors~\cite{Arute_2019, zhang2022digital,jones2022small}, trapped ions~\cite{Zhang_2017, Huerta_Alderete_2020}, and neutral atom arrays~\cite{Evered_2025, ExperimentalPaper}.\\
\indent Specifically, atoms in optical tweezers~\cite{saffman2010quantum, kaufman2021quantum} are a particularly appealing setting: thousands~\cite{Manetsch2025atoms6100, Pause2024atoms1000, Tao2024Large} of units can be arranged in arbitrary geometries, with native interacting regimes emerging via Rydberg couplings~\cite{kaufman2021quantum,browaeys2020many}. Engineering discrete local dynamics however requires local control of such interactions, while in these systems atom excitation is often driven by \emph{global} lasers, that address all the atoms uniformly. Correspondingly, many such experiments work in an analog quantum simulation mode, investigating the dynamics explored by the native Rydberg Hamiltonian~\cite{schauss2015crystallization,labuhn2016tunable,bernien2017probing,deleseleuc2018accurate,schauss2018quantum,guardado2018probing,lienhard2018observing,Leseleuc2019topological, Keesling2019QPT,surace2020lattice, scholl2021quantum,bluvstein2021controlling,lippe2021experimental, ebadi2021quantum, scholl2021quantum, bluvstein2021controlling, semeghini2021probing, Choi2023random, Chen2024breaking, Shaw2024analogue,  GonzalezCuadra2024String, manovitz2025quantum, Fang2025critical, senoo2025high, Tsai2025benchmarking, Emperauger2025Luttinger}. While recent experiments achieved effective local control through mid-circuit rearrangement~\cite{bluvstein2022quantum, evered2025probing}, designing protocols based on static atomic positions and global driving would drastically simplify the technological requirements. \\
\indent In this work, we introduce a method for engineering a variety of quantum many-body discrete dynamical models with \emph{dual-species} atom arrays~\cite{Zeng2017dual, Sheng2022dual, singh2022dual, singh2023mid, anand2024dual, ExperimentalPaper}, based solely on global driving and static processor layouts. Dual-species arrays constitute a new experimental frontier in neutral atom platforms~\cite{singh2022dual,singh2023mid}. Our protocol leverages the unique possibilities offered by the fact that the laser beams, while spatially uniform, are \emph{species-selective}. 
We show how to leverage this to natively implement a large class of translation-invariant discrete local dynamics in a globally-driven, yet effectively discretized setting.
As relevant examples, we provide detailed constructions and analyze near-term implementations of the paradigmatic kicked-Ising model~\cite{prosen2007chaos,prosen2002general} and the Floquet Kitaev honeycomb model~\cite{kitaev2006anyons}, as well as the digitization of arbitrary translation-invariant $2-$local Hamiltonians, which offers a route towards Trotterized evolution and more generic Floquet models~\cite{lloyd1996universal,trotter1959product,suzuki1991general}. Importantly, our approach, while rather general, has constant overhead in both space and time, making our proposal well matched to forthcoming dual-species experiments.  \\
\begin{figure*}[t!]
    \begin{minipage}{1.0\textwidth}
        \includegraphics[width=\textwidth]{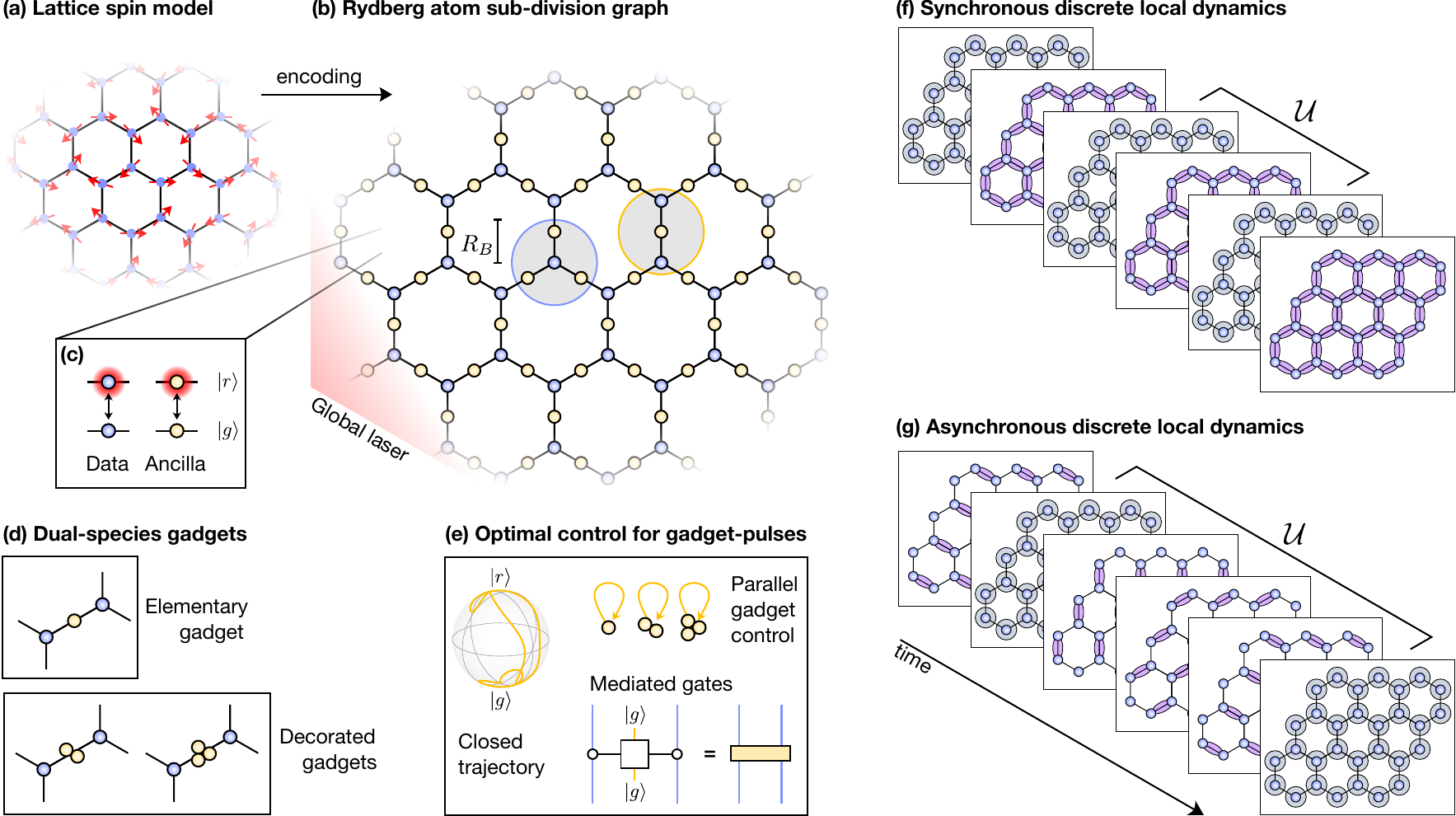}
        
    \end{minipage}%
\caption{(a, b) For a model defined on a 2D lattice, we construct a sub-division graph, by inserting one new vertex (in yellow) on each edge of the original lattice. We then arrange a dual-species atom array by positioning atoms of the data (blue) and ancillary (yellow) species on the vertices and edges, respectively. The role of the ancillas is to mediate gates between the data atoms. The atom array is then driven in the (nearest-neighbor) Rydberg blockade regime through a \emph{global} (species-selective) laser. (c) For each atom, we consider two levels: a ground state $\ket{g}$ and a highly excited Rydberg state $\ket{r}$. (d) Gadgets are constructed by placing $S\geq 1$ ancillas on a bond - i.e., between two data atoms. (e) We use optimal control methods to design \emph{mediated gates}: by controlling the ancillary species, we effectively implement entangling gates on the data atoms. (f) The Floquet driving realizes a discrete local dynamics, where each step corresponds to the application of an update rule $\mathcal{U}$. The latter is said to be \emph{synchronous} when the interacting updates can be simultaneously applied (i.e., if they commute). (g) Differently, $\mathcal{U}$ is said to be \emph{asynchronous} if the local interacting updates do not commute - imposing an ordering. }
\label{fig1}
\end{figure*}
\indent Given such ability to implement discrete local dynamics, a broad class of out-of-equilibrium many-body phenomena becomes accessible. To illustrate one direction, here we focus on probing signatures of quantum chaos~\cite{bohigas1984characterization,haake_quantum_1991,gutzwiller2013chaos}. While conventional diagnostics, such as out-of-time-ordered correlations (OTOCs)~\cite{shenker2014multiple,roberts2015diagnosing,roberts2015localized,maldacena2016bound,nahum2018operator,keyserlingk2018operator}, require access to complex and spatially non-uniform operators, we focus on a chaos indicator that can be measured leveraging only demonstrated state-of-the-art capabilities of globally-driven dual-species Rydberg experiments~\cite{ExperimentalPaper}. Focusing on representative examples, we show that our indicator successfully discriminates chaotic from non-ergodic dynamics in relevant regimes. Our results therefore suggest that our approach will enable informative near-term studies of non-equilibrium many-body physics through discrete local dynamics in Rydberg arrays, allowing for experimental explorations of quantum-chaotic dynamics and QCA. \\
\indent The rest of this work is organized as follows. We begin in Sec.~\ref{sec: discrete local dynamics in Rydberg arrays}, where we review discrete local dynamics. In Sec.~\ref{sec: Dual-species architecture for globally-driven QCA}, after introducing the physics of globally-driven dual-species Rydberg arrays, we present our core constructions, dubbed `mediated gates', which enable the realization of discrete local dynamics. Then, in Sec.~\ref{sec: Implementing QCA}, we detail our architecture for implementing QCA, by presenting explicit constructions for representative examples. In Sec.~\ref{sec: Observing quantum chaos}, we specialize to the task of observing signatures of quantum chaos within our architecture, introducing a dedicated protocol and discussing its near-term potential. Finally, in Sec.~\ref{sec: Discussion and outlook}, we conclude by commenting our results and their experimental feasibility, and providing an outlook for future research.  

\section{Discrete local dynamics}\label{sec: discrete local dynamics in Rydberg arrays}

We begin by describing the types of discrete local dynamics that we will consider. We focus on models defined on 1D or 2D lattices, where each vertex hosts a qubit. The evolution is defined by a \emph{local update rule}: the state is updated with an operation $\mathcal{U}$ in discrete time steps as
\begin{equation}\label{eq:update_rule}
	\ket{\psi_{t+1}}=\mathcal{U}\ket{\psi_t}\,,
\end{equation}
with $\mathcal{U}$ satisfying certain locality constraints~\cite{farrelly2020review}. For instance, those constraints are satisfied if $\mathcal{U}$ can be written as a shallow quantum circuit composed of local gates. \\
\indent We are specifically interested in updates of the form exemplified in Fig.~\ref{fig1}(f,g): $\mathcal{U}$ is composed by $D=O(1)$ \emph{translation-invariant} layers, each consisting of single- or (nearest-neighbor) two-qubit gates. Such models are particular types of QCA~\cite{arrighi2019overview,farrelly2020review,piroli2020quantumcellular}. \\ 
\indent To appreciate the expressivity of such dynamics, it is insightful to interpret $\mathcal{U}$ as realizing a Floquet protocol,
\begin{equation}
    \mathcal{U} = e^{-i\tau H_D}e^{-i\tau H_{D-1}}\dots e^{-i\tau H_1} = e^{-i\tau H_\text{eff}[\tau]},
\end{equation}
where $H_j$ are $2-$local Hamiltonians and $\tau$ is a Floquet time. Depending on $\tau$, the Floquet Hamiltonian 
\begin{equation}\label{eq: Floquet Hamiltonian}
    H_\text{eff}[\tau] = \sum_{j=1}^DH_j - \frac{i\tau}{2}\sum_{j>k}[H_j,H_k] + O(\tau^2)
\end{equation}
can have very different physical properties. For small $\tau$, $H_\text{eff}$ approximates the local Hamiltonian $H=\sum_jH_j$, and the QCA can be understood as a Trotterized evolution under $H$. In fact, \emph{any} local Hamiltonian dynamics can be Trotterized via such discrete local dynamics.\\
\indent For larger $\tau$, $H_\text{eff}$ typically contains longer-range many-body terms, allowing one to reproduce complex interacting dynamics and, in fact, intrinsically Floquet models~\cite{Abanin_2017,ishii2018heating,heyl2019quantum,chinni2022trotter,vernier2023integrable}. In these cases, $\tau$ becomes a model parameter, controlling physical properties and the emergence of Floquet phases, including, e.g., topological edge dynamics~\cite{PhysRevX.3.031005} and time-crystalline order~\cite{PhysRevX.7.011026, Kyprianidis_2021}. The $O(\tau)$ correction in Eq.~\eqref{eq: Floquet Hamiltonian} can also change and enrich the properties of $H_\text{eff}$, e.g., introducing gapped non-Abelian topological phases in Floquet spin liquids~\cite{PhysRevX.13.031008}.\\ 
\indent It is worth stressing, however, that for any finite $\tau$ the Floquet dynamics only generates non-trivial correlations within a sharp light cone~\cite{farrelly2020review},  mimicking the famous Lieb-Robinson bound characterizing a local Hamiltonian evolution~\cite{lieb1972finite,hastings2006spectral,hastings2010locality}. This feature makes it possible to view finite-$\tau$ discrete local dynamics as toy models for many-body physics: even if they do not coincide with a local-Hamiltonian evolution, they still display local evolution of correlations. In addition, as the geometrically local structure makes them easier to analyze in many situations, they provide  concrete theoretical handles to tackle several questions in non-equilibrium physics, including quantum-information spreading~\cite{nahum2018operator,vonKeyserlingk2018operator,hosur2016chaos}, entanglement dynamics~\cite{nahum2017quantum,gong2022coarse,zhou2020entanglement, bertini2019entanglement,klobas2021exact}  and phase transitions~\cite{fisher2023random,potter_entanglement_2022,gillman2020nonequilibrium,noel2022measurement,koh2023measurement}.\\
\indent In summary, while simple and discretized, the physics underlying such discrete local dynamics is extremely rich  and broad: in some regimes, it embeds out-of-equilibrium many-body phenomena in circuit-like dynamics; in other settings, it constitutes a separate paradigm, with a wide spectrum of interesting features \emph{per se}.  

\section{Dual-species architecture for globally-driven discrete dynamics}\label{sec: Dual-species architecture for globally-driven QCA}

\begin{figure}
    \includegraphics[width=1\linewidth]{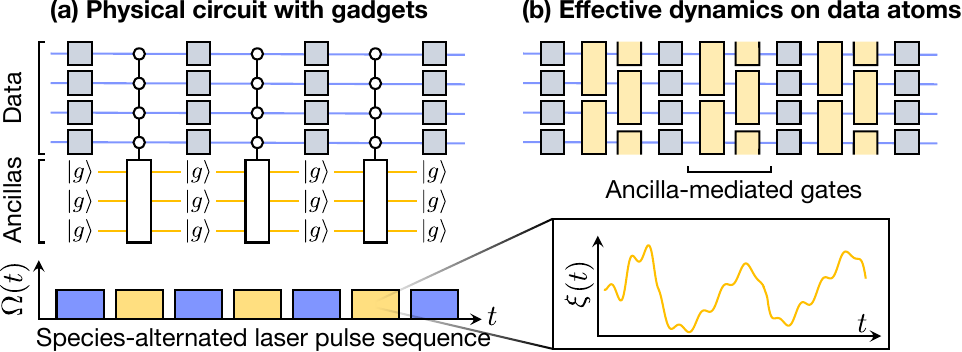}
      
	\caption{(a) Circuit depiction of our protocol, where we alternate driving on data and ancillary atoms. The first implements single-qubit gates on the data, while the second  physically acts as controlled gates, with the data controlling the evolution of the ancillas - which always return back to $\ket{g}$. Our control knob is the phase profile $\xi(t)$ of the global laser. (b) Since the ancillas are always deterministically mapped back to $\ket{g}$, this effectively results in a circuit on the data only, with entangling gates mediated by the ancillas.}
	\label{fig: fig2}
\end{figure}

\indent We are interested in engineering the discrete local dynamics described in the previous section in dual-species Rydberg experiments~\cite{Zeng2017dual, Sheng2022dual, singh2022dual, singh2023mid, anand2024dual, ExperimentalPaper}. To this end, after positioning the atoms, we use species-selective, but spatially uniform lasers (alternating the addressed species), and exploit strong dynamical constraints imposed by always-on interactions. Section~\ref{sec: setup} overviews this setup.\\
\indent Our construction is based on the prescription in Fig.~\ref{fig1}: (i) Atoms of the two species (blue and yellow in the figures) encode `data' and `ancilla' qubits, respectively. (ii) Data atoms occupy the vertices of the target model. (iii) Ancillas are positioned at the bonds, forming `gadgets' that effectively mediate interactions. (iv) A sequence of global laser pulses, alternating the two species, activates the gadgets to implement $\mathcal{U}$. We provide the technical implementation details in Section~\ref{sec: Implementing QCA}.\\
\indent Our key insight are dual-species \emph{mediated} gates [Section~\ref{sec: gadgets}], which enable entangling two noninteracting data atoms, through a mediator ancilla [Fig.~\ref{fig1}(e)]. In circuit language, our method can be understood as in Fig.~\ref{fig: fig2}(a,b): as we will detail, the drivings on the ancillary species only have a \emph{net} effect on the data atoms, leaving the ancillas stroboscopically disentangled. Crucially, such mediated gates are both intrinsically local \emph{and} globally-driven - allowing us to implement discretized evolutions even in such analog-like setting.

\subsection{Physical setup: dual-species Rydberg arrays}\label{sec: setup}

For each atom, we consider a low energy state $\ket{g}$, and an excited Rydberg state $\ket{r}$. The transitions $\ket{g}\leftrightarrow \ket{r}$ are well separated for the two species, allowing independent addressing of data and ancillary atoms. Denoting by $\mathcal{X}\in\left\{\text{data}, \text{ancillas}\right\}$ the two species, a laser driving on species $\mathcal{X}$ is described by the Hamiltonian
\begin{equation}
    H_\mathcal{X} = \frac{\Omega}{2}\sum_{i\in\mathcal{X}} \Big[e^{-i\xi(t)}\ket{g_i}\bra{r_i} + \text{h.c.} \Big] + \sum_{i > j} V_{i,j} \ket{r_i r_j}\bra{r_ir_j},
\end{equation}
with $\Omega$ the Rabi frequency and $\xi(t)$ a time-dependent laser phase - which will be our only control knob. \\
\indent In paradigmatic experiments~\cite{bernien2017probing,scholl2021quantum,ebadi2021quantum,semeghini2021probing}, $V_{i,j}$ is very strong at short distances ($V_{i,j}\gg \Omega$) and decays rapidly beyond a `blockade radius' $R_B$ ($\Omega\gg V_{i,j}$). This allows us to digitize the interactions: denoting by $\mathcal{B}_i$ the set of atoms $j$ within a radius $R_B$ from atom $i$, and defining $\mathbb{P}_i\equiv \bigotimes_{j\in \mathcal{B}_i} \ket{g_j}\bra{g_j}$ as the projector to the subspace where all atoms in $\mathcal{B}_i$ are in $\ket{g}$, one can approximate
\begin{equation}\label{eq: PXP hamiltonian}
    H_\mathcal{X} = \frac{\Omega}{2}\sum_{i\in\mathcal{X}} \mathbb{P}_i \Big[e^{-i\xi(t)}\ket{g_i}\bra{r_i} + \text{h.c.} \Big] \mathbb{P}_i,
\end{equation}
known as the PXP Hamiltonian~\cite{serbyn2021quantum}. In practice, an atom only evolves if all its neighbors are in $\ket{g}$, and is frozen otherwise - a standard regime in Rydberg atom experiments~\cite{saffman2010quantum, bernien2017probing, serbyn2021quantum}. We remark that, for our specific constructions, such approximation can potentially be realized with high precision in dual-species arrays, e.g., by exploiting F\"orster resonances~\cite{beterov2015rydberg, anand2024dual} to suppress intra-species interactions. In Appendix~\ref{sec:violations} we quantify the deviations from Eq.~\eqref{eq: PXP hamiltonian}, showing that violations of the blockade regime can be strongly suppressed, and are substantially smaller than for single-species implementations.\\
\indent Finally, we anticipate that in all our constructions, data atoms will always be positioned such that they are noninteracting with each other. In most cases, also the ancillas will not interact with each other, resulting in models defined on bipartite data-ancilla graphs, with only inter-species interactions. In some cases, we will introduce small clusters of interacting ancillary atoms; however, these will effectively behave as a unique entity [Sec.~\ref{sec: decorated gadgets}]. Thus, also in the latter case we will end up with an effectively bipartite model.  

\begin{figure}
    \includegraphics[width=1\linewidth]{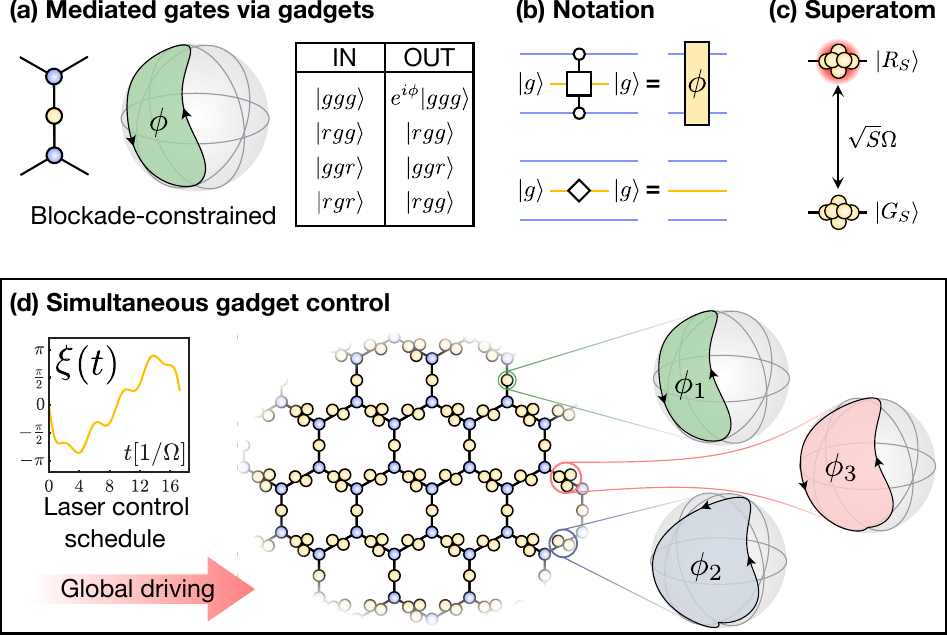}  
	\caption{(a) To implement mediated gates , we drive the ancillary species only, inducing a closed trajectory on the Bloch sphere. This trajectory is conditioned on both the neighboring data atoms being in $\ket{g}$: the geometrical phase $\phi$ is only acquired in this case. This results in an effective entangling gate on the data, as specified by the truth table. (b) Depiction of our circuit notation. Note that we use a diamond symbol to denote the special case where the mediated gate is an effective identity, i.e., $\phi = 2\pi n$ . (c) A cluster of $S$ atoms within a blockade radius behaves as a \emph{superatom}, oscillating between two effective states $\ket{G_S}\leftrightarrow\ket{R_S}$ with enhanced Rabi frequency $\sqrt{S}\Omega$. (d) Optimal control techniques allow us to simultaneously implement mediated gates through gadgets of various sizes, imprinting arbitrary phases $\phi_S$ to superatoms of size $S$, in parallel.}
	\label{fig: fig3}
\end{figure}

\subsection{Dual-species gadgets for mediated gates}\label{sec: gadgets}

We now introduce our mediated gates, which will be at the core of our implementations in Section~\ref{sec: Implementing QCA}.

\subsubsection{Mediated gates and elementary gadgets}\label{sec: elementary gadgets}

\indent First, consider the \emph{elementary gadget} in Fig.~\ref{fig1}(d,~top). As detailed in Fig.~\ref{fig: fig3}(a), this enables implementing mediated gates between two data atoms, through a mediator ancilla placed within the blockade radius of both. \\
\indent We now describe the implementation of a mediated gate. Initially, the mediator is in $\ket{g}$, while the data atoms are in any state. We then drive the ancillary species with a pulse that induces a \emph{closed trajectory} in the Bloch sphere [Fig.~\ref{fig: fig3}(a)]: the mediator returns to $\ket{g}$, up to a geometrical phase $\phi$~\cite{aharonov1987phase},
\begin{equation}\label{eq: geometrical phase}
    \ket{g} \rightarrow e^{i\phi}\ket{g}. 
\end{equation}
Crucially, due to the blockade constraint imposed on the ancilla by the data atoms, this only occurs if the latter are both in $\ket{g}$. The phase $\phi$ is therefore conditional~\cite{jaksch2000fast}, and reabsorbed in the wavefunction as an entangling phase acquired by the data atoms. Specifically, as summarized by the truth table in Fig.~\ref{fig: fig3}(a), this results in the gate
\begin{equation}\label{eq:elementary_gadget}
    CZ(\phi) = \id + (e^{i\phi}-1)\ket{gg}\bra{gg}
\end{equation}
on the data, which is maximally entangling for $\phi = \pi$. \\
\indent The simplest way to engineer an arbitrary phase $\phi$ is by linearly shaping the driving on the ancilla, $\xi(t) = -\Delta t$. Then, we drive for a total time $t_\text{f}=2\pi/\sqrt{\Delta^2+\Omega^2}$, with $\Delta=2(\pi-\phi)/t_f$. However, \emph{any} driving achieving Eq.~\eqref{eq: geometrical phase} activates the gadget: in settings that we discuss below, we will need more sophisticated designs of $\xi(t)$. \\
\indent The mediated gate construction extends to the many-body setting of Fig.~\ref{fig1}(b), featuring several elementary gadgets and $N$ data atoms. Indeed, let $\ket{\Psi}\otimes \ket{G}$ be the initial data-ancilla state, where 
\begin{equation}
    \ket{\Psi} = \sum_{x_1,\dots, x_N} \Psi_{x_1,\dots,x_N} \ket{x_1,\dots,x_N}
\end{equation}
is an arbitrary data state and $\ket{G}=\bigotimes_{j}\ket{g}_j$ features all the ancillas in $\ket{g}$. From Eq.~\eqref{eq: PXP hamiltonian} one can see that, under the driving above, \emph{each} ancilla picks up the phase $\phi$, only if both its neighbors are in $\ket{g}$. The output state is therefore again of the form $\ket{\Psi'}\otimes\ket{G}$, with the phases reabsorbed in the data wavefunction as 
\begin{equation}
    \Psi'_{x_1,\dots,x_N} = e^{i\phi \sum_{\langle j,k\rangle} \delta_{x_jx_k,x_j+x_k}} \Psi_{x_1,\dots,x_N};
\end{equation}
here, $\sum_{\langle j,k\rangle}$ iterates over all data neighboring pairs (for the algebra, here we identify $g\equiv 0$ and $r\equiv 1$). This corresponds to a $CZ(\phi)$ on all those pairs in parallel.

\subsubsection{Decorated gadgets}\label{sec: decorated gadgets}

For a \emph{decorated gadget}, we position $S\geq 2$ atoms on a bond [Fig.~\ref{fig1}(d), bottom]. As detailed below, this introduces additional control knobs for mediated gates. \\
\indent First, consider an isolated decorated gadget: a pair of data atoms, with a cluster of $S$ ancillary atoms placed between them. Each ancilla is within the blockade radius of (i) all the other ancillas composing the gadget, and (ii) both the data atoms. The ancillas are initially in $\ket{G_S}=\bigotimes_{j=1}^S\ket{g_j}$. Then, by symmetry and due to the intra-species blockade constraint, the Hamiltonian~\eqref{eq: PXP hamiltonian} only couples $\ket{G_S}$ to the W-state $\ket{R_S} = \sum_{j=1}^S\ket{g_1\dots r_j\dots g_S}/\sqrt{S}$, where a single Rydberg excitation is shared among the atoms. Thus, the $S$ atoms behave as a single two-level system - a `superatom'~\cite{cesa2023universal}.\\
\indent Crucially, the coupling $\ket{G_S}\leftrightarrow\ket{R_S}$ has an enhanced Rabi frequency $\sqrt{S}\Omega$~\cite{cesa2023universal}. Then, our discussion above straightoforwardly generalizes to decorated gadgets, simply by rescaling $\Omega\rightarrow\sqrt{S}\Omega$. Specifically, we can apply a $CZ(\phi)$ between the data atoms by implementing Eq.~\eqref{eq: geometrical phase} on the superatom, $\ket{G_S} \to e^{i\phi} \ket{G_S}$. To realize this, we can still choose the driving $\xi(t)=-\Delta t $ and $\Delta = 2(\pi-\phi)/t_f$, but now with $t_f = 2\pi/\sqrt{\Delta^2 +S\Omega^2}$. \\
\indent Second, consider positioning decorated gadgets of arbitrary sizes, with size $S_{i,j}\in\left\{1,\dots,S_\text{max}\right\}$ on bond $\langle i,j\rangle$ - as, e.g., in Fig.~\ref{fig: fig3}(c) on the example of a honeycomb lattice. We now have several superatom sizes at the same time. In Sec.~\ref{sec: quantum control} we discuss the following: given any set of target phases $\left\{\phi_1,\dots,\phi_{S_\text{max}}\right\}$, we can find a driving schedule $\xi(t)$ such that all superatoms, in parallel, complete a closed trajectory in the Bloch sphere - that is, for all $S\in \{1\dots S_{\rm max}\}$ we have
\begin{equation}\label{eq:loop}
    \ket{G_S} \xrightarrow[]{\;\;\;\xi(t)\;\;\;} e^{i\phi_S}\ket{G_S}.
\end{equation}
This is illustrated in Fig.~\ref{fig: fig3}(c). Thus, for each pair $\langle i,j\rangle$, the corresponding gadget implements a different mediated gate, realizing the desired $CZ(\phi_{S_{i,j}})$.\\
\indent This shows that we can leverage superatoms of different cardinality to implement $CZ(\phi)$ between neighboring data atoms $(i,j)$, with $\phi$ controlled by the size of the superatom on the bond $\langle i,j\rangle$. In practice, $S_\text{max}$ is physically constrained, thus limiting such construction. While larger $S$ may be feasible, for the constructions considered in Sec.~\ref{sec: Implementing QCA}, we will only need $S_\text{max}\leq 3$.

\subsubsection{Parallel quantum control of decorated gadgets}\label{sec: quantum control}

\begin{figure*}[t!]
    \begin{minipage}{1.0\textwidth}
        \includegraphics[width=\textwidth]{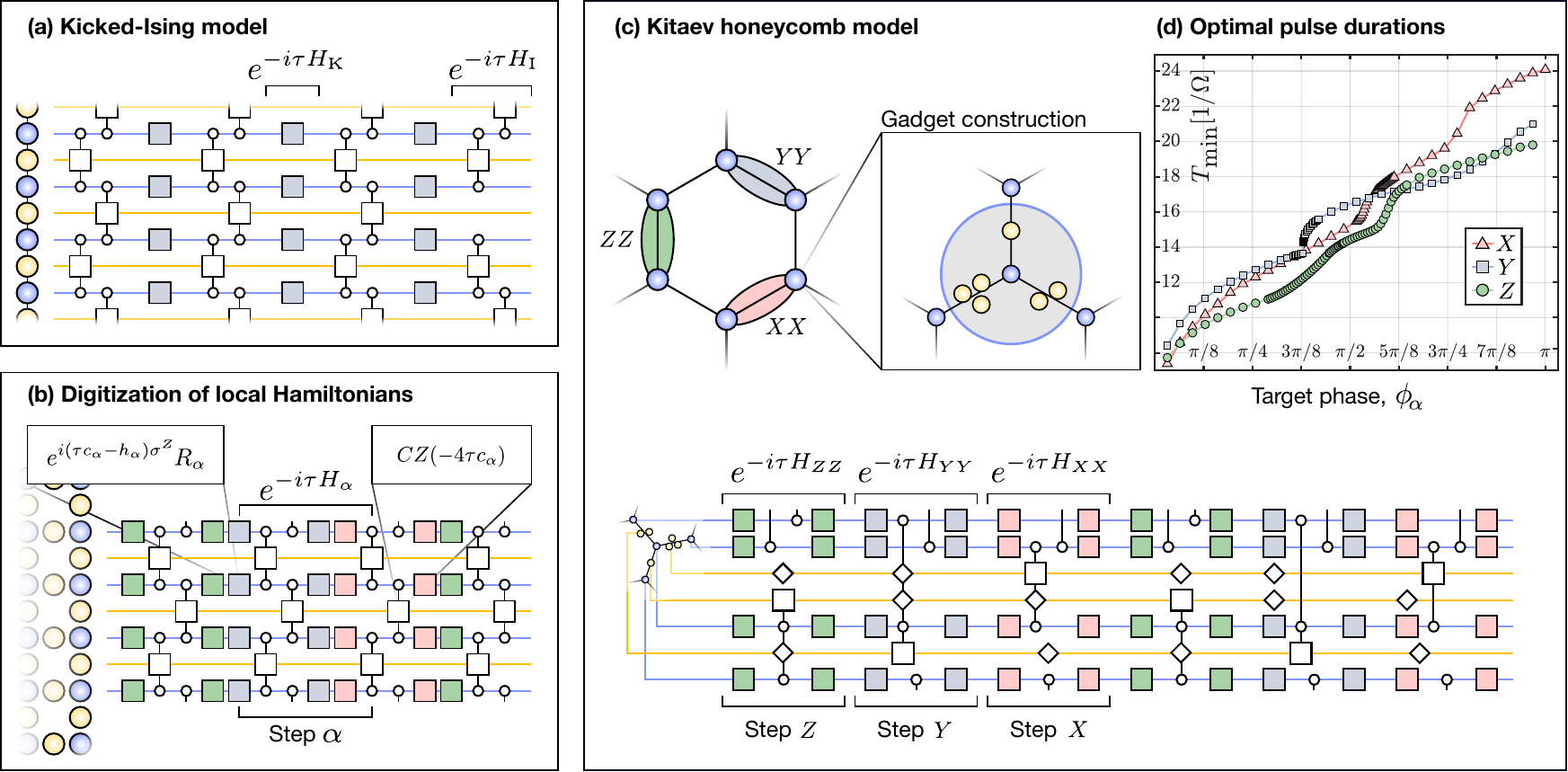}        
    \end{minipage}%
\caption{(a) For the kicked-Ising model, we alternate the application of single-atom rotations on the data species (implementing the kick term), and gadget pulses on the ancillary species, implementing the Ising layer via mediated gates. (b) For the digitization of an arbitrary $2-$local Hamiltonian, we proceed in three main steps, each operating in a different Pauli basis. (c) The Floquet Kitaev Honeycomb model is implemented by also employing decorated gadgets. (d) We use optimal control to derive time-optimal pulses for the mediated gates of the three steps of our implementation of the Floquet Kitaev honeycomb model. The figure shows the optimal durations $T_\text{min}$ as a function of the target phases $\phi_\alpha=-4\tau J_\alpha$, with $\alpha=X,Y,Z$. These values are shown for $\phi\in[0,\pi]$, holding $T_\text{min}(-\phi_\alpha)=T_\text{min}(\phi_\alpha)$. See Appendix~\ref{sec:optimal_control} for further details on the numerical procedure.}
\label{fig_examples}
\end{figure*}

For our gadget construction, a key feature is the ability to find proper driving schedules to realize the dynamics described in Eq.~\eqref{eq:loop}. The existence of such driving schedules is ensured by a \emph{complete controllability} property of such system. Specifically, given a set of $S_{\rm max}$ superatoms of different cardinality, for every choice of two product states $\ket{\psi_1}$ and $\ket{\psi_2}$, there exists a finite time $T>0$ and a control phase $\xi(t)$, with $t\in[0,T]$, which drives the system from $\ket{\psi_1}$ to $\ket{\psi_2}$. In Appendix~\ref{appendix: Controllability}, we provide a proof of this result.\\
\indent Furthermore, to minimize errors associated to the finite lifetime of the Rydberg states~\cite{bluvstein2022quantum, pagano2022error}, it is convenient to design these processes in a time-optimal fashion~\cite{fromonteil2024HJB,Astolfi2025}. 
In Fig.~\ref{fig_examples}(d), we report the minimum time required to assign a target phase to a specific superatom, while realizing the identity on the others. We consider the case of $S_{\rm max}=3$ superatoms, which is instrumental to implement the Floquet Kitaev honeycomb in our setup as described in Sec.~\ref{sec: Implementing QCA}. In Appendix~\ref{sec:optimal_control}, we show how to apply the Gradient Ascent Pulse Engineering (GRAPE) algorithm to compute the time-optimal controls~\cite{jandura2022control}.

\section{Implementing QCA}\label{sec: Implementing QCA} 

In this section, we present our construction for engineering QCA. Specifically, we now describe in detail several concrete examples which, altogether, illustrate all the mechanisms at play. In the following, we define the single-atom Pauli algebra through $\sigma^Z = \ket{r}\bra{r} - \ket{g}\bra{g}$, $\sigma^X=\ket{r}\bra{g}+\ket{g}\bra{r}$ and $\sigma^Y = i \sigma^X\sigma^Z$.

\subsection{Synchronous dynamics}

We begin by illustrating the simplest version of our scheme, which allows us to implement so-called \emph{synchronous} QCA. Broadly speaking, these are such that all the interacting local updates commute - i.e., they can be applied simultaneously [see Fig.~\ref{fig1}(f)]. In our architecture, this translates to the fact that we can implement $\mathcal{U}$ through two pulses only: the first on the data species (to update the local frames), and the second on the ancillas (to activate the gadgets). Although very simple, synchronous dynamics of this type include some interesting examples, such as the kicked-Ising model.

\subsubsection{The kicked-Ising model}
\label{sec:kicked_ising}
The kicked-Ising model, introduced in Refs.~\cite{prosen2007chaos,prosen2002general}, motivated intense research in the last decade. Notably, the 1D model provides a minimal setting featuring non-ergodicity and quantum chaos, allowing for rare analytic investigations in non-integrable systems~\cite{bertini2019entanglement, bertini2019exact}. Beyond 1D, the dynamics is classically challenging, also providing a benchmark for quantum computers~\cite{daley2022practical,kim2023evidence,tindall2024efficient,liao2023simulation,patra2024efficient}.\\
\indent Given a (regular) graph, the update rule reads
\begin{equation}\label{eq:ki_u}
    \mathcal{U}_{KI} = e^{-i\tau H_K}e^{-i\tau H_I},
\end{equation}
where we defined the standard Ising Hamiltonian
\begin{equation}\label{eq:h_I}
    H_I = J\sum_{\langle j,k\rangle} \sigma_j^Z\sigma_k^Z  + h \sum_j \sigma^Z_j
\end{equation}
with interaction $J$ and longitudinal field $h$, while
\begin{equation}\label{eq:h_K}
    H_K = b\sum_j\sigma_j^X
\end{equation}
is a `kick' term, rotating the local reference frame. \\
\indent Before proceeding, it is useful to understand this in circuit language as follows. Note that, for any $\phi$, one has
\begin{equation}\label{eq: CZ_rewriting}
    CZ(\phi) = \Big[e^{-i\phi\sigma^Z/4}\otimes e^{-i\phi\sigma^Z/4}\Big] e^{i\phi \sigma^Z\sigma^Z/4}.
\end{equation}
Thus, the Ising layer can be decomposed into nearest-neighbor gates $CZ(\phi)$ followed by a layer of single-qubit unitaries of the form $e^{-i\alpha\sigma^Z}$, by choosing $\phi = - 4\tau J$ and $\alpha = \tau(h +gJ)$, with $g$ the graph degree. The kick layer simply applies $e^{-i b \sigma^X }$ to all qubits.\\
\indent To realize $\mathcal{U}_\text{KI}$,
we proceed as in Fig.~\ref{fig_examples}(a), only employing elementary gadgets and the following two steps:\begin{enumerate}
    \item We drive the data species via a detuned pulse that directly applies $e^{-i\alpha\sigma^Z}e^{-i b\sigma^X}$ to all data atoms.
    \item We drive the ancillary species to realize the gate $CZ(\phi)$ between any pair of neighboring atoms $\langle j,k\rangle$ via mediated gates, as in Eq.~\eqref{eq:elementary_gadget}. 
\end{enumerate}  
Considering Eq.~\eqref{eq: CZ_rewriting}, it is straightforward to recognize that this exactly implements $\mathcal{U}_\text{KI}$. 
\subsubsection{Inhomogeneous kicked-Ising model}

We now consider an inhomogeneous variation of the kicked Ising model above, where the the coupling breaks some symmetry. As an example, let us focus on a 2D square lattice. The update rule is still of the form $\mathcal{U}'_{KI}=e^{-i\tau H_K}e^{-i\tau H'_I}$, with the kick term still as in Eq.~\eqref{eq:h_K}. However, the Ising term is modified as
\begin{equation}
    H'_I = J_x\sum_{\langle j,k\rangle_x}\sigma_j^Z\sigma_k^Z + J_y\sum_{\langle j,k\rangle_y}\sigma_j^Z\sigma_k^Z + h\sum_j\sigma_j^Z,
\end{equation}
where e.g. $\langle \cdot,\cdot\rangle_x$ denotes nearest-neigbors along the $x$ direction. Thus, $H'_I$ breaks isotropy, as different interaction strengths $J_x\neq J_y$ are associated with the two directions.\\
\indent To implement $\mathcal{U}'_{KI}$, we decorate bonds in the $y$ direction with superatoms of size $S=2$ (while $x$ bonds still host elementary gadgets). Then, for the inhomogeneous kicked Ising we modify the two steps above as follows: 
\begin{enumerate}
    \item We drive the data species via a detuned pulse that directly applies $e^{-i\alpha'\sigma^Z}e^{-ib\sigma^Z}$, where now we set $\alpha ' = h\tau +2\tau(J_x+J_y)$. 
    \item We drive the ancillary species with a pulse such that superatoms of size $S=1$ and $S=2$ acquire, respectively, geometrical phases $\phi_x=-4\tau J_x$ and $\phi_y=-4\tau J_y$. Through the gadgets, this realizes the gates $CZ(-4\tau J_x)$ and $CZ(-4\tau J_y)$ between nearest neighbors along the $x$ and $y$ directions, respectively. 
\end{enumerate}
Similarly to the simplest model above, one can recognize that the combination of these two steps indeed realizes $\mathcal{U}'_{KI}$. As a final note, we remark that the same concept extends naively to more complicated models and graph constructions, even when the interaction strength inhomogeneities are multiple and more disordered.

\subsection{Asynchronous QCA}

We now consider more general QCA, which are \emph{asynchronous}. That is, the local updates at different links do not commute and cannot be applied in parallel [Fig.~\ref{fig1}(g)]. Below, we detail two relevant examples. First, we discuss the Floquet Kitaev honeycomb model, a manifestly asynchronous QCA, which requires the use of decorated gadgets. Second, we discuss the digitization of arbitrary $2-$local (symmetric) Hamiltonians, which can be achieved with elementary gadgets only.

\subsubsection{The Floquet Kitaev honeycomb}\label{sec:kitae_qca}

First, we consider the Floquet Kitaev honeycomb model~\cite{kitaev2006anyons}, defined on a honeycomb lattice through 
\begin{equation}\label{eq:kitaev_hamiltonian_0}
    \mathcal{U}_{KH} = e^{-i\tau H_{XX}} e^{-i\tau H_{YY}} e^{-i\tau H_{ZZ}},
\end{equation}
where for $\alpha \in\left\{X, Y, Z\right\}$ the term $H_\alpha$ is defined as
\begin{equation}\label{eq:kitaev_hamiltonian}
    H_{\alpha\alpha} = J_\alpha\sum_{\langle j,k\rangle_\alpha}\sigma_j^\alpha\sigma_k^\alpha + h_\alpha \sum_j\sigma_j^\alpha , 
\end{equation}
Therein, $\langle\cdot,\cdot\rangle_l$ are bonds with orientation $\alpha$ [Fig.~\ref{fig_examples}(c)].  \\
\indent This QCA is linked to the paradigmatic Kitaev honeycomb Hamiltonian, $H_{KH} = H_X+H_Y+H_Z$~\cite{kitaev2006anyons}. However, since $[H_{\alpha\alpha},H_{\beta\beta'}]\neq 0$, the effective Floquet Hamiltonian
\begin{align}
    H_\text{eff}[\tau] = & H_{KH}  -\frac{i\tau}{2}[H_{XX},H_{YY}] -\frac{i\tau}{2}[H_{XX},H_{ZZ}] +\\
    & - \frac{i\tau}{2}[H_{YY},H_{ZZ}] + O(\tau^2)
\end{align}
only approximates it for small $\tau$. Interestingly, for finite $\tau$, the emerging $O(\tau)$ terms can significantly enrich the physics of this Floquet model: as shown in Ref.~\cite{PhysRevX.13.031008}, such first-order contributions, by breaking time-reversal symmetry, introduce a gap in the otherwise gapless non-Abelian phase, thereby enabling the realization of a dynamical Floquet spin liquid.   \\
\indent To realize this QCA, we dispose the data atoms on the vertices of a honeycomb lattice. We use three types of gadgets, of sizes $S=1,2,3$, in correspondence with the three lattice orientations [Fig.~\ref{fig_examples}(c)]. Specifically, bonds of orientation $\alpha$ hosts superatoms of size $S_\alpha$, with $S_{Z}=1$, $S_{Y}=2$ and $S_{X}=3$. To implement $\mathcal{U}_{KH}$, we proceed in three steps, labeled by $\alpha\in\left\{X, Y, Z\right\}$, where step $\alpha$ realizes $e^{-i\tau H_{\alpha\alpha}}$. As illustrated in Fig~\ref{fig_examples}(c), step $\alpha$ comprises the following three sub-steps:
\begin{enumerate}
    \item We drive the data species to implement the unitary $e^{-i\tau(h_\alpha+ J_\alpha) \sigma^Z} R_\alpha$ on each data atom, where $R_\alpha$ maps the $Z$ basis to the $\alpha$ basis (e.g., $R_{X}=e^{-i\pi\sigma^Y/4}$). 
    \item We drive the ancillary species with a pulse such that superatoms of size $S_\alpha$ acquire a phase $\phi_{\alpha} \equiv -4\tau J_\alpha $, while superatoms of size $\neq S_\alpha$ effectively acquire no geometrical phase (more precisely, they acquire a phase $2\pi n$ for $n\in\mathbb{Z}$). This therefore realizes $CZ(-4\tau J_\alpha)$ on the bonds of orientation $\alpha$, and an identity on all the other bonds. 
    \item We apply $R_\alpha^\dag$ to the data species. 
\end{enumerate}
To understand how $\mathcal{U}_{KH}$ emerges, note that, for any $\phi$,
\begin{equation}
    \Big(R_\alpha^\dag\otimes R_\alpha^\dag \Big) CZ(4\phi) e^{-i\phi \sigma^Z} \otimes e^{-i\phi \sigma^Z} \Big( R_\alpha \otimes R_\alpha \Big) = e^{i\phi \sigma^\alpha\otimes\sigma^\alpha},
\end{equation}
which can be derived from Eq.~\eqref{eq: CZ_rewriting} and the fact that, by definition, $R_\alpha^\dag \sigma^Z R_\alpha = \sigma^\alpha$. By choosing $\phi = -4\tau J_\alpha$ in the expression above, it is then straightforward to recognize that the three steps above implement the $e^{-i\tau H_{\alpha\alpha}}$. We also note that the scheme above can be simplified, e.g., by merging sub-step $3$ of one step with sub-step $1$ of the following step [which corresponds to merging two consecutive layers of single-qubit gates in Fig.~\ref{fig_examples}(c)]. 

\subsubsection{Digitization of arbitrary local Hamiltonians}\label{sec: Digitization of arbitrary local Hamiltonians}

Finally, we consider the generic digitization of local nearest-neighbor Hamiltonians. More precisely, as starting point, consider a $2-$body Hamiltonian of the form
\begin{equation}
    H = \sum_{\langle j,k\rangle} h_{j,k}.
\end{equation}
We assume that the model is defined on a locally-connected lattice, and that $h_{j,k}$ is symmetric ($h_{j,k}=h_{k,j}$) and the same on every bond. The simplest digitization of $H$ is a Trotter-type decomposition: an update rule $\mathcal{U}$ which, given a minimal step $\tau$ and an error $\varepsilon$, is $\varepsilon -$close to the evolution under $H$ for a time $\tau$,
\begin{equation}\label{eq: trotter error}
    \Big|\Big | \mathcal{U} - e^{-i\tau H} \Big|\Big| \leq \varepsilon,
\end{equation}
where $||\cdot||$ denotes the conventional operator norm. Then, the repeated application of $\mathcal{U}$ simulates evolution under $H$. However, as we discuss below, our digitization can also realize more generic discrete local dynamics derived from $H$, even way beyond such Trotterization. \\
\indent To construct $\mathcal{U}$, first we decompose
\begin{equation}
    h_{j,k} = \sum_{\alpha=X,Y,Z} \Big(c_\alpha \sigma_j^\alpha \sigma_k^\alpha + \frac{1}{2}h_\alpha(\sigma_j^\alpha +\sigma_k^\alpha)  \Big), 
\end{equation}
which can always be done by symmetry.\footnote{This can be derived, using that under adjoint local rotation $u$ the Pauli matrices transform as $\sigma^\alpha \to u\sigma^\alpha u^\dagger  = \sum_{\beta=X,Y,Z} R_{\alpha,\beta}\sigma^\beta$ where $R$ is a symmetric orthogonal matrix.} This allows us to rewrite the Hamiltonian as $H=\sum_\alpha H_\alpha$, where 
\begin{equation}
    H_\alpha = c_\alpha \sum_{\langle j,k\rangle} \sigma_j^\alpha\sigma_k^\alpha + h_\alpha \sum_j\sigma_j^\alpha.
\end{equation}
Then, we target the following digitization:
\begin{equation}\label{eq:trotter_single_step_kitaev}
    \mathcal{U}_H = e^{-i\tau H_Z} e^{-i\tau H_Y}e^{-i\tau H_X}.
\end{equation}
For small $\tau$, this realizes Eq.~\eqref{eq: trotter error} with error $\varepsilon = O(N\tau^2)$, where $N$ is the system size~\cite{childs2021theory}. Thus, it enables standard Trotterized simulation~\cite{lloyd1996universal}. However, as said, our implementation can realize $\mathcal{U}_H$ for arbitrary $\tau$. \\
\indent In our architecture, we implement $\mathcal{U}_H$ as shown in Fig.~\ref{fig_examples}(b). We arrange the data atoms on the vertices of the given lattice, and dispose elementary gadgets on the bonds (i.e., we do not need superatoms). The implementation proceeds in three steps, labeled by $\alpha\in\left\{X,Y,Z\right\}$, wherein step $\alpha$ comprises the following sub-steps:
\begin{enumerate}
    \item We apply a pulse $R_\alpha$ on the data species, mapping the $Z$ basis to the $\alpha$ basis; e.g., $R_X =e^{-i\pi \sigma^Y/4}$.
    \item We apply $e^{-i\tau(gc_\alpha+h_\alpha) \sigma^Z}$ on the data species, where $g$ is the graph degree of the lattice.
    \item We drive the ancillas to activate the gadgets and implement $CZ(-4c_\alpha\tau)$ at the bonds.
    \item Finally, we apply the inverse of step $1$, i.e. $R_\alpha^{-1}$. 
\end{enumerate}
It is straightforward to recognize that this indeed leads to the implementation of $\mathcal{U}_H$, with step $\alpha$ realizing exactly $e^{-i\tau H_\alpha}$. In practice, the sub-steps $1$ and $2$ above can be merged in a single pulse with sub-step $4$ of the previous step. Overall, this therefore requires $6$ species-alternated laser pulses to implement one application of $\mathcal{U}_H$.  

\subsection{Summary of concepts}

We now briefly summarize our construction in light of the examples above, which concretely capture all the ingredients. To this end, we identify two different classifications of the dynamical models, which highlight how generic features impact our implementations.\\
\indent On the one hand, as explicitly done above, we can divide the models in synchronous and asynchronous. Specifically, recall that $\mathcal{U}$ is synchronous if, when written as a discrete local circuit, the interacting local updates commute, and therefore can be applied simultaneously. Such characterization has a fundamental consequence (not specific to our method): if $\mathcal{U}$ is asynchronous, several interacting layers are needed to implement it.  \\
\indent On the other hand, we can divide the models in \emph{homogeneous} and \emph{inhomogeneous}. In this case, the main consequence is specific to our implementation, which relies on global drivings: decorated gadgets are needed in order to break the symmetry and render the realized dynamics spatially inhomogeneous. \\
\indent Importantly, we stress that these two characterizations are orthogonal: whether a dynamics is asynchronous is not fundamentally linked to its homogeneity. This is exemplified in Sec.~\ref{sec: Digitization of arbitrary local Hamiltonians}, where the model is homogeneous, yet asynchronous, as in general $[h_{i,j},h_{j,k}]\neq 0$. In practice, the circuit depth of our implementation is linked exclusively to the fundamental parallelizability of the model, while the space overhead (i.e., the superatom sizes) is related exclusively to its homogeneity.   

\section{Observing quantum chaos}\label{sec: Observing quantum chaos}

Quantum chaos is a important concept in many-body physics, encompassing, e.g., information scrambling~\cite{xu2024scrambling}, eigenstate thermalization~\cite{d2016quantum}, and the emergence of random matrix theory~\cite{bohigas1984characterization,haake_quantum_1991,gutzwiller2013chaos, kos2018many}. Generic coherent quantum many-body systems tend to naturally display chaotic dynamics; however, diagnosing such behavior is generally experimentally challenging, requiring the measurement of nontrivial quantities such as OTOCs~\cite{shenker2014multiple,shenker2014black,roberts2015localized}. \\
\indent In this section, we detail an approach to chaos detection that, by extracting coarse-grained information on the operator dynamics, allows us to discriminate different dynamical regimes of the QCA studied above. Importantly, our approach requires minimal experimental tools compared, e.g., to measuring OTOCs~\cite{joshi2020quantum,green2022experimental,braumuller2022probing,garttner2017measuring}. Moreover, such tools have already been demonstrated in dual-species globally-driven atom arrays~\cite{anand2024dual}. \\
\indent In the following, we will denote by $U(t)$ a generic unitary evolution operator. In our discretized setting, $t$ would be discrete, and $U(t)$ would describe $t$ applications of $\mathcal{U}$. However, we remark that all the results below do not assume the discretization of the dynamics, and also hold for any analog evolution.

\subsection{Coarse-grained signatures of quantum chaos}

In a chaotic quantum system, small perturbations rapidly spread across the system, redistributing local information over non-local degrees of freedom~\cite{xu2024scrambling}. This phenomenon can be quantified in terms of the evolution of local observables~\cite{shenker2014multiple,roberts2015diagnosing,roberts2015localized,maldacena2016bound,nahum2018operator,keyserlingk2018operator}: under chaotic dynamics, an initially localized operator rapidly increases its support over the whole system. Such an increase of the operator support can be probed by OTOCs and other measures~\cite{nahum2018operator,vonKeyserlingk2018operator,hosur2016chaos,qi2019measuring,fisher2023random}. Below, we consider a quantity that was originally introduced in Ref.~\cite{qi2019measuring} and can be seen as a coarse-grained signature of operator spreading. \\
\indent Specifically, for an observable $O$, with ${\rm Tr}[O]=0$ and ${\rm Tr}[O^2]=2^N$, we are interested in the quantity~\cite{qi2019measuring} 
\begin{equation}\label{def_g}
    g^O(t) = \mathbb{E}_\mu\Big[ \langle \psi | O(t) | \psi \rangle ^2\Big],
\end{equation}
where we set $O(t)=U^\dag(t) O U(t)$. Here, $\ket{\psi}=\ket{\psi_1,\dots,\psi_N}$ is a random \emph{product} state on $N$ qubits, and $\mathbb{E}_\mu[\cdot]\equiv\int d^N\mu(\psi)[\cdot]$ is the uniform average over such states - i.e., according to the measure
\begin{equation}
    d^N\mu(\psi_1,\dots,\psi_N) \equiv d\mu_H(\psi_1)\dots d\mu_H(\psi_N),
\end{equation}
where $d\mu_H(\psi)$ denotes the Haar measure on the single-qubit Hilbert space~\cite{mele2024introduction}. That is, $d\mu(\psi_1,\dots,\psi_N)$ is the uniform distribution of \emph{factorized} $N-$qubit states.\\

\indent We now elaborate on how $g^O(t)$ provides information on chaos~\cite{qi2019measuring}. First,  we recall that we may always write
\begin{equation}
    O(t) = \sum_{P\in\mathcal{P}} c_P(t) P,
\end{equation}
where $\mathcal{P}=\left\{\id, \sigma^X, \sigma^Y, \sigma^Z\right\}^{\otimes N}$ is the set of all Pauli strings, which form a complete basis for the space of $N$-qubit operators. Then, clearly the coefficients $c_P$ contain information about the spreading of $O(t)$ across the system. In particular, we can capture the increase of its support through its \emph{operator size distribution}, defined as 
\begin{equation}\label{eq:distribution_formula}
    p_\ell^O(t) = \sum_{P:|S(P)|=\ell} |c_P(t)|^2\,,
\end{equation}
where $|S(P)|$ denotes the number of qubits $P$ is supported on. Since $\sum_\ell p_\ell^O(t) = 1$, $p_\ell^O(t)$ can be interpreted as the probability that $P$ has support on $\ell$ sites. The connection to the quantity~\eqref{def_g} is given by the identity~\cite{qi2019measuring} 
\begin{equation}\label{chaos_equivalence}
    g^O(t) = \sum_{\ell = 0}^N \frac{p^O_\ell(t) }{3^\ell}. 
\end{equation}
Thus, by measuring $g^O(t)$, we gain useful - though incomplete - information on the operator spreading. \\
\indent More specifically, if the dynamics is chaotic, from Eq.~\eqref{chaos_equivalence} it follows that $g^{O}(t)$ obeys the late-time limit
\begin{equation}\label{eq:limit_haar}
    \lim_{t\rightarrow\infty} g^O(t) = \frac{1}{2^N+1}.
\end{equation}
This can be proved straightforwardly by noting that, for chaotic dynamics, $O(t)$ must approach a traceless random operator, resulting in $p_\ell^O(t\rightarrow\infty) = (1-4^{-N})^{-1}\binom{N}{\ell}(3/4)^\ell (1/4)^{N-\ell}$ for $\ell\neq 0$, and $p_0=0$; then, the limit above follows directly from Eq.~\eqref{chaos_equivalence}.\\
\indent Importantly, as we elaborate in Sec.~\ref{sec:numerical_analysys}, the behavior above is a specific feature of quantum chaos, and can serve to effectively distinguish chaotic and integrable dynamics. Its rather simple form makes it appealing as an experimental benchmark to probe chaotic regimes. \\
\indent Beyond the asymptotic behavior above, under chaotic dynamics one can estimate the scaling of $g^O(t)$ at finite times as follows. Let us specialize to the case when $O$ is local - e.g., supported on one single qubit. The key observation is that, for discrete local dynamics, the support of $O(t)$ only spreads inside a light cone propagating at a finite velocity $v$. More precisely, $O(t)$ can only have support on $A(t)\sim(vt)^D$ qubits, where $D$ is the dimension of the lattice. In the chaotic regime, the probability distribution of Pauli strings in $O(t)$ approaches that of a traceless random operator inside the light cone~\cite{nahum2018operator}, yielding $p_\ell(t)\simeq \left(1-4^{-A(t)}\right)^{-1} \binom{A(t)}{\ell} (3/4)^\ell (1/4)^{A(t)-\ell}\Theta(A(t)-\ell)$ for $\ell \neq 0$, and $p_0(t)=0$; here, $\Theta(\cdot)$ denotes the Heaviside step function. Thus, at finite times one expects \begin{equation}\label{eq: finite time decay}
    g^O(t) \sim 2^{-v^Dt^D} \;\;\; \text{for} \;\;\; vt\lesssim N^{1/D}.
\end{equation} 
For chaotic dynamics, $g^O(t)$ therefore decays exponentially in time (possibly after an initial small transient), approaching the stationary value in Eq.~\eqref{eq:limit_haar} as $A(t) \simeq N$, when information is scrambled across the entire lattice. \\
\indent We remark that such finite-time exponential decay of $g^O(t)$, alone, cannot unambiguously establish chaotic behavior (see Sec.~\ref{sec:numerical_analysys}). However, deviations from Eq.~\eqref{eq: finite time decay} \emph{do} signal the absence of quantum chaos: e.g., in many-body localized phases~\cite{abanin2019colloquium} the logarithmic growth of light cones~\cite{huang2017out,fan2017out,zhou2017operator} yields a polynomial decay of $g^{O}(t)$.

\begin{figure*}[t!]
    \begin{minipage}{1.0\textwidth}
        \includegraphics[width=\textwidth]{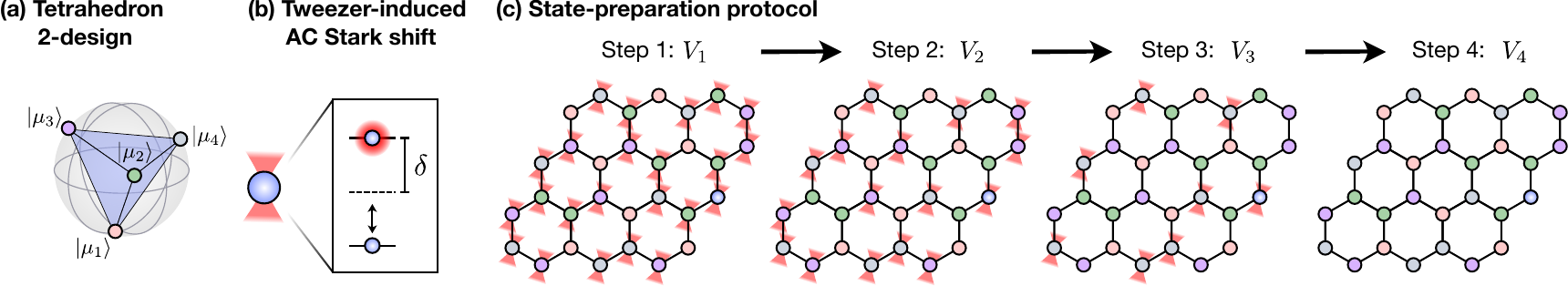}        
    \end{minipage}%
\caption{(a) A single-qubit $2-$design can be realized through four states $\mathcal{E}_T=\left\{\ket{\mu_1}, \ket{\mu_2}, \ket{\mu_3}, \ket{\mu_4}\right\}$ placed at the vertices of a tetrahedron in the Bloch sphere. (b) An optical tweezer can be used to shift the addressed atoms out of resonance by a detuning $\delta \gg \Omega$ (AC Stark shift). In this way, an atom initialized in $\ket{g}$ cannot be excited to $\ket{r}$ by the driving laser while addressed by the tweezer. (c) Our protocol to initialize the system in a randomly chosen state $\ket{\psi}\in\mathcal{E}_T^N$. The protocol begins by identifying $\ket{\psi}$ of Eq.~\eqref{eq: coloring decomposition} with a coloring of the lattice vertices, and initializing all the atoms in $\ket{g}$. Then, we proceed in four steps, wherein at step $k$ we use the global Rydberg laser to implement the unitary $V_k = U^\dag_{k+1}U_k$ uniformly on all the atoms, where $U_k\ket{g}=\ket{\mu_k}$. Crucially, during the steps we progressively turn off the tweezer light, such that more atoms are addressed. At the end of the protocol, each atom is prepared in the desired tetrahedron state, according to the initial coloring. }
\label{fig_protocol}
\end{figure*}

\subsection{Measuring signatures of quantum chaos in neutral atom experiments}
\label{sec:measuring_details}

We now introduce a method for measuring $g^O(t)$, which is compatible with simple, experimentally demonstrated capabilities~\cite{ExperimentalPaper}. From Eq.~\eqref{def_g}, it may seem that extracting $g^O(t)$ requires one to initialize the system in a fully disordered state, which can be prohibitive in globally-driven experiments. Our method gets around this by combining two ingredients, detailed below: First, we use standard properties of the Haar measure to reduce the initial state disorder; Second, we exploit this simplification to design an experimentally feasible protocol.  

\subsubsection{Minimal 2-design and formal measurement protocol}\label{sec:formal_measurement}

As mentioned, we need to initialize the system in a random state $\ket{\psi}=\ket{\psi_1,\dots,\psi_N}$, where the $\ket{\psi_i}$ are independently and identically Haar-distributed. This means that $\ket{\psi}$ features an extensive amount of disorder (i.e., the $\ket{\psi_i}$ are all different). Such extensively disordered states cannot be efficiently prepared with global controls. \\
\indent Crucially, however, we do not need to engineer the entire single-qubit Haar distribution: in fact, we only need to reproduce its second moment. This is achieved by a $2-$design~\cite{mele2024introduction}: an ensemble $\mathcal{E}=\left\{\ket{\mu_k}\right\}$ of (single-qubit) states such that $\sum_k\ket{\mu_k}\bra{\mu_k}=\mathbb{E}_H\left[\ket{\psi}\bra{\psi}\right]$ and
\begin{equation}\label{Harr_second_moment}
    \sum_k \ket{\mu_k, \mu_k}\bra{\mu_k, \mu_k} = \mathbb{E}_H\big[\ket{\psi, \psi}\bra{\psi, \psi}\big], 
\end{equation}
where $\mathbb{E}_H[\cdot]=\int d\mu_H(\psi)[\cdot]$ denotes the single-qubit Haar average. While many such ensembles exist, we are interested in the minimal one, $\mathcal{E}_T=\left\{\ket{\mu_1}, \ket{\mu_2}, \ket{\mu_3}, \ket{\mu_4}\right\}$, featuring $|\mathcal{E}_T|=4$ states~\cite{renes2004symmetric}. In the Bloch sphere, these are distributed at the vertices of a tetrahedron [Fig.~\ref{fig_protocol}(a)]; setting $\vartheta = \cos^{-1}(1/\sqrt{3})$ and $\varphi=\pi/4$, they are, explicitly, $\ket{\mu_{1,2}} = \cos(\vartheta/2)\ket{0} \pm e^{i\varphi}\sin(\vartheta /2)\ket{1}$ and $\ket{\mu_{3,4}} = \sin(\vartheta/2)\ket{0} \pm e^{-i\varphi}\cos(\vartheta /2)\ket{1}$. It can be straightforwardly checked that Eq.~\eqref{Harr_second_moment} holds for $\mathcal{E}_T$. We provide further details on $\mathcal{E}_T$ in Appendix \ref{appendix: moments}.\\
\indent Then, it is easy to show that the following protocol provides an unbiased estimator of $g^O(t)$: 
\begin{enumerate}
    \item For each qubit $i=1,\dots,N$, choose randomly a state $\ket{\psi_i}\in\mathcal{E}_T$, and set $\ket{\psi} = \ket{\psi_1,\dots,\psi_N}$. Then, extract an estimation of $\langle \psi | O(t) | \psi\rangle^2$ through $m$ measurements of $O$ after applying $U(t)$ to $\ket{\psi}$.
    \item Repeat the step above for $M$ different choices of $\ket{\psi}$, and extract an estimation of $\mathbb{E}_\mu[\langle \psi | O(t) | \psi\rangle^2]$ by uniformly averaging over the $M$ steps.  
\end{enumerate}
The advantage, here, is that the amount of initial disorder is not extensive anymore: still $\ket{\psi}$ is not homogeneous, but now each qubit is drawn from a small set of four states. Below, we leverage this simplification to design a protocol measuring $g^O(t)$. 

\subsubsection{Physical implementation with globally-driven Rydberg atom arrays}

We now show how to implement the protocol above, which requires preparing states from the factorized ensemble $\mathcal{E}_T^N$, with current experimental capabilities. First, we identify four unitaries $U_k$ for $k=1,\dots,4$, such that $U_k\ket{g}=\ket{\mu_k}$, and we define $V_k = U^\dag_{k+1}U_k$, with $V_4\equiv U_4$. \\
\indent Next, we rewrite the random state $\ket{\psi}\in\mathcal{E}_T^N$ as 
\begin{equation}\label{eq: coloring decomposition}
    \ket{\psi} = \bigotimes_{i=1}^4\bigotimes_{k\in A_i}\ket{\mu_k},
\end{equation}
where $\left\{A_1, A_2, A_3, A_4\right\}$ is a $4-$coloring of the lattice sites. As illustrated in Fig.~\ref{fig_protocol}(c), this corresponds to a $4-$partition of the data species: choosing a random $\ket{\psi}\in\mathcal{E}_T^N$ is equivalent to sampling a lattice coloring. \\
\indent Given one such coloring, we first initialize all the atoms in $\ket{g}$. Then, our protocol proceeds in four steps, as illustrated in Fig.~\ref{fig_protocol}(c). Specifically, at step $k$, we generate a pattern of focused optical tweezers, addressing all the atoms in coloring $A_{j>k}$. With the tweezers on, with the global Rydberg laser we uniformly apply $V_k$ on the array.\\
\indent Crucially, the tweezer light provides a strong AC Stark shift to the addressed atoms~\cite{ExperimentalPaper}, effectively shifting the transition $\ket{g}\leftrightarrow\ket{r}$ out of resonance with respect to the Rydberg laser [Fig.~\ref{fig_protocol}(b)]. In practice, during the driving, the tweezer-addressed atoms cannot evolve, and their dynamics is frozen - while the others are free to evolve. \\
\indent It is then easy to see that an atom in $A_k$ undergoes the unitary evolution $V_4\dots V_k = U_k$ (as it was frozen for $V_{j<k}$). Since it was initialized in $\ket{g}$, it therefore ends up in $U_k\ket{g}=\ket{\mu_k}$, as desired. We therefore end up with all the atoms prepared in the single-qubit state prescribed by Eq.~\eqref{eq: coloring decomposition}, thus preparing the desired initial state $\ket{\psi}$.\\
\indent After this state-preparation step, the protocol for extracting $g^O(t)$ is straightforward. Specifically we use the methods above to apply $U(t)$ to $\ket{\psi}$, and finally measure the observable $O$: as discussed in the Sec.~\ref{sec:formal_measurement}, repeating this $m$ times allows us to get an estimate of $\langle \psi |O(t)|\psi\rangle ^2$, and using $M$ different choices of $\ket{\psi}$ provides an estimate of $g^O(t)$. \\
\indent We conclude this section with three remarks on the protocol above. First, we note that the key ingredient (strong tweezer-induced AC Stark shifts) has been recently implemented in a globally-driven dual-species experiment~\cite{ExperimentalPaper}: in fact, it has been used for initializing the system according to a $2-$coloring. Extension to $4-$colorings to fully leverage our protocol above is straightforward, only requiring to reprogram the tweezers. Second, we remark that, with state-of-the-art capabilities, such technique cannot be used to achieve mid-circuit local control: atoms would decohere both due to strong fluctuations in the tweezer-induced detuning, as well as due to errors during the ramping of the tweezers, which is typically slow compared to the timescales of ground-Rydberg qubits. Moreover, e.g. for Alkali atoms, the tweezers would expel atoms in the Rydberg state, in fact measuring them in the $\left\{\ket{g}, \ket{r}\right\}$ basis. Third, we remark that our protocol enables efficiently estimating $g^O(t)$ through global controls as long as $O$ is \emph{uniform} across its support: that is, if it is of the form $O=\mathcal{O}_{q_1}\otimes \mathcal{O}_{q_2}\otimes \dots \otimes \mathcal{O}_{q_{|S(O)|}}$ for a single-qubit observable $\mathcal{O}$ and a set of qubits $q_1,\dots, q_{|S(O)|}$.

\subsection{Observing quantum chaos: numerical analysis}
\label{sec:numerical_analysys}

Here, we numerically analyze the method above to diagnose quantum chaos. We show that, in some relevant regimes of the considered QCA, $g^O(t)$ is able to distinguish between integrable and chaotic regimes.

\begin{figure}
	\includegraphics[width=1\linewidth]{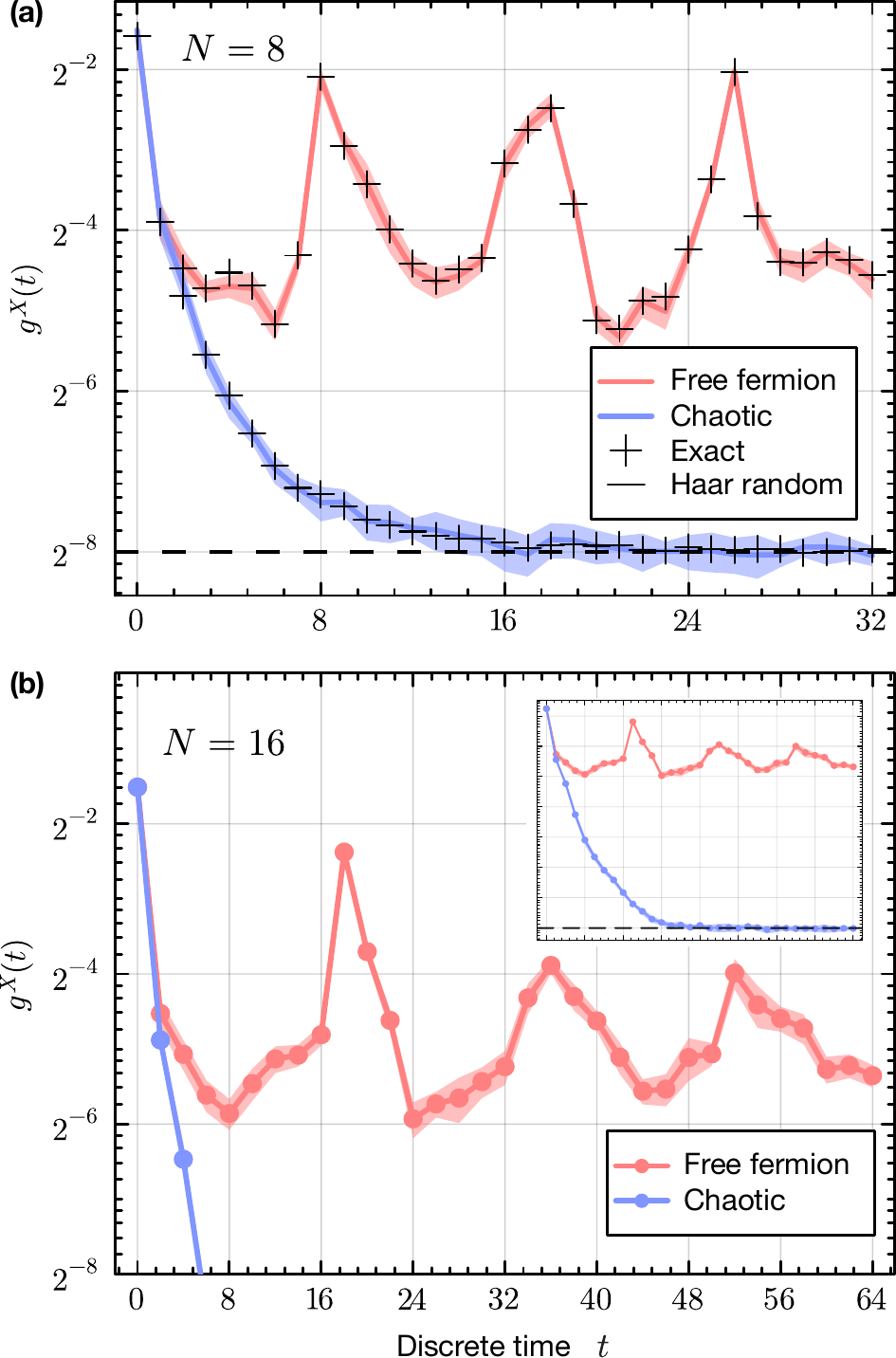}
           
	\caption{Generating function $g^X(t) =: g^O(t)$ for the Pauli matrix $O = \sigma^X_{N/2}$, for the $1$D kicked Ising models on $N=8$ sites (a) and $N=16$ sites (b) with open boundary conditions. The data are obtained from simulating the QCA dynamics, sampling over $1000$ initial random configurations. The curves correspond to the KIM with parameters $J=1$, $b=0.8$, and with $h=1.2$ (chaotic) and $h=0$ (free fermion). The parameters are defined in Eqs.~(\ref{eq:ki_u}-\ref{eq:h_K}) (with $\tau=1$). The dashed line is the late-time limit described in Eq.~\eqref{eq:limit_haar}. The shaded region indicates the uncertainty of the sample variance, as explained in  Appendix~\ref{sec:numerics_simulation}. The exact values for $N=8$  are obtained by calculating the exact operator size distribution $p_l$ in Eq.~\eqref{eq:distribution_formula}, of the time-evolved operator, which then determines $g^X(t)$ via~\eqref{chaos_equivalence}.
    In Fig.~\ref{fig:hist_n8} in the Appendix~\ref{sec:numerics_simulation}, we show the distribution of sampled values for $N=8$, for certain times.}
	\label{fig:KIM_1D}
\end{figure}

We start by studying the $1$D kicked Ising model discussed in Sec.~\ref{sec:kicked_ising}. In Fig.~\ref{fig:KIM_1D} we report the evolution of $g^O(t)$, where we choose $O=\sigma^X_{N/2}$. To obtain the numerical data, we have simulated the experimental protocol discussed above, sampling over $N_S=1000$ initial random configurations (cf. Appendix~\ref{sec:numerics_simulation} for details). Simulations are shown for $N=8$ and $N=16$ sites. For $N=8$, we additionally calculate the operator size distribution directly from the time evolved initial observable, confirming that Eq.~\eqref{chaos_equivalence} holds indeed.\\
\indent We consider two sets of the parameters $\left\{J, h, b\right\}$ defined in Eqs.~\eqref{eq:h_I} and~\eqref{eq:h_K}. One set corresponds to a chaotic point, while the other one to a `free fermionic' point. In the latter case, the system is integrable and can be mapped to a solvable model via a Jordan-Wigner transformation~\cite{jordan1993}. From Fig.~\ref{fig:KIM_1D}, we see that the dynamics of $g^O(t)$ is substantially different for the two sets of parameters. In particular, in the free-fermionic case $g^O(t)$ shows clear deviations from the expected chaotic behavior of Eq.~~\eqref{eq:limit_haar}, remaining much larger in magnitude and showing strong revivals - a hallmark of non-ergodicity.\\
\begin{figure}
	\includegraphics[width=1\linewidth]{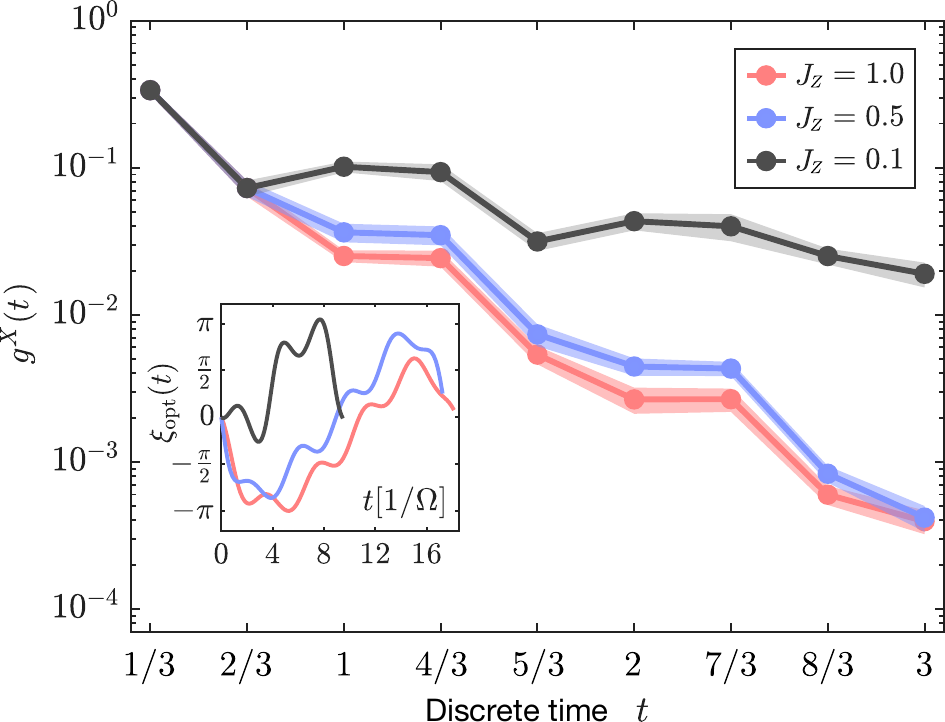}
           
	\caption{Generating function $g^O(t) =: g^X(t)$ of the Pauli matrix $O =\sigma^X$, for the $2$D Kitaev Floquet system for $J_X=J_Y =1$, $h_X=h_Y=h_Z =1$. The parameters are defined in Eqs.~(\ref{eq:kitaev_hamiltonian_0}-\ref{eq:kitaev_hamiltonian}) (with $\tau=1$). Note that we also include intermediate time steps as explained in Appendix~\ref{sec:numerics_simulation}. Thus, the time frame shown corresponds to three full cycles. We take the system size to be larger than the operator size at any time. The values and uncertainty region (shaded regions) for the two curves are calculated from 1000 samples, as described in Appendix~\ref{sec:numerics_simulation}.
    The inset shows the time-optimal pulse sequences which realizes the mediated gate of the $Z$ step, for the different values of $J_Z$. In the inset, $t$ represents the physical time and is expressed in units of $1/\Omega$. See Appendix~\ref{sec:optimal_control} for further details on the numerical procedure.}
	\label{fig:kitaev2D}
\end{figure}
\indent Next, we consider the Floquet Kitaev honeycomb model introduced in Sec.~\ref{sec:kitae_qca}. Here, numerical simulations are more challenging compared to the 1D case, and we only obtained reliable data for the first $\sim 10$ sub-steps - restricting our analysis to the behavior of $g^O(t)$ at early times, corresponding to $\sim3$ applications of $\mathcal{U}_{KH}$. In such not asymptotic regimes, we note that the initial behavior of $g^O(t)$ is already able to provide useful information about the QCA ergodicity properties. We show this in Fig.~\ref{fig:kitaev2D}, where we display data obtained by simulating the Kitaev QCA in qualitatively different regimes. Specifically, we consider sets of the parameters $\left\{J_X, J_Y, J_Z\right\}$ interpolating from a quasi-1D regime ($J_Z\ll J_X, J_Y$, suppressing the dynamics in one dimension) to a fully 2D one ($J_X\sim J_Y \sim J_Z$). In the quasi-1D case, the space accessible to the local operator spreading is effectively suppressed, and its support grows more slowly in time, compared to a genuine 2D dynamics. This is clearly visible from the behavior of the generating function $g^O(t)$ reported in Fig.~\ref{fig:kitaev2D}, even at the very small times we are able to simulate. While the late-time dynamics remains unexplored, this demonstrates that, for the Kitaev Floquet QCA, $g^O(t)$ is capable of capturing dynamical features and distinguishing different regimes also for relatively shallow experimental protocols.\\ 
\indent Finally, we also report that the early-time behavior of $g^O(t)$ is not always able to distinguish between chaotic and non-ergodic phases, displaying an initial exponential decay even for non-ergodic dynamics. We show this explicitly in Appendix~\ref{sec:numerics_simulation}: therein, we simulated the Kitaev honeycomb QCA at a point in the parameter space where it reduces to a Clifford circuit~\cite{nielsen2010quantum,gottesman1998theory}, corresponding to a non-ergodic evolution. There, at short times one observes an even faster decay of $g^O(t)$ than in the chaotic dynamics (such decay is followed by revivals). In this case, therefore, only the late-time behavior of $g^O(t)$ can unambiguously distinguish the chaotic regime.\\
\indent In summary, the results presented in this section can be viewed both as a benchmark of our proposed method, as well as a motivation for the experimental implementation of our constructions. In particular, in two dimensions we were not able to simulate the dynamics beyond small times, and an experimental implementation would allow us to perform a more thorough study of quantum chaos in 2D QCA dynamics.

\section{Discussions and outlook}\label{sec: Discussion and outlook}

In this work, we have introduced a method for engineering discrete local dynamics in dual-species atom arrays, based solely on global drivings. This shows that such platform, at the control level, can operate in a very efficient analog mode, while nevertheless inducing an effectively discretized, circuit-like evolution. In fact, our results pave the wave for the exciting near-term opportunity to study complex many-body dynamics in large atom arrays, beyond the native Hamiltonian, potentially in classically hard regimes. This can now be achieved with minimal experimental control requirements.\\
\indent We now discuss errors in our architecture. To this end, we note that our protocol does not suffer of major qualitative scaling disadvantages. Indeed, our implementation overhead is constant in both space and time: $O(N)$ atoms are sufficient to represent a model defined on $N$ qubits, and $T$ discrete steps are implemented in $O(T)$ physical time. In practice, our architecture is limited by physical constraints such as non-ideal blockade regimes. Such analog imperfections could be mitigated by exploiting asymmetric dual-species interactions to reduce the unwanted interactions~\cite{beterov2015rydberg, cesa2023universal, anand2024dual}. Moreover, in certain cases, small violations from the ideal regime could still allow us to observe the target physics at relevant times: this would be the case when the dynamics features prethermalization effects~\cite{moeckel2008interaction,rosch2008metastable,marcuzzi2013prethermalization,bertini2015prethermalization,brandino2015glimmers}, with the emergence of an intermediate time scale where the dynamics is dominated by the unperturbed Hamiltonian.\\
\indent To evaluate the near-term feasibility of our protocols, we note that globally-driven dual-species 1D arrays have been recently demonstrated with up to $35$ atoms, reporting mediated gates with fidelities up to $\sim 96.7\%$ (see Ref.~\cite{ExperimentalPaper}). Improved fidelities are possible, e.g., by exploiting already observed F\"orster resonances~\cite{beterov2015rydberg, anand2024dual}, and larger 2D arrays can be envisioned. Moreover, high-fidelity superatom operations are possible as shown in Ref.~\cite{evered2023high}, where the two- and three-atom blockade mechanism underlying our decorated gadgets was exploited for high-fidelity multi-qubit gates. \\ 
\indent Future directions may explore various generalized versions of our construction. For instance, our approach could also be combined with tools from quantum computing architectures. While here we have focused on static atom traps, one can envision hybrid operation modes, e.g., where globally-driven steps are interleaved by layout reconfigurations. Another interesting direction is to exploit the capability of dual-species experiments to measure one species without affecting the other~\cite{singh2023mid}, to introduce mid-circuit measurements in the evolution. One approach would be to reconsider our gadget constructions, letting the ancillas being entangled with the data atoms, to then measure them and project the data state. Potentially, this could allow one to realize generic circuit-like dynamics with measurements and feedback, e.g., to study measurement-induced phase transitions~\cite{skinner2019measurement,li2018quantum,potter_entanglement_2022,noel2022measurement,koh2023measurement}. More generally, the atoms at the bonds can be promoted from catalysts to data qubits. As an outlook consideration, we note that realizing proper local constraints at vertices and bonds could allow for the realization of lattice gauge theories~\cite{RevModPhys.51.659, surace2020lattice} within our setting. At a more basic level, we remark that even in our present proposal, introducing measurements on the ancillas would enable to remove errors, in fact implementing a simple form of error mitigation~\cite{cesa2023universal}. Finally, while in this work we have focused on implementing discretized dynamical models, a natural direction is to understand if the ideas presented here can allow one to implement QCA models serving as quantum computers~\cite{farrelly2020review, arrighi2019overview, raussendorf2005quantum} . \\
\\
\noindent \emph{Note added.} This manuscript is submitted simultaneously with a closely related experiment on Quantum Cellular Automata in globally-driven dual-species arrays~\cite{ExperimentalPaper}.

\section*{Acknowledgments}
We are grateful for insightful discussions with R. White and G. Giudici. LP and HP acknowledge hospitality from the Erwin
Schr\"odinger International Institute for Mathematics and
Physics (ESI) at the University of Vienna, during the workshop “Entangled matter out-of-equilibrium”, where part of
the research for this paper was performed. R. T. acknowledges support from the European Union’s Horizon Europe program under the Marie Sk\l{}odowska Curie Action
CODRA (Grant No. 101149823). The work of LP and DK was funded by the European Union (ERC, QUANTHEM, 101114881). We also acknowledge funding from the Office of Naval Research (N00014-23-1-2540), the Air Force Office of Scientific Research (FA9550-21-1-0209), Q-NEXT, a US Department of Energy Office of Science National Quantum Information Science Research Center, the European Union’s Horizon Europe research and innovation program under Grant Agreement No. 101113690 (PASQuanS2.1), the ERC Starting Grant QARA (Grant No. 101041435), and the Austrian Science Fund (FWF) (DOI 10.55776/COE1). Views and opinions expressed are however those of the author(s) only and do not necessarily reflect those of the European Union or the European Research Council Executive Agency. Neither the European Union nor the granting authority can be held responsible for them.

\newpage

\appendix

\section{Violations from the ideal blockade regime}
\label{sec:violations}

In our construction, each atom of one species should ideally interact only with its immediate neighboring atoms (which are chosen to be of the other species), according to the Hamiltonian~\eqref{eq: PXP hamiltonian}. In reality, each atom also experiences some `unwanted interactions', due to the tails of the interparticle potential. In this appendix, we provide a simple argument estimating the strongest among such unwanted interactions, thus justifying the ideal blockade-regime approximation.

Let us denote by $A$ and $B$ the data and ancillary species, respectively. Assuming that the interparticle interaction is dominant over the intra-particle one, the atom pairs experiencing the strongest unwanted interactions consist of one $A$ atom and the closest of the $B$ atoms which is not among its nearest neighbors. The distance between such atoms depends on the chosen lattice geometry, and in the following we will focus on the 2$D$ honeycomb lattice. In this geometry any $A$ atom has three neighboring $B$ atoms at a distance $d$; in addition it has $6$ $B$ atoms at a distance $\sqrt{7}d$, with which it experiences unwanted interactions. Therefore, assuming that the interparticle potential decays with the atomic separation $d_{i,j}$ as $V_{i,j}\sim d_{i,j}^{-6}$, we can estimate the ratio between the wanted and unwanted interactions as
\begin{equation}
\label{eq:ration_2D}
V_{\rm unwanted}/ V_{\rm blockade}\sim \frac{6}{3}\left(\frac{1}{\sqrt{7}}\right)^{6}\approx0.006\ll 1\,.
\end{equation}
Remarkably, the ratio~\eqref{eq:ration_2D} is smaller than  in the case  of established single-species 1D implementations of the PXP model~\cite{prosen2007chaos,prosen2002general}, for which  $V_{\rm unwanted}/ V_{\rm blockade}\simeq 0.0156$, thus justifying the Rydberg-blockade approximation. As mentioned, the physical setup described above refers to implementations on the 2D honeycomb lattice, but it extends straightforwardly to other geometries, including 2D square lattices and 1D arrangements. 

Some comments are in order. Above, we have estimated the unwanted interactions assuming that the interspecies interactions are dominant and using a van der Waals potential tail $V_{i,j}\sim d^{-6}_{i,j}$. However, while interspecies interactions may be indeed tuned via Förster resonances~\cite{anand2024dual}, this is a crude approximation. In the presence of two distinguishable types of atoms, the interspecies interaction features two regimes separated by a cross-over distance $R_c$. For distances $R$ smaller than $R_c$, the resonant interaction displays a $1/R^3$  decay, while for distances larger than $R_c$ the interactions decay becomes $1/R^6$~\cite{beterov2015rydberg,saffman2010quantum}. Therefore, a more quantitative estimation should take into account the full shape of inter-species potential. However, it is important to stress that we can choose the distance between the pair of atoms experiencing the unwanted interactions to be larger than $R_c$, so that their interaction can be approximated well by a van der Waals potential.

\section{Complete Controllability}\label{appendix: Controllability}
In this appendix, we discuss the controllability properties of a system of $S_{\rm max}$ superatoms of cardinality $S=1,\dots, S_{\rm max}$, evolving under the Hamiltonian ${H=\oplus_{S=1}^{S_{\rm max}}H_S}$, where 
\begin{equation}\label{eq: superatoms}
H_S(\xi)=\frac{\sqrt{S}\Omega}{2}\left[\exp(i\xi)\sigma^-_S + \exp(-i\xi)\sigma^+_S\right],   
\end{equation}
with $\sigma_S^{\pm}$ being the raising and lowering operators acting on the $S$-th superatom.
We show that this system is \emph{completely} controllable: for every choice of two product states $\ket{\psi^i}=\ket{\psi_1^i,\dots,\psi_{S_{\rm max}}^i}$ and $\ket{\psi^f}=\ket{\psi_1^f,\dots,\psi_{S_{\rm max}}^f}$, there exist a time $T>0$ and a time-dependent phase $\xi(t)$ (with $t\in[0,T]$), such that $\ket{\psi^f}=U(T)\ket{\psi^i}$, with ${U=\otimes_{S=1}^{S_{\rm max}}U_S(T)}$, where $U_S(T)=\mathcal{T}\exp\!\left(-\frac{i}{\hbar}\int_0^T H_S(t)\, dt\right)$. 

Due to the independent evolution of the superatoms, the propagator $U(t)$ evolves on the Lie group $G = SU(2)^{S_{\rm max}}$ and the Schr\"odinger equation generated by the Hamiltonian $H(\xi(t))$ defines a right-invariant control system on $G$, $\dot U(t) = -i H(\xi(t))\,U(t) = W(\xi(t))\,U(t)$ with $U(0)=\mathbb{I}$. The complete controllability of the system follows if the Lie algebra $\mathfrak{L}$ generated by the available control directions, i.e., by $\mbox{span}_{\xi\in\mathbb{R}}\left\{W(\xi)\right\}$, coincides with the full Lie algebra of the group $G$, namely $\mathfrak{g} := \bigoplus_{s=1}^{S_{\rm max}} \mathfrak{su}(2)$ (see \cite{Dalessandro2021}).  

To prove that $\mathfrak{L}=\mathfrak{g}$, the strategy is to show that the generators of $\mathfrak{g}$, i.e., the generators of the independent rotations, belong to $\mathfrak{L}$. Before proceeding to the proof, it is convenient to introduce the following operators $P_S=-i\sigma^x_S$, $Q_S=i\sigma^y_S$, and $R_S=i\sigma^z_S$, which verify the commutation relations $[P_S,Q_S]=2R_S$, $[Q_S,R_S]=2P_S$, $[R_S,P_S]=2Q_S$, and to introduce the following notation $W=\oplus_{S=1}^{S_{\rm max}} W_S=(W_1,\dots,W_{S_{\rm max}})$, where $W_S=a_S\left[\cos(\xi)P_S+\sin(\xi)Q_S\right]$, with  $a_S=\sqrt{S}\Omega/2$. Our goal is then to show that $\tilde{P}_S\equiv(0,\dots,0,P_S,0,\dots,0)$, $\tilde{Q}_S\equiv(0,\dots,0,Q_S,0,\dots,0)$, and $\tilde{R}_S\equiv(0,\dots,0,R_S,0,\dots,0)$ belong to $\mathfrak{L}$, for all $S=1,2,\dots,S_{\rm max}$. 

The first step of the proof consists in observing that the operators of the following form:
\begin{align}
W_P^{(2j-1)} &= (a_1^{2j-1}P_1,\dots,a_{S_{\rm max}}^{2j-1}P_{S_{\rm max}}) \nonumber\\
W_Q^{(2j-1)} &= (a_1^{2j-1}Q_1,\dots,a_{S_{\rm max}}^{2j-1}Q_{S_{\rm max}})
\end{align}
with $j\ge 1$, belong to $\mathfrak{L}$. They indeed can be obtained, up to a constant factor, from repeated commutators of two specific constant-phase operators: 
\begin{align}
&W_P^{(1)}=(a_1 P_1, a_2 P_2,\dots, a_{S_{\rm max}} P_{S_{\rm max}})=W(\xi=0)\nonumber\\ 
&W_Q^{(1)}=(a_1 Q_1, a_2 Q_2,\dots, a_{S_{\rm max}} Q_{S_{\rm max}})=W(\xi=\pi/2),
\end{align}
(for instance, $[W_Q^{(1)},[W_P^{(1)},W_Q^{(1)}]]=4W_P^{(3)}$).\\
The second step of the proof consists in showing that, for any $S=1,\dots,S_{\rm max}$, it exists a set of coefficients $\beta^{(S)}$, such that $\tilde{P}_S=\sum_{j=1}^{S_{\rm max}}\beta^{(S)}_jW_P^{(2j-1)}$ and $\tilde{Q}_S=\sum_{j=1}^{S_{\rm max}}\beta^{(S)}_jW_Q^{(2j-1)}$. Their commutators return then $\tilde{R}_S$. The coefficients $\beta^{(S)}$ are obtained by solving the following linear system ${M\cdot\beta^{(S)}=e_S}$. Here, $e_S$ is the the $S$-th element of the canonical basis and $M\in\mathbb{R}^{S_{\rm max}\times S_{\rm max}}$ has the following components $M_{S,j}=a_S^{2j-1}$, with $a_S>0$ by construction. From the properties of the Vandermonde matrices, it follows that $M$ is invertible \emph{iif} all $a_S$ are distinct from each other, i.e., iif all superatoms have different cardinality. Indeed, we have:
$\mbox{det}(M)=\prod_ia_i\prod_{1\le i<j\le S_{\rm max}}(a_i^2-a_j^2)$.

\section{Quantum optimal control} \label{sec:optimal_control}
In this appendix, we describe the optimal control method used to derive the time-optimal laser pulses discussed in Sec.~\ref{sec: quantum control}. Specifically, we show how to design a time-dependent global phase $\xi(t)$ to simultaneously drive $S_{\max}$ superatoms of different cardinality $S=1,\dots,S_{\rm max}$ (see Appendix \ref{appendix: Controllability}), from their ground state $\ket{G_S}$ to the target state $\ket{\bar{\psi}_S}\equiv e^{i\phi_S}\ket{G_S}$, with $\phi_S$ being specific target phases.  
In the PXP limit, each superatom evolves as an independent two-level system of enhanced Rabi frequency $\sqrt S\Omega$ and phase $\xi(t)$, as described in Eq.~\eqref{eq: superatoms}. Due to the different Rabi frequency enhancements, finding such a control function $\xi(t)$ is a non-trivial task~\cite{levine2019high} that requires numerical optimization strategies~\cite{jandura2022control, fromonteil2024HJB}. Moreover, among the possible solutions, time-optimal ones are preferred since they reduce errors caused by the finite lifetime of the Rydberg states~\cite{bluvstein2022quantum}, which are known to grow with the process duration~\cite{pagano2022error}. 

To accomplish this task, we resort to the Gradient Ascent Pulse Engineering (GRAPE) algorithm~\cite{khaneja2005GRAPE}, which has already been successfully applied in Rydberg platforms to design time-optimal gates as well as a broader range of control tasks~\cite{jandura2022control, ma2023high, evered2023high, giudici2025dipole, maskara2025programmable, Zeng2025AdiabaticEcho}. This method searches for a piecewise-constant approximation of the control function that minimizes a given error function encoding the target evolution. In formulas, for a fixed time-evolution $T$, we discretize the time-interval $[0,T]$ in small intervals over which the laser phase is assumed to be constant, and optimize over such constant values:  $\xi(t)=\xi(t_i)\equiv\xi_i$ for $t\in\left[t_i,t_{i+1}\right]$ with $i=1,\dots M$, $t_1=0$, and $t_M=T-dt$. As a cost function, we found numerically efficient to consider the following error:
\begin{equation}
\label{appendix:Err}
\mathrm{Err}(\boldsymbol{\xi})=\sum_{S=1}^{S_{\rm max}}\left|1- \braket{\bar{\psi}_S}{\psi_S(T)}\right|^2, \;\;\mbox{with}\;\boldsymbol{\xi}=\left(\xi_1,\dots,\xi_M\right).  
\end{equation}
Here, $\ket{\psi_s(T)}=U_{s,M}\dots U_{s,1}\ket{G_s}$ is the state of the $S$-th superatom evolved under the piecewise-constant control $U_{S,i}=\exp(-iH_S(\xi_i)dt)$.

The search for the minimum of $\mathrm{Err}(\boldsymbol{\xi})$ can be conducted efficiently by providing the following analytic expression of its gradient:
\begin{align}
\label{appendix:gradient}
\frac{\partial\,\mathrm{Err}(\boldsymbol{\xi})}{\partial \xi_i}
&= -2\,\Re\Biggl[\sum_{S=1}^{S_{\rm max}}
\bra{\bar{\psi}_S}\frac{\partial}{\partial \xi_i}\ket{\psi_S(T)}\nonumber\\
&\qquad\quad \times
\Bigl(1 - \braket{\bar{\psi}_S}{\psi_S(T)}\Bigr)^{*}
\Biggr].
\end{align}
where the superscript $\ast$ indicates complex conjugation. Here, 
\begin{align}
\bra{\bar{\psi}_S}\frac{\partial }{\partial \xi_i}\ket{\psi_S(T)}
&= \bra{\bar{\psi}_S}U_{S,M}\dots U_{S,i+1}\nonumber\\
&\times
\frac{\partial U_{S,i}}{\partial \xi_i}U_{S,i-1}\dots U_{S,1}\ket{\psi_S(T)},
\end{align}
and $\frac{\partial U_{S,i}}{\partial \xi_i}$ can be efficiently computed by approximating its integral expression:
\begin{align}
&\frac{\partial e^{V_S(\boldsymbol{\xi})}}{\partial \xi_i}=\int_0^1d\alpha e^{\alpha V_S(\boldsymbol{\xi})}\frac{\partial V_S(\boldsymbol{\xi})}{\partial \xi_i}e^{(1-\alpha)V_S(\boldsymbol{\xi})}\nonumber\\ &\mathrm{with}\:\:V_S(\boldsymbol{\xi})=-iH_S(\boldsymbol{\xi})dt,
\end{align}
using the 2-point Gauss quadrature formula $\int_0^1d\alpha f(\alpha)\approx [f(1/2-\sqrt{3}/6)+f(1/2+\sqrt{3}/6)]/2$. 

To perform the optimization, we choose $M=100$ and resort to MATLAB's fminunc optimizer with the quasi-Newton (BFGS) algorithm (setting $\mathrm{Tol X}=10^{-15}$, $\mathrm{TolFun}=10^{-15}$, $\mathrm{MaxIterations}=4000$, $\mathrm{MaxFunctionEvaluations}=10^5$). To determine the time-optimal solution, we perform this optimization for different values of $T$ and approximate the function $\mathrm{Err_{min}}(T)$, defined as the minimum error of Eq.~\eqref{appendix:Err} for a given $T$.  
This function typically exhibits a \textit{sharp} profile, rapidly dropping to 0 around a critical duration $T_{\mathrm{min}}$, which represents the quantum speed limit of the process. In this work, we define $T_{\mathrm{min}}$ as the smallest value at which $\mathrm{Err_{min}}<1.01\times\mathrm{E}_{\mathrm{th}}$, with $\mathrm{E}_{\mathrm{th}}=10^{-10}$. In principle, the precise value of $T_{\mathrm{min}}$ depends on the error threshold $\mathrm{E}_{\mathrm{th}}$ and the discretization step $dt$, but the intrinsically sharp profile of $\mathrm{Err_{min}}(T)$ makes it effectively independent of these choices. In general, we found numerically convenient to start with a large value of $T$, where convergence is easily reached from a random initial control and then reduce $T$ by taking the previous optimal control as an initial guess for optimization at a smaller $T$.  

In Fig.~\ref{fig_examples}(d), we consider $S_{\rm max}=3$ and target processes where $\phi_S=0$ for all superatoms but one, say $\bar{S}$, where $\phi_{\bar{S}}=\phi\in[0,\pi]$. We solve the time-optimal instance for different values of $\phi$, and assemble the functions $T_{\mathrm{min}}(\phi)$. As it has been shown in Ref.~\cite{fromonteil2024HJB}, leveraging the theory of Hamilton-Jacobi-Bellman, this function is in general continuous, but not differentiable. On the one hand, continuity can be efficiently exploited to improve the optimization scheme: the optimal-time solution at $\phi$ qualifies as a good candidate as initial condition for the optimization at $\phi\pm \Delta\phi$. On the other hand, there exist points where such function is not differentiable, and where different local minima branches meet. At around those points, a more careful study is required, and the optimization needs to be approached from multiple directions, e.g., $\phi\simeq\left(3\pi/8\right)^{\pm}$ for $\bar S=2$.

\section{Details on the numerical simulation of the QCA}
\label{sec:numerics_simulation}

The numerical results of the quantum quench protocol, shown in Fig.~\ref{fig:KIM_1D} for lattice dimension $D=1$ and in Fig.~\ref{fig:kitaev2D} for $D=2$, are obtained through a state-vector simulation based on tensor network states \cite{white1992density,vidal2003efficient,verstraete2004renormalization}, which we implement using the ITensor library in Julia \cite{ITensor}. Here, we give a detailed account and show additional data. 

Before going into the details of the simulations in $D=1,2$, let us give some general remarks that apply to all simulations shown. We emphasize that in all cases we use controlled methods and the results shown are in regimes where the simulations are numerically exact (within significant digits). Naturally, this limits us to small system sizes or few time steps.

Let us recall the quench protocol. We calculate $N_s$ samples
\be\label{eq:sample}
O_{t}(s) =\langle \psi_t(\vec{\theta}_s)|O|\psi_t(\vec{\theta}_s)\rangle,\quad s=1,\dots N_s
\ee
for different initial states $\ket{\psi_0(\vec{\theta}_s)}$ sampled from the tetrahedral or Haar ensemble. The function $g^O(t)$ in Eq.~\eqref{def_g} is approximated with the sample variance
\begin{align}
    g^O(t) &= \mathrm{Var}_{\vec{\theta}_s}\left[O_t(s)\right] \sim \frac{1}{N_s-1} \sum_s \left( O_{t}(s) - \overline{O_{t}}\right)^2\nonumber\\
    &+ r(N_s), \label{eq:sample_var}
\end{align}
where $\overline{O_{t}}$ denotes the sample average and $r(N_s)$ goes to zero as  $N_s \to \infty$. Its precise form depends on the unknown distribution, see also Fig.~\ref{fig:hist_n8}.
To estimate an uncertainty region of the sample variance independently of the underlying distribution, we calculate the sample variance for ten independent `batches'  of size $N_s/10$. This could, e.g.,~correspond to ten independent experimental runs. The uncertainty is then taken as the standard deviation of those ten data points, displayed as shaded regions in the plots. We benchmark this procedure for the cases where we can compare with exact values of $g^O(t)$, finding the expected agreement. We also note that there is no visible difference in the results if one samples over the Haar random distribution instead of the tetrahedral ensemble. 

\begin{figure}[!htb]
    \centering
\includegraphics[width=1\linewidth]{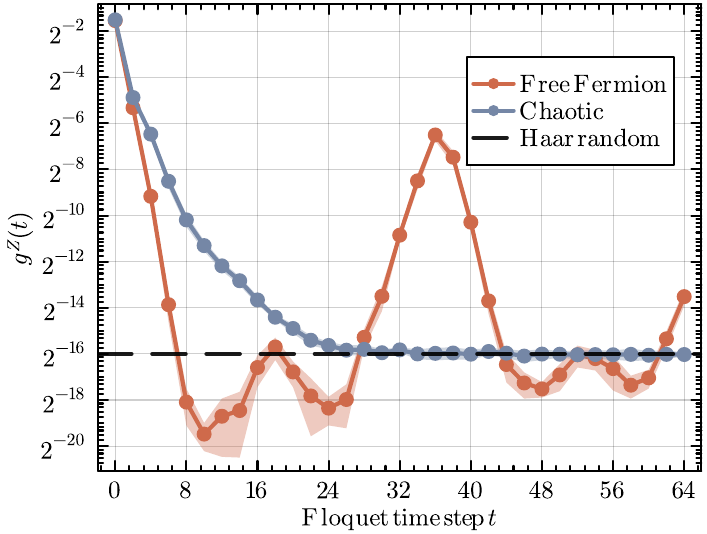}
    \includegraphics[width=1\linewidth]{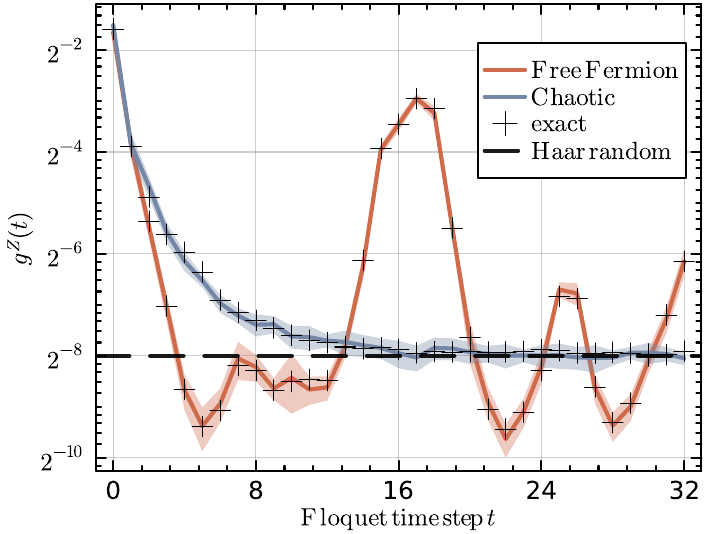}
    \caption{Generating function $g^Z(t) =: g^O(t)$ for the Pauli matrix $O = \sigma^Z_{N/2}$, for the $1$D kicked Ising model (KIM) on a system of $N=16$ (up) $N=8$ (down) sites with open boundary conditions. The curves correspond to the KIM with parameters $J=1,b=0.8$ and (i) $h=1.2$ (chaotic) and (ii) $h=0$ (free fermionic). The parameters are defined in Eqs.~(\ref{eq:ki_u}-\ref{eq:h_K}) (with $\tau=1$). The dashed line is the late-time limit described in Eq.~\eqref{eq:limit_haar}. The shaded region indicates the uncertainty of the sample variance, as explained in  Appendix~\ref{sec:numerics_simulation}. The exact values for $N=8$  are obtained by calculating the exact operator size distribution $p_l$ in Eq.~\eqref{eq:distribution_formula}, of the time-evolved operator, which then determines $g^Z(t)$ via~\eqref{chaos_equivalence}. The values and uncertainty region (shaded regions) for the two curves are calculated  from 1000 samples obtained using MPS as described in Appendix \ref{sec:numerics_simulation}.}
    \label{fig:mps_n8}
\end{figure}
For $D=1$, we use matrix product states (MPS) to evolve the initial product state with the Floquet operator $\mathcal{U}_\text{KI}$ of the kicked Ising model using open boundary conditions (see Sec.~\ref{sec:kicked_ising}). The unitary $\mathcal{U}_\text{KI}$ is decomposed into nearest neighbor gates $U_{i,j}$. Applying a unitary $U_{i,j}$ to an MPS $\ket{M}$ using a singular value decomposition (SVD) increases the bond dimension between sites $i$ and $j$ by a constant factor, which is at most $d^2=4$ for the local Hilbert space dimension $d=2$. Thus, an exact MPS representation of $\ket{\psi_t} = \mathcal{U}_\text{KI}^t\ket{\psi_0}$ has bond dimension $\chi_t$ upper bounded as $\chi_t\leq \min(4^t,\chi_{\max})$, where $\chi_{\max} =2^{N/2}$ is the bond dimension of a maximally-entangled state, and $N$ being the system size. Overall, the cost for calculating one sample of Eq.~\eqref{eq:sample} scales with $O(N\chi_t^3)$.

In practice, we set a truncation cutoff $\epsilon_0$, which is a fixed value for the maximally truncated weight: after applying a gate $U_{i,j}$ to an MPS $\ket{M}$ of bond dimension $\chi_{i,j}$ between sites $i,j$, we approximate the state with a new MPS $\ket{M'}$ with bond-dimension $\chi'_{i,j}<4\chi_{i,j}$,  such that 
\begin{equation}
    \| U_{i,j}\ket{M} - \ket{M'}\| \leq \epsilon_0.
\end{equation}
This implies that we calculate $O_{t}(s)$ up to an error $\epsilon = O(\epsilon_0,N,t)$. Ultimately, we want to calculate $g^O(t)$ using Eq.~\eqref{eq:sample_var}. For chaotic dynamics, the late time value is $g^O(t) \to 1/(2^{N}+1)$. Therefore, to obtain the correct result for $g^O(t)$ at late times, we generically need to ensure that $\epsilon \ll 2^{-N}$. With reasonable computational resources, we can do that for systems of size $N =O(20)$ by setting a small enough $\epsilon_0\ll 2^{-N}$. For a small system of $N=8$ sites, we compare the MPS simulation to an exact operator calculation of $g^O(t)$ using Eq.~\eqref{eq:distribution_formula}. This is shown in Figs.~\ref{fig:KIM_1D} and~\ref{fig:mps_n8}, where we find perfect agreement.

In Fig.~\ref{fig:hist_n8}, we show the distribution of the sampled values $\{O_{t}(s)\}_s$ for some specific times. For chaotic time-evolution, the distribution of $\{O_{t}(s)\}$ is well described by a Gaussian after a few time steps; becoming narrower with the increasing of time. For the non-ergodic time-evolution in the free fermion parameter regime, the distribution of  $\{O_{t}(s)\}_s$ shows deviations from the Gaussian distribution, especially at times where $g^{O}(t)$ shows recurrences.

\begin{figure}
    \centering
    \includegraphics[width=1\linewidth]{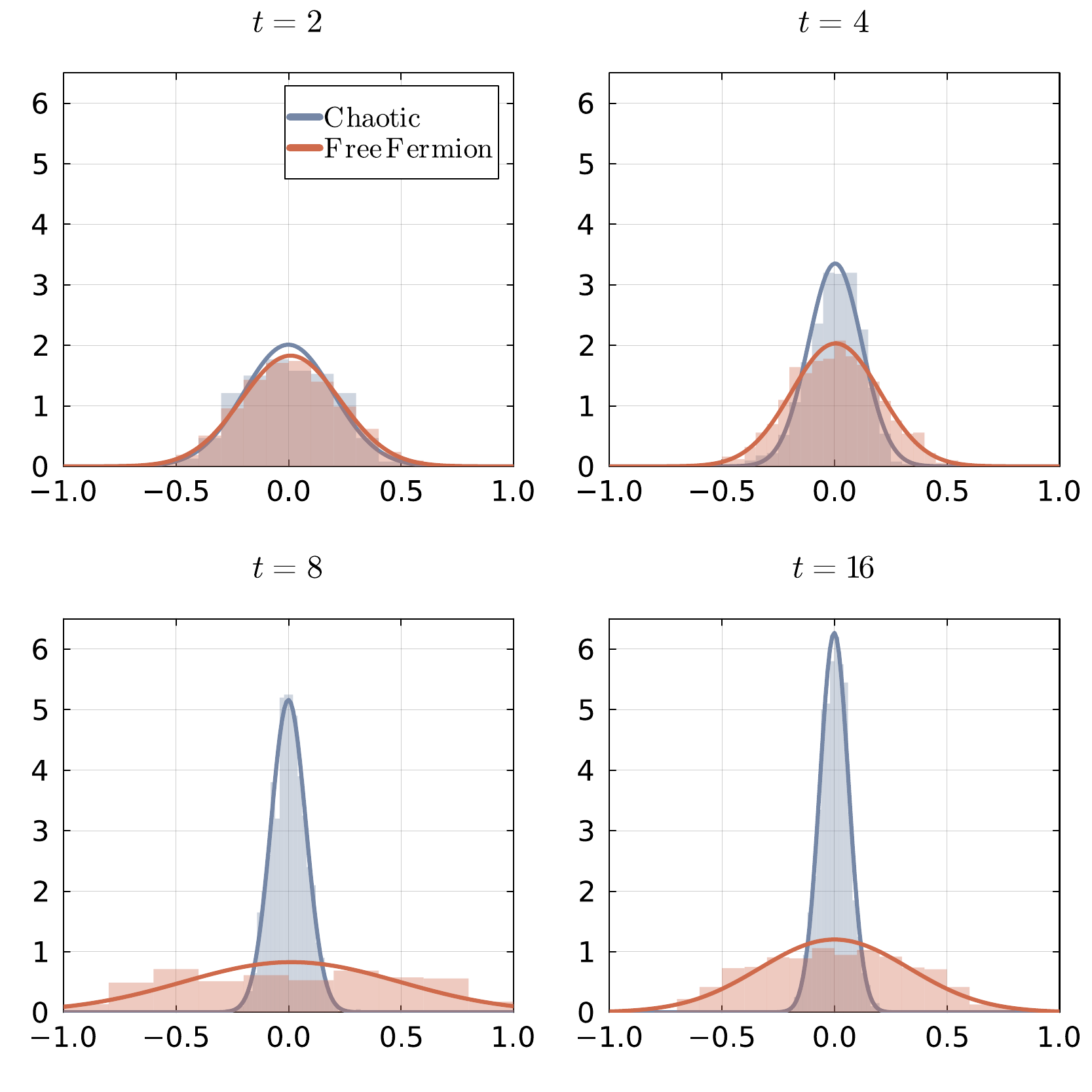}
    \caption{Histograms of simulated observables $\{O_{t}(s)\}_s$ for different time steps for the for the data in Fig.~\ref{fig:KIM_1D} for systems size $N=8$. The bold curves are Gaussian with the mean and variance of the corresponding data.}
    \label{fig:hist_n8}
\end{figure}

For $D=2$, we use projected entangled pair states (PEPS) \cite{verstraete2004renormalization} to simulate the time-evolution under the Kitaev Floquet operator in Eq.~\eqref{eq:trotter_single_step_kitaev}. As with MPS, the bond dimension grows exponentially with time. For our purposes, there are two main differences compared to the MPS simulation for $D=1$. 

First, due to the presence of loops in the tensor network, the local SVD truncation after applying a gate is no longer optimal. The globally optimal truncation is known as full update~\cite{phien2015infinite}, whereas the ad-hoc SVD approach we use is known as simple update~\cite{jiang2008accurate}. 

Second, the expectation value $\langle \psi_t(\vec{\theta}_s)|O|\psi_t(\vec{\theta}_s)\rangle$ can not  be computed exactly, and different controlled approximate methods have been developed for this task. Under some assumptions, the cost for evaluating expectation values as those in Eq.~\eqref{eq:sample} typically scales as $O(\chi^{14})$ \cite{corboz2020ipeps}. This severely restricts the number of time-steps that we can simulate. We opt for the double-layer boundary MPS method, see e.g.~\cite{lubasch2014unifying,lubasch2014algorithms}, which evaluates $\langle \psi_t(\vec{\theta}_s)|O|\psi_t(\vec{\theta}_s)\rangle$, the `double' layer, as a MPS-MPO product
\begin{equation}
    \langle \psi_t(\vec{\theta}_s)|O|\psi_t(\vec{\theta}_s)\rangle = \langle \mathrm{MPS}|\mathrm{MPO}\cdot \dots \cdot \mathrm{MPO}|\mathrm{MPS}\rangle.
\end{equation}
On a square lattice, the boundary MPSs are defined as the first and last row of double tensors along one spatial direction, while the MPOs are the intermediate rows. For the simulations on the honeycomb lattice, the MPOs are in a staircase form and we exploit the reduced connectivity to make the contraction more efficient compared to the square lattice. We set a dynamical cutoff $\epsilon$ for the boundary MPSs. As before, we can faithfully estimate the sample variance for small enough $\epsilon\ll 2^{-t^2}$. Here, we compare against the worst-case (i.e. smallest possible) early-time values of $g^O(t)$ - as we are limited to very small $t$ this is feasible.

Crucially, we can reduce the computational cost significantly by evolving the initial product state only within the light cone of the local observable we want to calculate. For the Kitaev Floquet model, the number $n$ of qubits within the light cone of a single-site operator grows as $n = 8,28,60,\dots$ for $t=1,2,3,\dots$. As the size of the light cone rapidly grows, it is instructive to also consider the intermediate time steps, corresponding to applying the three different $e^{-i\tau H_
\alpha}$, $\alpha=Z,X,Y$ from Eq.~\eqref{eq:trotter_single_step_kitaev}. For the simulation, we chose the order such that at $t=1$ we only apply the $X$ term. Thus, nine time steps correspond to three full cycles.

\begin{figure}
    \centering
    \includegraphics[width=1\linewidth]{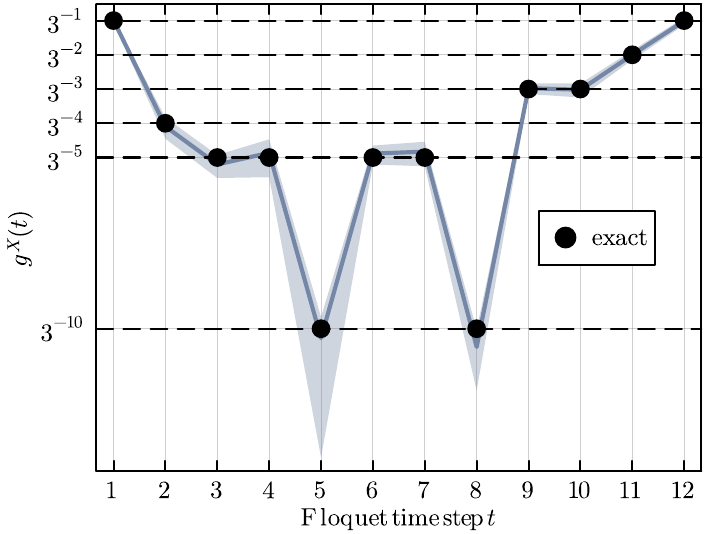}    
    \caption{Generating function $g^O(t) =:g^X(t)$ of the Pauli matrix $O =\sigma^X$, for the $2$D Kitaev Floquet system at the Clifford point $J_\alpha = \pi/4,h_\alpha=0$. The parameters are defined in Eqs.~(\ref{eq:kitaev_hamiltonian_0}-\ref{eq:kitaev_hamiltonian}) (with $\tau=1$). Note that we also include intermediate time steps as explained in Appendix~\ref{sec:numerics_simulation}. We take the system size to be larger then the operator size at any time. The values and uncertainty region (shaded regions) for the two curves are calculated  from 1000 samples obtained using PEPS as described in Appendix~\ref{sec:numerics_simulation}. Since the time-evolved operator $\sigma^X(t)$ is a single Pauli string at any time, we have $g^O(t)= 3^{-|O(t)|}$, where $|O(t)|$ is the Pauli weight.}
    \label{fig:kitaev_clifford_simulation}
\end{figure}

At the Clifford point $J_{\alpha} = \pi/4,h_\alpha = 0,\alpha = X,Y,Z$ we can calculate the time-evolution of an initial Pauli-string analytically (or numerically using the stabilizer formalism). We show the exact results and PEPS simulation in Fig.~\ref{fig:kitaev_clifford_simulation}.

\begin{figure}
    \centering
    \includegraphics[width=1\linewidth]{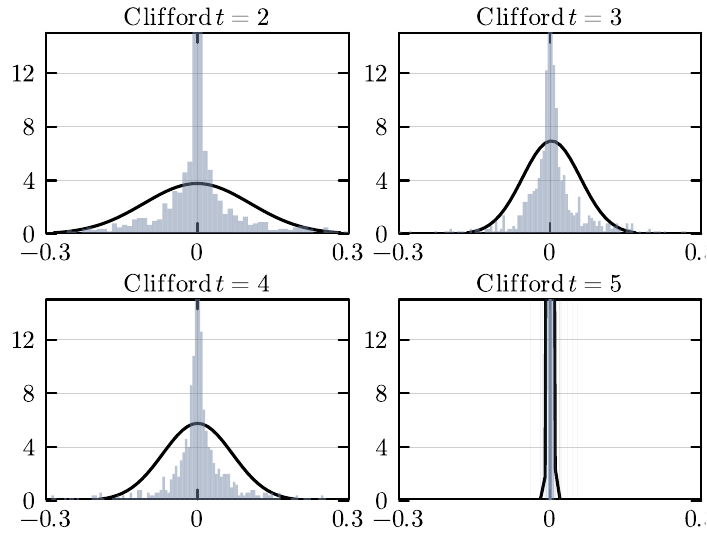}
\caption{Histograms of simulated observables $\{O_{t}(s)\}_s$ for different time steps for the for the data in Fig.~\ref{fig:kitaev_clifford_simulation}. The bold curves are Gaussian with the mean and variance of the corresponding data.}
    \label{fig:kitaev_clifford_histograms}
\end{figure}

In Fig.\ref{fig:kitaev_clifford_histograms} we show the distribution of the observables. They deviate strongly from a Gaussian and approach delta functions with growing operator size.

\section{Higher-order moments of the tetrahedral ensemble}
\label{appendix: moments}
In this appendix, we compute a general expression for the $k$-th moment of the tetrahedron ensemble $\mathcal{E}_T$ described in Sec.~\ref{sec: Observing quantum chaos}. Specifically, we compute a general expression for $N_k \equiv \frac{1}{4}\sum_{i=1}^4 \rho_i^{\otimes k}$, where $\rho_i = \ket{\phi_i}\!\bra{\phi_i} = \frac{1}{2}\bigl(\mathbb{I} + \vec n_i \cdot \vec \sigma\bigr)$ and the unit vectors $\vec n_i$ point to the vertices of a regular tetrahedron on the Bloch sphere.

We start by expanding the tensor product of each single-qubit state and observing that every term is made of a subset of sites $E$ on which a Pauli operator acts, while the identity acts on the remaining sites. It is convenient to order this sum with the cardinality $|E|$ of the subsets:
\begin{equation}\label{eq:tensor-expansion}
\bigl(\mathbb{I} + \vec n_i \cdot \vec \sigma\bigr)^{\otimes k}=\sum_{m=0}^k\sum_{\substack{E \subset \{1,\dots,k\} \\ |E|=m}}\Biggl(\bigotimes_{j \notin E} \mathbb{I}_j\Biggr)\Biggl(\bigotimes_{j \in E} \vec n_i \cdot \vec \sigma_j\Biggr),
\end{equation}
and then to explicitly expand the Pauli components on the sites in $E$:
\begin{equation}
\label{eq:pauli-expansion}
\bigotimes_{j \in E} \vec n_i \cdot \vec \sigma_j=\sum_{\{a_j\}_{j \in E}}\left( \prod_{j \in E} n_i^{a_j} \right)\bigotimes_{j \in E} \sigma_j^{a_j},
\end{equation}
with $a_j \in \{x,y,z\}$. 

We proceed by defining the \emph{tetrahedral moment tensors} of order $m$:
\begin{equation}
\label{eq:moment-tensors}
T^m_{a_1,\dots,a_m}=\frac{1}{4}\sum_{i=1}^4n_i^{a_1}\cdots n_i^{a_m},\qquad T^0=1,
\end{equation}
whose value depends only on $m$ (and not on the choice of sites). It can therefore always be evaluated for $E=\{1,2,\dots,m\}$.

Finally, combining Eqs.~\eqref{eq:tensor-expansion}--\eqref{eq:moment-tensors}, the moment
operator $N_k$ takes the following compact form:
\begin{align}\label{eq:Nk-final}
&N_k=\frac{1}{2^k}\sum_{m=0}^k\sum_{\substack{E \subset \{1,\dots,k\} \\ |E|=m}}\sum_{a_1,\dots,a_m}T^m_{a_1,\dots,a_m}\sigma^{a_1,\dots,a_m}(E)\nonumber\\
&\mbox{where}\quad
\sigma^{a_1,\dots,a_m}(E)=\sigma_{e_1}^{a_1}\otimes\sigma_{e_2}^{a_2}\otimes\dots\sigma_{e_m}^{a_m}\bigotimes_{j \notin E} \mathbb{I}_j,
\end{align}
with $E=\left\{e_1<e_2<\dots<e_m\right\}$.

Eq.~\eqref{eq:Nk-final} provides a decomposition of $N_k$ in terms of tetrahedral moment tensors and tensor products of Pauli operators.

\subsection{Explicit moments up to fourth order}
Here, we provide an explicit expression of the moment operators $N_k$, for $k\le4$, by direct evaluation of Eq.~\eqref{eq:Nk-final}. As anticipated, they are analogous to those of the Haar distribution when $k\le2$. We make the following explicit choice of the tetrahedron (analogous to that of Sec.~\ref{sec: Observing quantum chaos}): $\vec n_1=(1,1,1)/\sqrt{3}$ $\vec n_2=(1,-1,-1)/\sqrt{3}$, $\vec n_3=(-1,1,-1)/\sqrt{3}$, and $\vec n_4=(-1,-1,1)/\sqrt{3}$.
\paragraph*{First order ($k=1$).}
By symmetry, all first-order moment tensors vanish: $T^1_a = \frac{1}{4}\sum_{i=1}^4 n_i^a = 0$, with $a=x,y,z$. Therefore, we have $N_1 = \frac{\mathbb{I}}{2}$. 

\paragraph*{Second order ($k=2$).}
The only nonvanishing second-order moment tensor is diagonal: $T^2_{ab}=\frac{1}{4}\sum_{i=1}^4 n_i^a n_i^b=\frac{1}{3}\delta_{ab}$. Therefore, we have ${N_2= \frac{1}{4}\left(\mathbb{I}+ \frac{1}{3}\sum_{a=x,y,z}\sigma^a_1\sigma^a_2\right)}$.

\paragraph*{Third order ($k=3$).}
At third order, the moment operator contains contributions of tensor order $m=0,2,3$. It can therefore be written as $N_3=N_3^{(0)}+N_3^{(2)}+N_3^{(3)}$ where
\begin{align*}
&N_3^{(0)} = \frac{\mathbb{I}}{8},\qquad N_3^{(2)} = \frac{1}{24} \sum_{a=x,y,z}\sum_{1\le i<j\le 3} \sigma_i^a\sigma_j^a\\
&N_3^{(3)} = \frac{1}{8\sqrt{27}} \sum_{\pi\in \Sigma_3} \sigma_1^{\pi(x)} \sigma_2^{\pi(y)} \sigma_3^{\pi(z)},      
\end{align*}
with $\Sigma_3$ being the symmetric group of order $3$ (which we apply to $\left\{x,y,z\right\}$).

\paragraph*{Fourth order ($k=4$).}
At fourth order, the moment operator contains contributions of tensor order $m=0,2,3,4$, and it is therefor decomposed as $N_4 = N_4^{(0)} + N_4^{(2)} + N_4^{(3)} + N_4^{(4)}$, where
\begin{align}
&N_4^{(0)} = \frac{\mathbb{I}}{16},\qquad N_4^{(2)} = \frac{1}{48} \sum_{a=x,y,z}\sum_{1\le i<j\le 4} \sigma_i^a\sigma_j^a ,\nonumber\\[4pt]
&N_4^{(3)} = \frac{1}{16\sqrt{27}} \sum_{1\le i< j< l\le4} \sum_{\pi\in \Sigma_3} \sigma_i^{\pi(x)} \sigma_j^{\pi(y)} \sigma_l^{\pi(z)},\nonumber\\[4pt]
&N_4^{(4)}=\frac{1}{144}\left[
\sum_{a=x,y,z}\sigma_1^a\sigma_2^a\sigma_3^a\sigma_4^a\right.\nonumber \\ 
& \left. +\sum_{\substack{a,b=x,y,z\\ a\neq b}}
\;\sum_{(c_1,c_2,c_3,c_4)\in \mathrm{Perm}(a,a,b,b)}
\sigma_1^{c_1}\sigma_2^{c_2}\sigma_3^{c_3}\sigma_4^{c_4}
\right].
\end{align}
Here, $\mathrm{Perm}(a,a,b,b)$ denotes the set of all distinct permutations of the multiset $\{a,a,b,b\}$.

\FloatBarrier
\bibliography{bibliography}

\begin{thebibliography}{150}%
\makeatletter
\providecommand \@ifxundefined [1]{%
 \@ifx{#1\undefined}
}%
\providecommand \@ifnum [1]{%
 \ifnum #1\expandafter \@firstoftwo
 \else \expandafter \@secondoftwo
 \fi
}%
\providecommand \@ifx [1]{%
 \ifx #1\expandafter \@firstoftwo
 \else \expandafter \@secondoftwo
 \fi
}%
\providecommand \natexlab [1]{#1}%
\providecommand \enquote  [1]{``#1''}%
\providecommand \bibnamefont  [1]{#1}%
\providecommand \bibfnamefont [1]{#1}%
\providecommand \citenamefont [1]{#1}%
\providecommand \href@noop [0]{\@secondoftwo}%
\providecommand \href [0]{\begingroup \@sanitize@url \@href}%
\providecommand \@href[1]{\@@startlink{#1}\@@href}%
\providecommand \@@href[1]{\endgroup#1\@@endlink}%
\providecommand \@sanitize@url [0]{\catcode `\\12\catcode `\$12\catcode `\&12\catcode `\#12\catcode `\^12\catcode `\_12\catcode `\%12\relax}%
\providecommand \@@startlink[1]{}%
\providecommand \@@endlink[0]{}%
\providecommand \url  [0]{\begingroup\@sanitize@url \@url }%
\providecommand \@url [1]{\endgroup\@href {#1}{\urlprefix }}%
\providecommand \urlprefix  [0]{URL }%
\providecommand \Eprint [0]{\href }%
\providecommand \doibase [0]{https://doi.org/}%
\providecommand \selectlanguage [0]{\@gobble}%
\providecommand \bibinfo  [0]{\@secondoftwo}%
\providecommand \bibfield  [0]{\@secondoftwo}%
\providecommand \translation [1]{[#1]}%
\providecommand \BibitemOpen [0]{}%
\providecommand \bibitemStop [0]{}%
\providecommand \bibitemNoStop [0]{.\EOS\space}%
\providecommand \EOS [0]{\spacefactor3000\relax}%
\providecommand \BibitemShut  [1]{\csname bibitem#1\endcsname}%
\let\auto@bib@innerbib\@empty
%</preamble>
\bibitem [{\citenamefont {Preskill}(2018)}]{preskill2018quantum}%
  \BibitemOpen
  \bibfield  {author} {\bibinfo {author} {\bibfnamefont {J.}~\bibnamefont {Preskill}},\ }\bibfield  {title} {\bibinfo {title} {Quantum computing in the nisq era and beyond},\ }\href {https://doi.org/10.22331/q-2018-08-06-79} {\bibfield  {journal} {\bibinfo  {journal} {Quantum}\ }\textbf {\bibinfo {volume} {2}},\ \bibinfo {pages} {79} (\bibinfo {year} {2018})}\BibitemShut {NoStop}%
\bibitem [{\citenamefont {Fisher}\ \emph {et~al.}(2023)\citenamefont {Fisher}, \citenamefont {Khemani}, \citenamefont {Nahum},\ and\ \citenamefont {Vijay}}]{fisher2023random}%
  \BibitemOpen
  \bibfield  {author} {\bibinfo {author} {\bibfnamefont {M.~P.}\ \bibnamefont {Fisher}}, \bibinfo {author} {\bibfnamefont {V.}~\bibnamefont {Khemani}}, \bibinfo {author} {\bibfnamefont {A.}~\bibnamefont {Nahum}},\ and\ \bibinfo {author} {\bibfnamefont {S.}~\bibnamefont {Vijay}},\ }\bibfield  {title} {\bibinfo {title} {Random quantum circuits},\ }\href {https://doi.org/10.1146/annurev-conmatphys-031720-030658} {\bibfield  {journal} {\bibinfo  {journal} {Ann. Rev. Cond. Matt. Phys.}\ }\textbf {\bibinfo {volume} {14}},\ \bibinfo {pages} {335} (\bibinfo {year} {2023})}\BibitemShut {NoStop}%
\bibitem [{\citenamefont {Bertini}\ \emph {et~al.}(2025)\citenamefont {Bertini}, \citenamefont {Claeys},\ and\ \citenamefont {Prosen}}]{bertini2025exactly}%
  \BibitemOpen
  \bibfield  {author} {\bibinfo {author} {\bibfnamefont {B.}~\bibnamefont {Bertini}}, \bibinfo {author} {\bibfnamefont {P.~W.}\ \bibnamefont {Claeys}},\ and\ \bibinfo {author} {\bibfnamefont {T.}~\bibnamefont {Prosen}},\ }\bibfield  {title} {\bibinfo {title} {Exactly solvable many-body dynamics from space-time duality},\ }\href {https://arxiv.org/abs/2505.11489} {\bibfield  {journal} {\bibinfo  {journal} {arXiv:2505.11489}\ } (\bibinfo {year} {2025})}\BibitemShut {NoStop}%
\bibitem [{\citenamefont {Nahum}\ \emph {et~al.}(2018)\citenamefont {Nahum}, \citenamefont {Vijay},\ and\ \citenamefont {Haah}}]{nahum2018operator}%
  \BibitemOpen
  \bibfield  {author} {\bibinfo {author} {\bibfnamefont {A.}~\bibnamefont {Nahum}}, \bibinfo {author} {\bibfnamefont {S.}~\bibnamefont {Vijay}},\ and\ \bibinfo {author} {\bibfnamefont {J.}~\bibnamefont {Haah}},\ }\bibfield  {title} {\bibinfo {title} {Operator spreading in random unitary circuits},\ }\href {https://doi.org/10.1103/PhysRevX.8.021014} {\bibfield  {journal} {\bibinfo  {journal} {Phys. Rev. X}\ }\textbf {\bibinfo {volume} {8}},\ \bibinfo {pages} {021014} (\bibinfo {year} {2018})}\BibitemShut {NoStop}%
\bibitem [{\citenamefont {von Keyserlingk}\ \emph {et~al.}(2018{\natexlab{a}})\citenamefont {von Keyserlingk}, \citenamefont {Rakovszky}, \citenamefont {Pollmann},\ and\ \citenamefont {Sondhi}}]{vonKeyserlingk2018operator}%
  \BibitemOpen
  \bibfield  {author} {\bibinfo {author} {\bibfnamefont {C.~W.}\ \bibnamefont {von Keyserlingk}}, \bibinfo {author} {\bibfnamefont {T.}~\bibnamefont {Rakovszky}}, \bibinfo {author} {\bibfnamefont {F.}~\bibnamefont {Pollmann}},\ and\ \bibinfo {author} {\bibfnamefont {S.~L.}\ \bibnamefont {Sondhi}},\ }\bibfield  {title} {\bibinfo {title} {Operator hydrodynamics, otocs, and entanglement growth in systems without conservation laws},\ }\href {https://doi.org/10.1103/PhysRevX.8.021013} {\bibfield  {journal} {\bibinfo  {journal} {Phys. Rev. X}\ }\textbf {\bibinfo {volume} {8}},\ \bibinfo {pages} {021013} (\bibinfo {year} {2018}{\natexlab{a}})}\BibitemShut {NoStop}%
\bibitem [{\citenamefont {Zhou}\ and\ \citenamefont {Nahum}(2020)}]{zhou2020entanglement}%
  \BibitemOpen
  \bibfield  {author} {\bibinfo {author} {\bibfnamefont {T.}~\bibnamefont {Zhou}}\ and\ \bibinfo {author} {\bibfnamefont {A.}~\bibnamefont {Nahum}},\ }\bibfield  {title} {\bibinfo {title} {Entanglement membrane in chaotic many-body systems},\ }\href {https://doi.org/10.1103/PhysRevX.10.031066} {\bibfield  {journal} {\bibinfo  {journal} {Phys. Rev. X}\ }\textbf {\bibinfo {volume} {10}},\ \bibinfo {pages} {031066} (\bibinfo {year} {2020})}\BibitemShut {NoStop}%
\bibitem [{\citenamefont {Nahum}\ \emph {et~al.}(2017)\citenamefont {Nahum}, \citenamefont {Ruhman}, \citenamefont {Vijay},\ and\ \citenamefont {Haah}}]{nahum2017quantum}%
  \BibitemOpen
  \bibfield  {author} {\bibinfo {author} {\bibfnamefont {A.}~\bibnamefont {Nahum}}, \bibinfo {author} {\bibfnamefont {J.}~\bibnamefont {Ruhman}}, \bibinfo {author} {\bibfnamefont {S.}~\bibnamefont {Vijay}},\ and\ \bibinfo {author} {\bibfnamefont {J.}~\bibnamefont {Haah}},\ }\bibfield  {title} {\bibinfo {title} {Quantum entanglement growth under random unitary dynamics},\ }\href {https://doi.org/10.1103/PhysRevX.7.031016} {\bibfield  {journal} {\bibinfo  {journal} {Phys. Rev. X}\ }\textbf {\bibinfo {volume} {7}},\ \bibinfo {pages} {031016} (\bibinfo {year} {2017})}\BibitemShut {NoStop}%
\bibitem [{\citenamefont {Bertini}\ \emph {et~al.}(2019{\natexlab{a}})\citenamefont {Bertini}, \citenamefont {Kos},\ and\ \citenamefont {Prosen}}]{bertini2019entanglement}%
  \BibitemOpen
  \bibfield  {author} {\bibinfo {author} {\bibfnamefont {B.}~\bibnamefont {Bertini}}, \bibinfo {author} {\bibfnamefont {P.}~\bibnamefont {Kos}},\ and\ \bibinfo {author} {\bibfnamefont {T.}~\bibnamefont {Prosen}},\ }\bibfield  {title} {\bibinfo {title} {Entanglement spreading in a minimal model of maximal many-body quantum chaos},\ }\href {https://doi.org/10.1103/PhysRevX.9.021033} {\bibfield  {journal} {\bibinfo  {journal} {Phys. Rev. X}\ }\textbf {\bibinfo {volume} {9}},\ \bibinfo {pages} {021033} (\bibinfo {year} {2019}{\natexlab{a}})}\BibitemShut {NoStop}%
\bibitem [{\citenamefont {Klobas}\ \emph {et~al.}(2021)\citenamefont {Klobas}, \citenamefont {Bertini},\ and\ \citenamefont {Piroli}}]{klobas2021exact}%
  \BibitemOpen
  \bibfield  {author} {\bibinfo {author} {\bibfnamefont {K.}~\bibnamefont {Klobas}}, \bibinfo {author} {\bibfnamefont {B.}~\bibnamefont {Bertini}},\ and\ \bibinfo {author} {\bibfnamefont {L.}~\bibnamefont {Piroli}},\ }\bibfield  {title} {\bibinfo {title} {Exact thermalization dynamics in the ``rule 54'' quantum cellular automaton},\ }\href {https://doi.org/10.1103/PhysRevLett.126.160602} {\bibfield  {journal} {\bibinfo  {journal} {Phys. Rev. Lett.}\ }\textbf {\bibinfo {volume} {126}},\ \bibinfo {pages} {160602} (\bibinfo {year} {2021})}\BibitemShut {NoStop}%
\bibitem [{\citenamefont {Chan}\ \emph {et~al.}(2018)\citenamefont {Chan}, \citenamefont {De~Luca},\ and\ \citenamefont {Chalker}}]{chan2018solution}%
  \BibitemOpen
  \bibfield  {author} {\bibinfo {author} {\bibfnamefont {A.}~\bibnamefont {Chan}}, \bibinfo {author} {\bibfnamefont {A.}~\bibnamefont {De~Luca}},\ and\ \bibinfo {author} {\bibfnamefont {J.~T.}\ \bibnamefont {Chalker}},\ }\bibfield  {title} {\bibinfo {title} {Solution of a minimal model for many-body quantum chaos},\ }\href {https://doi.org/10.1103/PhysRevX.8.041019} {\bibfield  {journal} {\bibinfo  {journal} {Phys. Rev. X}\ }\textbf {\bibinfo {volume} {8}},\ \bibinfo {pages} {041019} (\bibinfo {year} {2018})}\BibitemShut {NoStop}%
\bibitem [{\citenamefont {Giudici}\ \emph {et~al.}(2024)\citenamefont {Giudici}, \citenamefont {Surace},\ and\ \citenamefont {Pichler}}]{Giudici2024PXP}%
  \BibitemOpen
  \bibfield  {author} {\bibinfo {author} {\bibfnamefont {G.}~\bibnamefont {Giudici}}, \bibinfo {author} {\bibfnamefont {F.~M.}\ \bibnamefont {Surace}},\ and\ \bibinfo {author} {\bibfnamefont {H.}~\bibnamefont {Pichler}},\ }\bibfield  {title} {\bibinfo {title} {Unraveling pxp many-body scars through floquet dynamics},\ }\href {https://doi.org/10.1103/PhysRevLett.133.190404} {\bibfield  {journal} {\bibinfo  {journal} {Phys. Rev. Lett.}\ }\textbf {\bibinfo {volume} {133}},\ \bibinfo {pages} {190404} (\bibinfo {year} {2024})}\BibitemShut {NoStop}%
\bibitem [{\citenamefont {Farrelly}(2020)}]{farrelly2020review}%
  \BibitemOpen
  \bibfield  {author} {\bibinfo {author} {\bibfnamefont {T.}~\bibnamefont {Farrelly}},\ }\bibfield  {title} {\bibinfo {title} {A review of quantum cellular automata},\ }\href {https://doi.org/10.22331/q-2020-11-30-368} {\bibfield  {journal} {\bibinfo  {journal} {Quantum}\ }\textbf {\bibinfo {volume} {4}},\ \bibinfo {pages} {368} (\bibinfo {year} {2020})}\BibitemShut {NoStop}%
\bibitem [{\citenamefont {Arrighi}(2019)}]{arrighi2019overview}%
  \BibitemOpen
  \bibfield  {author} {\bibinfo {author} {\bibfnamefont {P.}~\bibnamefont {Arrighi}},\ }\bibfield  {title} {\bibinfo {title} {An overview of quantum cellular automata},\ }\href {https://doi.org/10.1007/s11047-019-09762-6} {\bibfield  {journal} {\bibinfo  {journal} {Natural Comp.}\ }\textbf {\bibinfo {volume} {18}},\ \bibinfo {pages} {885} (\bibinfo {year} {2019})}\BibitemShut {NoStop}%
\bibitem [{\citenamefont {Arute}\ \emph {et~al.}(2019)\citenamefont {Arute}, \citenamefont {Arya}, \citenamefont {Babbush}, \citenamefont {Bacon}, \citenamefont {Bardin}, \citenamefont {Barends}, \citenamefont {Biswas}, \citenamefont {Boixo}, \citenamefont {Brandao}, \citenamefont {Buell}, \citenamefont {Burkett}, \citenamefont {Chen}, \citenamefont {Chen}, \citenamefont {Chiaro}, \citenamefont {Collins}, \citenamefont {Courtney}, \citenamefont {Dunsworth}, \citenamefont {Farhi}, \citenamefont {Foxen}, \citenamefont {Fowler}, \citenamefont {Gidney}, \citenamefont {Giustina}, \citenamefont {Graff}, \citenamefont {Guerin}, \citenamefont {Habegger}, \citenamefont {Harrigan}, \citenamefont {Hartmann}, \citenamefont {Ho}, \citenamefont {Hoffmann}, \citenamefont {Huang}, \citenamefont {Humble}, \citenamefont {Isakov}, \citenamefont {Jeffrey}, \citenamefont {Jiang}, \citenamefont {Kafri}, \citenamefont {Kechedzhi}, \citenamefont {Kelly}, \citenamefont {Klimov}, \citenamefont {Knysh}, \citenamefont {Korotkov},
  \citenamefont {Kostritsa}, \citenamefont {Landhuis}, \citenamefont {Lindmark}, \citenamefont {Lucero}, \citenamefont {Lyakh}, \citenamefont {Mandrà}, \citenamefont {McClean}, \citenamefont {McEwen}, \citenamefont {Megrant}, \citenamefont {Mi}, \citenamefont {Michielsen}, \citenamefont {Mohseni}, \citenamefont {Mutus}, \citenamefont {Naaman}, \citenamefont {Neeley}, \citenamefont {Neill}, \citenamefont {Niu}, \citenamefont {Ostby}, \citenamefont {Petukhov}, \citenamefont {Platt}, \citenamefont {Quintana}, \citenamefont {Rieffel}, \citenamefont {Roushan}, \citenamefont {Rubin}, \citenamefont {Sank}, \citenamefont {Satzinger}, \citenamefont {Smelyanskiy}, \citenamefont {Sung}, \citenamefont {Trevithick}, \citenamefont {Vainsencher}, \citenamefont {Villalonga}, \citenamefont {White}, \citenamefont {Yao}, \citenamefont {Yeh}, \citenamefont {Zalcman}, \citenamefont {Neven},\ and\ \citenamefont {Martinis}}]{Arute_2019}%
  \BibitemOpen
  \bibfield  {author} {\bibinfo {author} {\bibfnamefont {F.}~\bibnamefont {Arute}}, \bibinfo {author} {\bibfnamefont {K.}~\bibnamefont {Arya}}, \bibinfo {author} {\bibfnamefont {R.}~\bibnamefont {Babbush}}, \bibinfo {author} {\bibfnamefont {D.}~\bibnamefont {Bacon}}, \bibinfo {author} {\bibfnamefont {J.~C.}\ \bibnamefont {Bardin}}, \bibinfo {author} {\bibfnamefont {R.}~\bibnamefont {Barends}}, \bibinfo {author} {\bibfnamefont {R.}~\bibnamefont {Biswas}}, \bibinfo {author} {\bibfnamefont {S.}~\bibnamefont {Boixo}}, \bibinfo {author} {\bibfnamefont {F.~G. S.~L.}\ \bibnamefont {Brandao}}, \bibinfo {author} {\bibfnamefont {D.~A.}\ \bibnamefont {Buell}}, \bibinfo {author} {\bibfnamefont {B.}~\bibnamefont {Burkett}}, \bibinfo {author} {\bibfnamefont {Y.}~\bibnamefont {Chen}}, \bibinfo {author} {\bibfnamefont {Z.}~\bibnamefont {Chen}}, \bibinfo {author} {\bibfnamefont {B.}~\bibnamefont {Chiaro}}, \bibinfo {author} {\bibfnamefont {R.}~\bibnamefont {Collins}}, \bibinfo {author} {\bibfnamefont {W.}~\bibnamefont
  {Courtney}}, \bibinfo {author} {\bibfnamefont {A.}~\bibnamefont {Dunsworth}}, \bibinfo {author} {\bibfnamefont {E.}~\bibnamefont {Farhi}}, \bibinfo {author} {\bibfnamefont {B.}~\bibnamefont {Foxen}}, \bibinfo {author} {\bibfnamefont {A.}~\bibnamefont {Fowler}}, \bibinfo {author} {\bibfnamefont {C.}~\bibnamefont {Gidney}}, \bibinfo {author} {\bibfnamefont {M.}~\bibnamefont {Giustina}}, \bibinfo {author} {\bibfnamefont {R.}~\bibnamefont {Graff}}, \bibinfo {author} {\bibfnamefont {K.}~\bibnamefont {Guerin}}, \bibinfo {author} {\bibfnamefont {S.}~\bibnamefont {Habegger}}, \bibinfo {author} {\bibfnamefont {M.~P.}\ \bibnamefont {Harrigan}}, \bibinfo {author} {\bibfnamefont {M.~J.}\ \bibnamefont {Hartmann}}, \bibinfo {author} {\bibfnamefont {A.}~\bibnamefont {Ho}}, \bibinfo {author} {\bibfnamefont {M.}~\bibnamefont {Hoffmann}}, \bibinfo {author} {\bibfnamefont {T.}~\bibnamefont {Huang}}, \bibinfo {author} {\bibfnamefont {T.~S.}\ \bibnamefont {Humble}}, \bibinfo {author} {\bibfnamefont {S.~V.}\ \bibnamefont
  {Isakov}}, \bibinfo {author} {\bibfnamefont {E.}~\bibnamefont {Jeffrey}}, \bibinfo {author} {\bibfnamefont {Z.}~\bibnamefont {Jiang}}, \bibinfo {author} {\bibfnamefont {D.}~\bibnamefont {Kafri}}, \bibinfo {author} {\bibfnamefont {K.}~\bibnamefont {Kechedzhi}}, \bibinfo {author} {\bibfnamefont {J.}~\bibnamefont {Kelly}}, \bibinfo {author} {\bibfnamefont {P.~V.}\ \bibnamefont {Klimov}}, \bibinfo {author} {\bibfnamefont {S.}~\bibnamefont {Knysh}}, \bibinfo {author} {\bibfnamefont {A.}~\bibnamefont {Korotkov}}, \bibinfo {author} {\bibfnamefont {F.}~\bibnamefont {Kostritsa}}, \bibinfo {author} {\bibfnamefont {D.}~\bibnamefont {Landhuis}}, \bibinfo {author} {\bibfnamefont {M.}~\bibnamefont {Lindmark}}, \bibinfo {author} {\bibfnamefont {E.}~\bibnamefont {Lucero}}, \bibinfo {author} {\bibfnamefont {D.}~\bibnamefont {Lyakh}}, \bibinfo {author} {\bibfnamefont {S.}~\bibnamefont {Mandrà}}, \bibinfo {author} {\bibfnamefont {J.~R.}\ \bibnamefont {McClean}}, \bibinfo {author} {\bibfnamefont {M.}~\bibnamefont {McEwen}},
  \bibinfo {author} {\bibfnamefont {A.}~\bibnamefont {Megrant}}, \bibinfo {author} {\bibfnamefont {X.}~\bibnamefont {Mi}}, \bibinfo {author} {\bibfnamefont {K.}~\bibnamefont {Michielsen}}, \bibinfo {author} {\bibfnamefont {M.}~\bibnamefont {Mohseni}}, \bibinfo {author} {\bibfnamefont {J.}~\bibnamefont {Mutus}}, \bibinfo {author} {\bibfnamefont {O.}~\bibnamefont {Naaman}}, \bibinfo {author} {\bibfnamefont {M.}~\bibnamefont {Neeley}}, \bibinfo {author} {\bibfnamefont {C.}~\bibnamefont {Neill}}, \bibinfo {author} {\bibfnamefont {M.~Y.}\ \bibnamefont {Niu}}, \bibinfo {author} {\bibfnamefont {E.}~\bibnamefont {Ostby}}, \bibinfo {author} {\bibfnamefont {A.}~\bibnamefont {Petukhov}}, \bibinfo {author} {\bibfnamefont {J.~C.}\ \bibnamefont {Platt}}, \bibinfo {author} {\bibfnamefont {C.}~\bibnamefont {Quintana}}, \bibinfo {author} {\bibfnamefont {E.~G.}\ \bibnamefont {Rieffel}}, \bibinfo {author} {\bibfnamefont {P.}~\bibnamefont {Roushan}}, \bibinfo {author} {\bibfnamefont {N.~C.}\ \bibnamefont {Rubin}}, \bibinfo
  {author} {\bibfnamefont {D.}~\bibnamefont {Sank}}, \bibinfo {author} {\bibfnamefont {K.~J.}\ \bibnamefont {Satzinger}}, \bibinfo {author} {\bibfnamefont {V.}~\bibnamefont {Smelyanskiy}}, \bibinfo {author} {\bibfnamefont {K.~J.}\ \bibnamefont {Sung}}, \bibinfo {author} {\bibfnamefont {M.~D.}\ \bibnamefont {Trevithick}}, \bibinfo {author} {\bibfnamefont {A.}~\bibnamefont {Vainsencher}}, \bibinfo {author} {\bibfnamefont {B.}~\bibnamefont {Villalonga}}, \bibinfo {author} {\bibfnamefont {T.}~\bibnamefont {White}}, \bibinfo {author} {\bibfnamefont {Z.~J.}\ \bibnamefont {Yao}}, \bibinfo {author} {\bibfnamefont {P.}~\bibnamefont {Yeh}}, \bibinfo {author} {\bibfnamefont {A.}~\bibnamefont {Zalcman}}, \bibinfo {author} {\bibfnamefont {H.}~\bibnamefont {Neven}},\ and\ \bibinfo {author} {\bibfnamefont {J.~M.}\ \bibnamefont {Martinis}},\ }\bibfield  {title} {\bibinfo {title} {Quantum supremacy using a programmable superconducting processor},\ }\href {https://doi.org/10.1038/s41586-019-1666-5} {\bibfield  {journal}
  {\bibinfo  {journal} {Nature}\ }\textbf {\bibinfo {volume} {574}},\ \bibinfo {pages} {505–510} (\bibinfo {year} {2019})}\BibitemShut {NoStop}%
\bibitem [{\citenamefont {Zhang}\ \emph {et~al.}(2022)\citenamefont {Zhang}, \citenamefont {Jiang}, \citenamefont {Deng}, \citenamefont {Wang}, \citenamefont {Chen}, \citenamefont {Zhang}, \citenamefont {Ren}, \citenamefont {Dong}, \citenamefont {Xu}, \citenamefont {Gao} \emph {et~al.}}]{zhang2022digital}%
  \BibitemOpen
  \bibfield  {author} {\bibinfo {author} {\bibfnamefont {X.}~\bibnamefont {Zhang}}, \bibinfo {author} {\bibfnamefont {W.}~\bibnamefont {Jiang}}, \bibinfo {author} {\bibfnamefont {J.}~\bibnamefont {Deng}}, \bibinfo {author} {\bibfnamefont {K.}~\bibnamefont {Wang}}, \bibinfo {author} {\bibfnamefont {J.}~\bibnamefont {Chen}}, \bibinfo {author} {\bibfnamefont {P.}~\bibnamefont {Zhang}}, \bibinfo {author} {\bibfnamefont {W.}~\bibnamefont {Ren}}, \bibinfo {author} {\bibfnamefont {H.}~\bibnamefont {Dong}}, \bibinfo {author} {\bibfnamefont {S.}~\bibnamefont {Xu}}, \bibinfo {author} {\bibfnamefont {Y.}~\bibnamefont {Gao}}, \emph {et~al.},\ }\bibfield  {title} {\bibinfo {title} {Digital quantum simulation of floquet symmetry-protected topological phases},\ }\href {https://doi.org/10.1038/s41586-022-04854-3} {\bibfield  {journal} {\bibinfo  {journal} {Nature}\ }\textbf {\bibinfo {volume} {607}},\ \bibinfo {pages} {468} (\bibinfo {year} {2022})}\BibitemShut {NoStop}%
\bibitem [{\citenamefont {Jones}\ \emph {et~al.}(2022)\citenamefont {Jones}, \citenamefont {Hillberry}, \citenamefont {Jones}, \citenamefont {Fasihi}, \citenamefont {Roushan}, \citenamefont {Jiang}, \citenamefont {Ho}, \citenamefont {Neill}, \citenamefont {Ostby}, \citenamefont {Graf} \emph {et~al.}}]{jones2022small}%
  \BibitemOpen
  \bibfield  {author} {\bibinfo {author} {\bibfnamefont {E.~B.}\ \bibnamefont {Jones}}, \bibinfo {author} {\bibfnamefont {L.~E.}\ \bibnamefont {Hillberry}}, \bibinfo {author} {\bibfnamefont {M.~T.}\ \bibnamefont {Jones}}, \bibinfo {author} {\bibfnamefont {M.}~\bibnamefont {Fasihi}}, \bibinfo {author} {\bibfnamefont {P.}~\bibnamefont {Roushan}}, \bibinfo {author} {\bibfnamefont {Z.}~\bibnamefont {Jiang}}, \bibinfo {author} {\bibfnamefont {A.}~\bibnamefont {Ho}}, \bibinfo {author} {\bibfnamefont {C.}~\bibnamefont {Neill}}, \bibinfo {author} {\bibfnamefont {E.}~\bibnamefont {Ostby}}, \bibinfo {author} {\bibfnamefont {P.}~\bibnamefont {Graf}}, \emph {et~al.},\ }\bibfield  {title} {\bibinfo {title} {Small-world complex network generation on a digital quantum processor},\ }\href {https://doi.org/10.1038/s41467-022-32056-y} {\bibfield  {journal} {\bibinfo  {journal} {Nature Comm.}\ }\textbf {\bibinfo {volume} {13}},\ \bibinfo {pages} {4483} (\bibinfo {year} {2022})}\BibitemShut {NoStop}%
\bibitem [{\citenamefont {Zhang}\ \emph {et~al.}(2017)\citenamefont {Zhang}, \citenamefont {Hess}, \citenamefont {Kyprianidis}, \citenamefont {Becker}, \citenamefont {Lee}, \citenamefont {Smith}, \citenamefont {Pagano}, \citenamefont {Potirniche}, \citenamefont {Potter}, \citenamefont {Vishwanath}, \citenamefont {Yao},\ and\ \citenamefont {Monroe}}]{Zhang_2017}%
  \BibitemOpen
  \bibfield  {author} {\bibinfo {author} {\bibfnamefont {J.}~\bibnamefont {Zhang}}, \bibinfo {author} {\bibfnamefont {P.~W.}\ \bibnamefont {Hess}}, \bibinfo {author} {\bibfnamefont {A.}~\bibnamefont {Kyprianidis}}, \bibinfo {author} {\bibfnamefont {P.}~\bibnamefont {Becker}}, \bibinfo {author} {\bibfnamefont {A.}~\bibnamefont {Lee}}, \bibinfo {author} {\bibfnamefont {J.}~\bibnamefont {Smith}}, \bibinfo {author} {\bibfnamefont {G.}~\bibnamefont {Pagano}}, \bibinfo {author} {\bibfnamefont {I.-D.}\ \bibnamefont {Potirniche}}, \bibinfo {author} {\bibfnamefont {A.~C.}\ \bibnamefont {Potter}}, \bibinfo {author} {\bibfnamefont {A.}~\bibnamefont {Vishwanath}}, \bibinfo {author} {\bibfnamefont {N.~Y.}\ \bibnamefont {Yao}},\ and\ \bibinfo {author} {\bibfnamefont {C.}~\bibnamefont {Monroe}},\ }\bibfield  {title} {\bibinfo {title} {Observation of a discrete time crystal},\ }\href {https://doi.org/10.1038/nature21413} {\bibfield  {journal} {\bibinfo  {journal} {Nature}\ }\textbf {\bibinfo {volume} {543}},\ \bibinfo
  {pages} {217–220} (\bibinfo {year} {2017})}\BibitemShut {NoStop}%
\bibitem [{\citenamefont {Huerta~Alderete}\ \emph {et~al.}(2020)\citenamefont {Huerta~Alderete}, \citenamefont {Singh}, \citenamefont {Nguyen}, \citenamefont {Zhu}, \citenamefont {Balu}, \citenamefont {Monroe}, \citenamefont {Chandrashekar},\ and\ \citenamefont {Linke}}]{Huerta_Alderete_2020}%
  \BibitemOpen
  \bibfield  {author} {\bibinfo {author} {\bibfnamefont {C.}~\bibnamefont {Huerta~Alderete}}, \bibinfo {author} {\bibfnamefont {S.}~\bibnamefont {Singh}}, \bibinfo {author} {\bibfnamefont {N.~H.}\ \bibnamefont {Nguyen}}, \bibinfo {author} {\bibfnamefont {D.}~\bibnamefont {Zhu}}, \bibinfo {author} {\bibfnamefont {R.}~\bibnamefont {Balu}}, \bibinfo {author} {\bibfnamefont {C.}~\bibnamefont {Monroe}}, \bibinfo {author} {\bibfnamefont {C.~M.}\ \bibnamefont {Chandrashekar}},\ and\ \bibinfo {author} {\bibfnamefont {N.~M.}\ \bibnamefont {Linke}},\ }\bibfield  {title} {\bibinfo {title} {Quantum walks and dirac cellular automata on a programmable trapped-ion quantum computer},\ }\href {https://doi.org/10.1038/s41467-020-17519-4} {\bibfield  {journal} {\bibinfo  {journal} {Nature Comm.}\ }\textbf {\bibinfo {volume} {11}} (\bibinfo {year} {2020})}\BibitemShut {NoStop}%
\bibitem [{\citenamefont {Evered}\ \emph {et~al.}(2025{\natexlab{a}})\citenamefont {Evered}, \citenamefont {Kalinowski}, \citenamefont {Geim}, \citenamefont {Manovitz}, \citenamefont {Bluvstein}, \citenamefont {Li}, \citenamefont {Maskara}, \citenamefont {Zhou}, \citenamefont {Ebadi}, \citenamefont {Xu}, \citenamefont {Campo}, \citenamefont {Cain}, \citenamefont {Ostermann}, \citenamefont {Yelin}, \citenamefont {Sachdev}, \citenamefont {Greiner}, \citenamefont {Vuletić},\ and\ \citenamefont {Lukin}}]{Evered_2025}%
  \BibitemOpen
  \bibfield  {author} {\bibinfo {author} {\bibfnamefont {S.~J.}\ \bibnamefont {Evered}}, \bibinfo {author} {\bibfnamefont {M.}~\bibnamefont {Kalinowski}}, \bibinfo {author} {\bibfnamefont {A.~A.}\ \bibnamefont {Geim}}, \bibinfo {author} {\bibfnamefont {T.}~\bibnamefont {Manovitz}}, \bibinfo {author} {\bibfnamefont {D.}~\bibnamefont {Bluvstein}}, \bibinfo {author} {\bibfnamefont {S.~H.}\ \bibnamefont {Li}}, \bibinfo {author} {\bibfnamefont {N.}~\bibnamefont {Maskara}}, \bibinfo {author} {\bibfnamefont {H.}~\bibnamefont {Zhou}}, \bibinfo {author} {\bibfnamefont {S.}~\bibnamefont {Ebadi}}, \bibinfo {author} {\bibfnamefont {M.}~\bibnamefont {Xu}}, \bibinfo {author} {\bibfnamefont {J.}~\bibnamefont {Campo}}, \bibinfo {author} {\bibfnamefont {M.}~\bibnamefont {Cain}}, \bibinfo {author} {\bibfnamefont {S.}~\bibnamefont {Ostermann}}, \bibinfo {author} {\bibfnamefont {S.~F.}\ \bibnamefont {Yelin}}, \bibinfo {author} {\bibfnamefont {S.}~\bibnamefont {Sachdev}}, \bibinfo {author} {\bibfnamefont {M.}~\bibnamefont
  {Greiner}}, \bibinfo {author} {\bibfnamefont {V.}~\bibnamefont {Vuletić}},\ and\ \bibinfo {author} {\bibfnamefont {M.~D.}\ \bibnamefont {Lukin}},\ }\bibfield  {title} {\bibinfo {title} {Probing the kitaev honeycomb model on a neutral-atom quantum computer},\ }\href {https://doi.org/10.1038/s41586-025-09475-0} {\bibfield  {journal} {\bibinfo  {journal} {Nature}\ }\textbf {\bibinfo {volume} {645}},\ \bibinfo {pages} {341–347} (\bibinfo {year} {2025}{\natexlab{a}})}\BibitemShut {NoStop}%
\bibitem [{\citenamefont {White}\ \emph {et~al.}(2026)\citenamefont {White}, \citenamefont {Ramesh}, \citenamefont {Impertro}, \citenamefont {Anand}, \citenamefont {Cesa}, \citenamefont {Giudici}, \citenamefont {Iadecola}, \citenamefont {Pichler},\ and\ \citenamefont {Bernien}}]{ExperimentalPaper}%
  \BibitemOpen
  \bibfield  {author} {\bibinfo {author} {\bibfnamefont {R.}~\bibnamefont {White}}, \bibinfo {author} {\bibfnamefont {V.}~\bibnamefont {Ramesh}}, \bibinfo {author} {\bibfnamefont {A.}~\bibnamefont {Impertro}}, \bibinfo {author} {\bibfnamefont {S.}~\bibnamefont {Anand}}, \bibinfo {author} {\bibfnamefont {F.}~\bibnamefont {Cesa}}, \bibinfo {author} {\bibfnamefont {G.}~\bibnamefont {Giudici}}, \bibinfo {author} {\bibfnamefont {T.}~\bibnamefont {Iadecola}}, \bibinfo {author} {\bibfnamefont {H.}~\bibnamefont {Pichler}},\ and\ \bibinfo {author} {\bibfnamefont {H.}~\bibnamefont {Bernien}},\ }\bibfield  {title} {\bibinfo {title} {Quantum cellular automata on a dual-species rydberg processor},\ }\href@noop {} {\bibfield  {journal} {\bibinfo  {journal} {arXiv Preprint}\ } (\bibinfo {year} {2026})}\BibitemShut {NoStop}%
\bibitem [{\citenamefont {Saffman}\ \emph {et~al.}(2010)\citenamefont {Saffman}, \citenamefont {Walker},\ and\ \citenamefont {M\o{}lmer}}]{saffman2010quantum}%
  \BibitemOpen
  \bibfield  {author} {\bibinfo {author} {\bibfnamefont {M.}~\bibnamefont {Saffman}}, \bibinfo {author} {\bibfnamefont {T.~G.}\ \bibnamefont {Walker}},\ and\ \bibinfo {author} {\bibfnamefont {K.}~\bibnamefont {M\o{}lmer}},\ }\bibfield  {title} {\bibinfo {title} {Quantum information with rydberg atoms},\ }\href {https://doi.org/10.1103/RevModPhys.82.2313} {\bibfield  {journal} {\bibinfo  {journal} {Rev. Mod. Phys.}\ }\textbf {\bibinfo {volume} {82}},\ \bibinfo {pages} {2313} (\bibinfo {year} {2010})}\BibitemShut {NoStop}%
\bibitem [{\citenamefont {Kaufman}\ and\ \citenamefont {Ni}(2021)}]{kaufman2021quantum}%
  \BibitemOpen
  \bibfield  {author} {\bibinfo {author} {\bibfnamefont {A.~M.}\ \bibnamefont {Kaufman}}\ and\ \bibinfo {author} {\bibfnamefont {K.-K.}\ \bibnamefont {Ni}},\ }\bibfield  {title} {\bibinfo {title} {Quantum science with optical tweezer arrays of ultracold atoms and molecules},\ }\href {https://doi.org/10.1038/s41567-021-01357-2} {\bibfield  {journal} {\bibinfo  {journal} {Nature Phys.}\ }\textbf {\bibinfo {volume} {17}},\ \bibinfo {pages} {1324} (\bibinfo {year} {2021})}\BibitemShut {NoStop}%
\bibitem [{\citenamefont {Manetsch}\ \emph {et~al.}(2025)\citenamefont {Manetsch}, \citenamefont {Nomura}, \citenamefont {Bataille}, \citenamefont {Lv}, \citenamefont {Leung},\ and\ \citenamefont {Endres}}]{Manetsch2025atoms6100}%
  \BibitemOpen
  \bibfield  {author} {\bibinfo {author} {\bibfnamefont {H.~J.}\ \bibnamefont {Manetsch}}, \bibinfo {author} {\bibfnamefont {G.}~\bibnamefont {Nomura}}, \bibinfo {author} {\bibfnamefont {E.}~\bibnamefont {Bataille}}, \bibinfo {author} {\bibfnamefont {X.}~\bibnamefont {Lv}}, \bibinfo {author} {\bibfnamefont {K.~H.}\ \bibnamefont {Leung}},\ and\ \bibinfo {author} {\bibfnamefont {M.}~\bibnamefont {Endres}},\ }\bibfield  {title} {\bibinfo {title} {A tweezer array with 6,100 highly coherent atomic qubits},\ }\href {https://doi.org/10.1038/s41586-025-09641-4} {\bibfield  {journal} {\bibinfo  {journal} {Nature}\ }\textbf {\bibinfo {volume} {647}},\ \bibinfo {pages} {60} (\bibinfo {year} {2025})}\BibitemShut {NoStop}%
\bibitem [{\citenamefont {Pause}\ \emph {et~al.}(2024)\citenamefont {Pause}, \citenamefont {Sturm}, \citenamefont {Mittenb\"{u}hler}, \citenamefont {Amann}, \citenamefont {Preuschoff}, \citenamefont {Sch\"{a}ffner}, \citenamefont {Schlosser},\ and\ \citenamefont {Birkl}}]{Pause2024atoms1000}%
  \BibitemOpen
  \bibfield  {author} {\bibinfo {author} {\bibfnamefont {L.}~\bibnamefont {Pause}}, \bibinfo {author} {\bibfnamefont {L.}~\bibnamefont {Sturm}}, \bibinfo {author} {\bibfnamefont {M.}~\bibnamefont {Mittenb\"{u}hler}}, \bibinfo {author} {\bibfnamefont {S.}~\bibnamefont {Amann}}, \bibinfo {author} {\bibfnamefont {T.}~\bibnamefont {Preuschoff}}, \bibinfo {author} {\bibfnamefont {D.}~\bibnamefont {Sch\"{a}ffner}}, \bibinfo {author} {\bibfnamefont {M.}~\bibnamefont {Schlosser}},\ and\ \bibinfo {author} {\bibfnamefont {G.}~\bibnamefont {Birkl}},\ }\bibfield  {title} {\bibinfo {title} {Supercharged two-dimensional tweezer array with more than 1000 atomic qubits},\ }\href {https://doi.org/10.1364/OPTICA.513551} {\bibfield  {journal} {\bibinfo  {journal} {Optica}\ }\textbf {\bibinfo {volume} {11}},\ \bibinfo {pages} {222} (\bibinfo {year} {2024})}\BibitemShut {NoStop}%
\bibitem [{\citenamefont {Tao}\ \emph {et~al.}(2024)\citenamefont {Tao}, \citenamefont {Ammenwerth}, \citenamefont {Gyger}, \citenamefont {Bloch},\ and\ \citenamefont {Zeiher}}]{Tao2024Large}%
  \BibitemOpen
  \bibfield  {author} {\bibinfo {author} {\bibfnamefont {R.}~\bibnamefont {Tao}}, \bibinfo {author} {\bibfnamefont {M.}~\bibnamefont {Ammenwerth}}, \bibinfo {author} {\bibfnamefont {F.}~\bibnamefont {Gyger}}, \bibinfo {author} {\bibfnamefont {I.}~\bibnamefont {Bloch}},\ and\ \bibinfo {author} {\bibfnamefont {J.}~\bibnamefont {Zeiher}},\ }\bibfield  {title} {\bibinfo {title} {High-fidelity detection of large-scale atom arrays in an optical lattice},\ }\href {https://doi.org/10.1103/PhysRevLett.133.013401} {\bibfield  {journal} {\bibinfo  {journal} {Phys. Rev. Lett.}\ }\textbf {\bibinfo {volume} {133}},\ \bibinfo {pages} {013401} (\bibinfo {year} {2024})}\BibitemShut {NoStop}%
\bibitem [{\citenamefont {Browaeys}\ and\ \citenamefont {Lahaye}(2020)}]{browaeys2020many}%
  \BibitemOpen
  \bibfield  {author} {\bibinfo {author} {\bibfnamefont {A.}~\bibnamefont {Browaeys}}\ and\ \bibinfo {author} {\bibfnamefont {T.}~\bibnamefont {Lahaye}},\ }\bibfield  {title} {\bibinfo {title} {Many-body physics with individually controlled rydberg atoms},\ }\href {https://doi.org/10.1038/s41567-019-0733-z} {\bibfield  {journal} {\bibinfo  {journal} {Nature Physics}\ }\textbf {\bibinfo {volume} {16}},\ \bibinfo {pages} {132} (\bibinfo {year} {2020})}\BibitemShut {NoStop}%
\bibitem [{\citenamefont {Schau{\ss}}\ \emph {et~al.}(2015)\citenamefont {Schau{\ss}}, \citenamefont {Zeiher}, \citenamefont {Fukuhara}, \citenamefont {Hild}, \citenamefont {Cheneau}, \citenamefont {Macr{\`\i}}, \citenamefont {Pohl}, \citenamefont {Bloch},\ and\ \citenamefont {Gro{\ss}}}]{schauss2015crystallization}%
  \BibitemOpen
  \bibfield  {author} {\bibinfo {author} {\bibfnamefont {P.}~\bibnamefont {Schau{\ss}}}, \bibinfo {author} {\bibfnamefont {J.}~\bibnamefont {Zeiher}}, \bibinfo {author} {\bibfnamefont {T.}~\bibnamefont {Fukuhara}}, \bibinfo {author} {\bibfnamefont {S.}~\bibnamefont {Hild}}, \bibinfo {author} {\bibfnamefont {M.}~\bibnamefont {Cheneau}}, \bibinfo {author} {\bibfnamefont {T.}~\bibnamefont {Macr{\`\i}}}, \bibinfo {author} {\bibfnamefont {T.}~\bibnamefont {Pohl}}, \bibinfo {author} {\bibfnamefont {I.}~\bibnamefont {Bloch}},\ and\ \bibinfo {author} {\bibfnamefont {C.}~\bibnamefont {Gro{\ss}}},\ }\bibfield  {title} {\bibinfo {title} {Crystallization in ising quantum magnets},\ }\href {https://doi.org/10.1126/science.1258351} {\bibfield  {journal} {\bibinfo  {journal} {Science}\ }\textbf {\bibinfo {volume} {347}},\ \bibinfo {pages} {1455} (\bibinfo {year} {2015})}\BibitemShut {NoStop}%
\bibitem [{\citenamefont {Labuhn}\ \emph {et~al.}(2016)\citenamefont {Labuhn}, \citenamefont {Barredo}, \citenamefont {Ravets}, \citenamefont {De~L{\'e}s{\'e}leuc}, \citenamefont {Macr{\`\i}}, \citenamefont {Lahaye},\ and\ \citenamefont {Browaeys}}]{labuhn2016tunable}%
  \BibitemOpen
  \bibfield  {author} {\bibinfo {author} {\bibfnamefont {H.}~\bibnamefont {Labuhn}}, \bibinfo {author} {\bibfnamefont {D.}~\bibnamefont {Barredo}}, \bibinfo {author} {\bibfnamefont {S.}~\bibnamefont {Ravets}}, \bibinfo {author} {\bibfnamefont {S.}~\bibnamefont {De~L{\'e}s{\'e}leuc}}, \bibinfo {author} {\bibfnamefont {T.}~\bibnamefont {Macr{\`\i}}}, \bibinfo {author} {\bibfnamefont {T.}~\bibnamefont {Lahaye}},\ and\ \bibinfo {author} {\bibfnamefont {A.}~\bibnamefont {Browaeys}},\ }\bibfield  {title} {\bibinfo {title} {Tunable two-dimensional arrays of single rydberg atoms for realizing quantum ising models},\ }\href {https://doi.org/10.1038/nature18274} {\bibfield  {journal} {\bibinfo  {journal} {Nature}\ }\textbf {\bibinfo {volume} {534}},\ \bibinfo {pages} {667} (\bibinfo {year} {2016})}\BibitemShut {NoStop}%
\bibitem [{\citenamefont {Bernien}\ \emph {et~al.}(2017)\citenamefont {Bernien}, \citenamefont {Schwartz}, \citenamefont {Keesling}, \citenamefont {Levine}, \citenamefont {Omran}, \citenamefont {Pichler}, \citenamefont {Choi}, \citenamefont {Zibrov}, \citenamefont {Endres}, \citenamefont {Greiner} \emph {et~al.}}]{bernien2017probing}%
  \BibitemOpen
  \bibfield  {author} {\bibinfo {author} {\bibfnamefont {H.}~\bibnamefont {Bernien}}, \bibinfo {author} {\bibfnamefont {S.}~\bibnamefont {Schwartz}}, \bibinfo {author} {\bibfnamefont {A.}~\bibnamefont {Keesling}}, \bibinfo {author} {\bibfnamefont {H.}~\bibnamefont {Levine}}, \bibinfo {author} {\bibfnamefont {A.}~\bibnamefont {Omran}}, \bibinfo {author} {\bibfnamefont {H.}~\bibnamefont {Pichler}}, \bibinfo {author} {\bibfnamefont {S.}~\bibnamefont {Choi}}, \bibinfo {author} {\bibfnamefont {A.~S.}\ \bibnamefont {Zibrov}}, \bibinfo {author} {\bibfnamefont {M.}~\bibnamefont {Endres}}, \bibinfo {author} {\bibfnamefont {M.}~\bibnamefont {Greiner}}, \emph {et~al.},\ }\bibfield  {title} {\bibinfo {title} {Probing many-body dynamics on a 51-atom quantum simulator},\ }\href {https://doi.org/10.1038/nature24622} {\bibfield  {journal} {\bibinfo  {journal} {Nature}\ }\textbf {\bibinfo {volume} {551}},\ \bibinfo {pages} {579} (\bibinfo {year} {2017})}\BibitemShut {NoStop}%
\bibitem [{\citenamefont {de~L\'es\'eleuc}\ \emph {et~al.}(2018)\citenamefont {de~L\'es\'eleuc}, \citenamefont {Weber}, \citenamefont {Lienhard}, \citenamefont {Barredo}, \citenamefont {B\"uchler}, \citenamefont {Lahaye},\ and\ \citenamefont {Browaeys}}]{deleseleuc2018accurate}%
  \BibitemOpen
  \bibfield  {author} {\bibinfo {author} {\bibfnamefont {S.}~\bibnamefont {de~L\'es\'eleuc}}, \bibinfo {author} {\bibfnamefont {S.}~\bibnamefont {Weber}}, \bibinfo {author} {\bibfnamefont {V.}~\bibnamefont {Lienhard}}, \bibinfo {author} {\bibfnamefont {D.}~\bibnamefont {Barredo}}, \bibinfo {author} {\bibfnamefont {H.~P.}\ \bibnamefont {B\"uchler}}, \bibinfo {author} {\bibfnamefont {T.}~\bibnamefont {Lahaye}},\ and\ \bibinfo {author} {\bibfnamefont {A.}~\bibnamefont {Browaeys}},\ }\bibfield  {title} {\bibinfo {title} {Accurate mapping of multilevel rydberg atoms on interacting spin-$1/2$ particles for the quantum simulation of ising models},\ }\href {https://doi.org/10.1103/PhysRevLett.120.113602} {\bibfield  {journal} {\bibinfo  {journal} {Phys. Rev. Lett.}\ }\textbf {\bibinfo {volume} {120}},\ \bibinfo {pages} {113602} (\bibinfo {year} {2018})}\BibitemShut {NoStop}%
\bibitem [{\citenamefont {Schauss}(2018)}]{schauss2018quantum}%
  \BibitemOpen
  \bibfield  {author} {\bibinfo {author} {\bibfnamefont {P.}~\bibnamefont {Schauss}},\ }\bibfield  {title} {\bibinfo {title} {Quantum simulation of transverse ising models with rydberg atoms},\ }\href {https://doi.org/10.1088/2058-9565/aa9c59} {\bibfield  {journal} {\bibinfo  {journal} {Quantum Science Tech.}\ }\textbf {\bibinfo {volume} {3}},\ \bibinfo {pages} {023001} (\bibinfo {year} {2018})}\BibitemShut {NoStop}%
\bibitem [{\citenamefont {Guardado-Sanchez}\ \emph {et~al.}(2018)\citenamefont {Guardado-Sanchez}, \citenamefont {Brown}, \citenamefont {Mitra}, \citenamefont {Devakul}, \citenamefont {Huse}, \citenamefont {Schau\ss{}},\ and\ \citenamefont {Bakr}}]{guardado2018probing}%
  \BibitemOpen
  \bibfield  {author} {\bibinfo {author} {\bibfnamefont {E.}~\bibnamefont {Guardado-Sanchez}}, \bibinfo {author} {\bibfnamefont {P.~T.}\ \bibnamefont {Brown}}, \bibinfo {author} {\bibfnamefont {D.}~\bibnamefont {Mitra}}, \bibinfo {author} {\bibfnamefont {T.}~\bibnamefont {Devakul}}, \bibinfo {author} {\bibfnamefont {D.~A.}\ \bibnamefont {Huse}}, \bibinfo {author} {\bibfnamefont {P.}~\bibnamefont {Schau\ss{}}},\ and\ \bibinfo {author} {\bibfnamefont {W.~S.}\ \bibnamefont {Bakr}},\ }\bibfield  {title} {\bibinfo {title} {Probing the quench dynamics of antiferromagnetic correlations in a 2d quantum ising spin system},\ }\href {https://doi.org/10.1103/PhysRevX.8.021069} {\bibfield  {journal} {\bibinfo  {journal} {Phys. Rev. X}\ }\textbf {\bibinfo {volume} {8}},\ \bibinfo {pages} {021069} (\bibinfo {year} {2018})}\BibitemShut {NoStop}%
\bibitem [{\citenamefont {Lienhard}\ \emph {et~al.}(2018)\citenamefont {Lienhard}, \citenamefont {de~L\'es\'eleuc}, \citenamefont {Barredo}, \citenamefont {Lahaye}, \citenamefont {Browaeys}, \citenamefont {Schuler}, \citenamefont {Henry},\ and\ \citenamefont {L\"auchli}}]{lienhard2018observing}%
  \BibitemOpen
  \bibfield  {author} {\bibinfo {author} {\bibfnamefont {V.}~\bibnamefont {Lienhard}}, \bibinfo {author} {\bibfnamefont {S.}~\bibnamefont {de~L\'es\'eleuc}}, \bibinfo {author} {\bibfnamefont {D.}~\bibnamefont {Barredo}}, \bibinfo {author} {\bibfnamefont {T.}~\bibnamefont {Lahaye}}, \bibinfo {author} {\bibfnamefont {A.}~\bibnamefont {Browaeys}}, \bibinfo {author} {\bibfnamefont {M.}~\bibnamefont {Schuler}}, \bibinfo {author} {\bibfnamefont {L.-P.}\ \bibnamefont {Henry}},\ and\ \bibinfo {author} {\bibfnamefont {A.~M.}\ \bibnamefont {L\"auchli}},\ }\bibfield  {title} {\bibinfo {title} {Observing the space- and time-dependent growth of correlations in dynamically tuned synthetic ising models with antiferromagnetic interactions},\ }\href {https://doi.org/10.1103/PhysRevX.8.021070} {\bibfield  {journal} {\bibinfo  {journal} {Phys. Rev. X}\ }\textbf {\bibinfo {volume} {8}},\ \bibinfo {pages} {021070} (\bibinfo {year} {2018})}\BibitemShut {NoStop}%
\bibitem [{\citenamefont {de~L{\'e}s{\'e}leuc}\ \emph {et~al.}(2019)\citenamefont {de~L{\'e}s{\'e}leuc}, \citenamefont {Lienhard}, \citenamefont {Scholl}, \citenamefont {Barredo}, \citenamefont {Weber}, \citenamefont {Lang}, \citenamefont {B{\"u}chler}, \citenamefont {Lahaye},\ and\ \citenamefont {Browaeys}}]{Leseleuc2019topological}%
  \BibitemOpen
  \bibfield  {author} {\bibinfo {author} {\bibfnamefont {S.}~\bibnamefont {de~L{\'e}s{\'e}leuc}}, \bibinfo {author} {\bibfnamefont {V.}~\bibnamefont {Lienhard}}, \bibinfo {author} {\bibfnamefont {P.}~\bibnamefont {Scholl}}, \bibinfo {author} {\bibfnamefont {D.}~\bibnamefont {Barredo}}, \bibinfo {author} {\bibfnamefont {S.}~\bibnamefont {Weber}}, \bibinfo {author} {\bibfnamefont {N.}~\bibnamefont {Lang}}, \bibinfo {author} {\bibfnamefont {H.~P.}\ \bibnamefont {B{\"u}chler}}, \bibinfo {author} {\bibfnamefont {T.}~\bibnamefont {Lahaye}},\ and\ \bibinfo {author} {\bibfnamefont {A.}~\bibnamefont {Browaeys}},\ }\bibfield  {title} {\bibinfo {title} {{Observation of a symmetry-protected topological phase of interacting bosons with Rydberg atoms}},\ }\href {https://doi.org/10.1126/science.aav9105} {\bibfield  {journal} {\bibinfo  {journal} {Science}\ }\textbf {\bibinfo {volume} {365}},\ \bibinfo {pages} {775} (\bibinfo {year} {2019})}\BibitemShut {NoStop}%
\bibitem [{\citenamefont {Keesling}\ \emph {et~al.}(2019)\citenamefont {Keesling}, \citenamefont {Omran}, \citenamefont {Levine}, \citenamefont {Bernien}, \citenamefont {Pichler}, \citenamefont {Choi}, \citenamefont {Samajdar}, \citenamefont {Schwartz}, \citenamefont {Silvi}, \citenamefont {Sachdev}, \citenamefont {Zoller}, \citenamefont {Endres}, \citenamefont {Greiner}, \citenamefont {Vuleti{\'c}},\ and\ \citenamefont {Lukin}}]{Keesling2019QPT}%
  \BibitemOpen
  \bibfield  {author} {\bibinfo {author} {\bibfnamefont {A.}~\bibnamefont {Keesling}}, \bibinfo {author} {\bibfnamefont {A.}~\bibnamefont {Omran}}, \bibinfo {author} {\bibfnamefont {H.}~\bibnamefont {Levine}}, \bibinfo {author} {\bibfnamefont {H.}~\bibnamefont {Bernien}}, \bibinfo {author} {\bibfnamefont {H.}~\bibnamefont {Pichler}}, \bibinfo {author} {\bibfnamefont {S.}~\bibnamefont {Choi}}, \bibinfo {author} {\bibfnamefont {R.}~\bibnamefont {Samajdar}}, \bibinfo {author} {\bibfnamefont {S.}~\bibnamefont {Schwartz}}, \bibinfo {author} {\bibfnamefont {P.}~\bibnamefont {Silvi}}, \bibinfo {author} {\bibfnamefont {S.}~\bibnamefont {Sachdev}}, \bibinfo {author} {\bibfnamefont {P.}~\bibnamefont {Zoller}}, \bibinfo {author} {\bibfnamefont {M.}~\bibnamefont {Endres}}, \bibinfo {author} {\bibfnamefont {M.}~\bibnamefont {Greiner}}, \bibinfo {author} {\bibfnamefont {V.}~\bibnamefont {Vuleti{\'c}}},\ and\ \bibinfo {author} {\bibfnamefont {M.~D.}\ \bibnamefont {Lukin}},\ }\bibfield  {title} {\bibinfo {title} {{Quantum
  Kibble--Zurek mechanism and critical dynamics on a programmable Rydberg simulator}},\ }\href {https://doi.org/10.1038/s41586-019-1070-1} {\bibfield  {journal} {\bibinfo  {journal} {Nature}\ }\textbf {\bibinfo {volume} {568}},\ \bibinfo {pages} {207} (\bibinfo {year} {2019})}\BibitemShut {NoStop}%
\bibitem [{\citenamefont {Surace}\ \emph {et~al.}(2020)\citenamefont {Surace}, \citenamefont {Mazza}, \citenamefont {Giudici}, \citenamefont {Lerose}, \citenamefont {Gambassi},\ and\ \citenamefont {Dalmonte}}]{surace2020lattice}%
  \BibitemOpen
  \bibfield  {author} {\bibinfo {author} {\bibfnamefont {F.~M.}\ \bibnamefont {Surace}}, \bibinfo {author} {\bibfnamefont {P.~P.}\ \bibnamefont {Mazza}}, \bibinfo {author} {\bibfnamefont {G.}~\bibnamefont {Giudici}}, \bibinfo {author} {\bibfnamefont {A.}~\bibnamefont {Lerose}}, \bibinfo {author} {\bibfnamefont {A.}~\bibnamefont {Gambassi}},\ and\ \bibinfo {author} {\bibfnamefont {M.}~\bibnamefont {Dalmonte}},\ }\bibfield  {title} {\bibinfo {title} {Lattice gauge theories and string dynamics in rydberg atom quantum simulators},\ }\href {https://doi.org/10.1103/PhysRevX.10.021041} {\bibfield  {journal} {\bibinfo  {journal} {Phys. Rev. X}\ }\textbf {\bibinfo {volume} {10}},\ \bibinfo {pages} {021041} (\bibinfo {year} {2020})}\BibitemShut {NoStop}%
\bibitem [{\citenamefont {Scholl}\ \emph {et~al.}(2021)\citenamefont {Scholl}, \citenamefont {Schuler}, \citenamefont {Williams}, \citenamefont {Eberharter}, \citenamefont {Barredo}, \citenamefont {Schymik}, \citenamefont {Lienhard}, \citenamefont {Henry}, \citenamefont {Lang}, \citenamefont {Lahaye} \emph {et~al.}}]{scholl2021quantum}%
  \BibitemOpen
  \bibfield  {author} {\bibinfo {author} {\bibfnamefont {P.}~\bibnamefont {Scholl}}, \bibinfo {author} {\bibfnamefont {M.}~\bibnamefont {Schuler}}, \bibinfo {author} {\bibfnamefont {H.~J.}\ \bibnamefont {Williams}}, \bibinfo {author} {\bibfnamefont {A.~A.}\ \bibnamefont {Eberharter}}, \bibinfo {author} {\bibfnamefont {D.}~\bibnamefont {Barredo}}, \bibinfo {author} {\bibfnamefont {K.-N.}\ \bibnamefont {Schymik}}, \bibinfo {author} {\bibfnamefont {V.}~\bibnamefont {Lienhard}}, \bibinfo {author} {\bibfnamefont {L.-P.}\ \bibnamefont {Henry}}, \bibinfo {author} {\bibfnamefont {T.~C.}\ \bibnamefont {Lang}}, \bibinfo {author} {\bibfnamefont {T.}~\bibnamefont {Lahaye}}, \emph {et~al.},\ }\bibfield  {title} {\bibinfo {title} {Quantum simulation of 2d antiferromagnets with hundreds of rydberg atoms},\ }\href {https://doi.org/10.1038/s41586-021-03585-1} {\bibfield  {journal} {\bibinfo  {journal} {Nature}\ }\textbf {\bibinfo {volume} {595}},\ \bibinfo {pages} {233} (\bibinfo {year} {2021})}\BibitemShut {NoStop}%
\bibitem [{\citenamefont {Bluvstein}\ \emph {et~al.}(2021)\citenamefont {Bluvstein}, \citenamefont {Omran}, \citenamefont {Levine}, \citenamefont {Keesling}, \citenamefont {Semeghini}, \citenamefont {Ebadi}, \citenamefont {Wang}, \citenamefont {Michailidis}, \citenamefont {Maskara}, \citenamefont {Ho} \emph {et~al.}}]{bluvstein2021controlling}%
  \BibitemOpen
  \bibfield  {author} {\bibinfo {author} {\bibfnamefont {D.}~\bibnamefont {Bluvstein}}, \bibinfo {author} {\bibfnamefont {A.}~\bibnamefont {Omran}}, \bibinfo {author} {\bibfnamefont {H.}~\bibnamefont {Levine}}, \bibinfo {author} {\bibfnamefont {A.}~\bibnamefont {Keesling}}, \bibinfo {author} {\bibfnamefont {G.}~\bibnamefont {Semeghini}}, \bibinfo {author} {\bibfnamefont {S.}~\bibnamefont {Ebadi}}, \bibinfo {author} {\bibfnamefont {T.~T.}\ \bibnamefont {Wang}}, \bibinfo {author} {\bibfnamefont {A.~A.}\ \bibnamefont {Michailidis}}, \bibinfo {author} {\bibfnamefont {N.}~\bibnamefont {Maskara}}, \bibinfo {author} {\bibfnamefont {W.~W.}\ \bibnamefont {Ho}}, \emph {et~al.},\ }\bibfield  {title} {\bibinfo {title} {Controlling quantum many-body dynamics in driven rydberg atom arrays},\ }\href {https://doi.org/10.1126/science.abg2530} {\bibfield  {journal} {\bibinfo  {journal} {Science}\ }\textbf {\bibinfo {volume} {371}},\ \bibinfo {pages} {1355} (\bibinfo {year} {2021})}\BibitemShut {NoStop}%
\bibitem [{\citenamefont {Lippe}\ \emph {et~al.}(2021)\citenamefont {Lippe}, \citenamefont {Klas}, \citenamefont {Bender}, \citenamefont {Mischke}, \citenamefont {Niederpr{\"u}m},\ and\ \citenamefont {Ott}}]{lippe2021experimental}%
  \BibitemOpen
  \bibfield  {author} {\bibinfo {author} {\bibfnamefont {C.}~\bibnamefont {Lippe}}, \bibinfo {author} {\bibfnamefont {T.}~\bibnamefont {Klas}}, \bibinfo {author} {\bibfnamefont {J.}~\bibnamefont {Bender}}, \bibinfo {author} {\bibfnamefont {P.}~\bibnamefont {Mischke}}, \bibinfo {author} {\bibfnamefont {T.}~\bibnamefont {Niederpr{\"u}m}},\ and\ \bibinfo {author} {\bibfnamefont {H.}~\bibnamefont {Ott}},\ }\bibfield  {title} {\bibinfo {title} {Experimental realization of a 3d random hopping model},\ }\href {https://doi.org/10.1038/s41467-021-27243-2} {\bibfield  {journal} {\bibinfo  {journal} {Nature Comm.}\ }\textbf {\bibinfo {volume} {12}},\ \bibinfo {pages} {6976} (\bibinfo {year} {2021})}\BibitemShut {NoStop}%
\bibitem [{\citenamefont {Ebadi}\ \emph {et~al.}(2021)\citenamefont {Ebadi}, \citenamefont {Wang}, \citenamefont {Levine}, \citenamefont {Keesling}, \citenamefont {Semeghini}, \citenamefont {Omran}, \citenamefont {Bluvstein}, \citenamefont {Samajdar}, \citenamefont {Pichler}, \citenamefont {Ho} \emph {et~al.}}]{ebadi2021quantum}%
  \BibitemOpen
  \bibfield  {author} {\bibinfo {author} {\bibfnamefont {S.}~\bibnamefont {Ebadi}}, \bibinfo {author} {\bibfnamefont {T.~T.}\ \bibnamefont {Wang}}, \bibinfo {author} {\bibfnamefont {H.}~\bibnamefont {Levine}}, \bibinfo {author} {\bibfnamefont {A.}~\bibnamefont {Keesling}}, \bibinfo {author} {\bibfnamefont {G.}~\bibnamefont {Semeghini}}, \bibinfo {author} {\bibfnamefont {A.}~\bibnamefont {Omran}}, \bibinfo {author} {\bibfnamefont {D.}~\bibnamefont {Bluvstein}}, \bibinfo {author} {\bibfnamefont {R.}~\bibnamefont {Samajdar}}, \bibinfo {author} {\bibfnamefont {H.}~\bibnamefont {Pichler}}, \bibinfo {author} {\bibfnamefont {W.~W.}\ \bibnamefont {Ho}}, \emph {et~al.},\ }\bibfield  {title} {\bibinfo {title} {Quantum phases of matter on a 256-atom programmable quantum simulator},\ }\href {https://doi.org/10.1038/s41586-021-03582-4} {\bibfield  {journal} {\bibinfo  {journal} {Nature}\ }\textbf {\bibinfo {volume} {595}},\ \bibinfo {pages} {227} (\bibinfo {year} {2021})}\BibitemShut {NoStop}%
\bibitem [{\citenamefont {Semeghini}\ \emph {et~al.}(2021)\citenamefont {Semeghini}, \citenamefont {Levine}, \citenamefont {Keesling}, \citenamefont {Ebadi}, \citenamefont {Wang}, \citenamefont {Bluvstein}, \citenamefont {Verresen}, \citenamefont {Pichler}, \citenamefont {Kalinowski}, \citenamefont {Samajdar} \emph {et~al.}}]{semeghini2021probing}%
  \BibitemOpen
  \bibfield  {author} {\bibinfo {author} {\bibfnamefont {G.}~\bibnamefont {Semeghini}}, \bibinfo {author} {\bibfnamefont {H.}~\bibnamefont {Levine}}, \bibinfo {author} {\bibfnamefont {A.}~\bibnamefont {Keesling}}, \bibinfo {author} {\bibfnamefont {S.}~\bibnamefont {Ebadi}}, \bibinfo {author} {\bibfnamefont {T.~T.}\ \bibnamefont {Wang}}, \bibinfo {author} {\bibfnamefont {D.}~\bibnamefont {Bluvstein}}, \bibinfo {author} {\bibfnamefont {R.}~\bibnamefont {Verresen}}, \bibinfo {author} {\bibfnamefont {H.}~\bibnamefont {Pichler}}, \bibinfo {author} {\bibfnamefont {M.}~\bibnamefont {Kalinowski}}, \bibinfo {author} {\bibfnamefont {R.}~\bibnamefont {Samajdar}}, \emph {et~al.},\ }\bibfield  {title} {\bibinfo {title} {Probing topological spin liquids on a programmable quantum simulator},\ }\href {https://doi.org/10.1126/science.abi8794} {\bibfield  {journal} {\bibinfo  {journal} {Science}\ }\textbf {\bibinfo {volume} {374}},\ \bibinfo {pages} {1242} (\bibinfo {year} {2021})}\BibitemShut {NoStop}%
\bibitem [{\citenamefont {Choi}\ \emph {et~al.}(2023)\citenamefont {Choi}, \citenamefont {Shaw}, \citenamefont {Madjarov}, \citenamefont {Xie}, \citenamefont {Finkelstein}, \citenamefont {Covey}, \citenamefont {Cotler}, \citenamefont {Mark}, \citenamefont {Huang}, \citenamefont {Kale}, \citenamefont {Pichler}, \citenamefont {Brand{\~a}o}, \citenamefont {Choi},\ and\ \citenamefont {Endres}}]{Choi2023random}%
  \BibitemOpen
  \bibfield  {author} {\bibinfo {author} {\bibfnamefont {J.}~\bibnamefont {Choi}}, \bibinfo {author} {\bibfnamefont {A.~L.}\ \bibnamefont {Shaw}}, \bibinfo {author} {\bibfnamefont {I.~S.}\ \bibnamefont {Madjarov}}, \bibinfo {author} {\bibfnamefont {X.}~\bibnamefont {Xie}}, \bibinfo {author} {\bibfnamefont {R.}~\bibnamefont {Finkelstein}}, \bibinfo {author} {\bibfnamefont {J.~P.}\ \bibnamefont {Covey}}, \bibinfo {author} {\bibfnamefont {J.~S.}\ \bibnamefont {Cotler}}, \bibinfo {author} {\bibfnamefont {D.~K.}\ \bibnamefont {Mark}}, \bibinfo {author} {\bibfnamefont {H.-Y.}\ \bibnamefont {Huang}}, \bibinfo {author} {\bibfnamefont {A.}~\bibnamefont {Kale}}, \bibinfo {author} {\bibfnamefont {H.}~\bibnamefont {Pichler}}, \bibinfo {author} {\bibfnamefont {F.~G. S.~L.}\ \bibnamefont {Brand{\~a}o}}, \bibinfo {author} {\bibfnamefont {S.}~\bibnamefont {Choi}},\ and\ \bibinfo {author} {\bibfnamefont {M.}~\bibnamefont {Endres}},\ }\bibfield  {title} {\bibinfo {title} {Preparing random states and benchmarking with many-body
  quantum chaos},\ }\href {https://doi.org/10.1038/s41586-022-05442-1} {\bibfield  {journal} {\bibinfo  {journal} {Nature}\ }\textbf {\bibinfo {volume} {613}},\ \bibinfo {pages} {468} (\bibinfo {year} {2023})}\BibitemShut {NoStop}%
\bibitem [{\citenamefont {Chen}\ \emph {et~al.}(2023)\citenamefont {Chen}, \citenamefont {Bornet}, \citenamefont {Bintz}, \citenamefont {Emperauger}, \citenamefont {Leclerc}, \citenamefont {Liu}, \citenamefont {Scholl}, \citenamefont {Barredo}, \citenamefont {Hauschild}, \citenamefont {Chatterjee}, \citenamefont {Schuler}, \citenamefont {L{\"a}uchli}, \citenamefont {Zaletel}, \citenamefont {Lahaye}, \citenamefont {Yao},\ and\ \citenamefont {Browaeys}}]{Chen2024breaking}%
  \BibitemOpen
  \bibfield  {author} {\bibinfo {author} {\bibfnamefont {C.}~\bibnamefont {Chen}}, \bibinfo {author} {\bibfnamefont {G.}~\bibnamefont {Bornet}}, \bibinfo {author} {\bibfnamefont {M.}~\bibnamefont {Bintz}}, \bibinfo {author} {\bibfnamefont {G.}~\bibnamefont {Emperauger}}, \bibinfo {author} {\bibfnamefont {L.}~\bibnamefont {Leclerc}}, \bibinfo {author} {\bibfnamefont {V.~S.}\ \bibnamefont {Liu}}, \bibinfo {author} {\bibfnamefont {P.}~\bibnamefont {Scholl}}, \bibinfo {author} {\bibfnamefont {D.}~\bibnamefont {Barredo}}, \bibinfo {author} {\bibfnamefont {J.}~\bibnamefont {Hauschild}}, \bibinfo {author} {\bibfnamefont {S.}~\bibnamefont {Chatterjee}}, \bibinfo {author} {\bibfnamefont {M.}~\bibnamefont {Schuler}}, \bibinfo {author} {\bibfnamefont {A.~M.}\ \bibnamefont {L{\"a}uchli}}, \bibinfo {author} {\bibfnamefont {M.~P.}\ \bibnamefont {Zaletel}}, \bibinfo {author} {\bibfnamefont {T.}~\bibnamefont {Lahaye}}, \bibinfo {author} {\bibfnamefont {N.~Y.}\ \bibnamefont {Yao}},\ and\ \bibinfo {author} {\bibfnamefont
  {A.}~\bibnamefont {Browaeys}},\ }\bibfield  {title} {\bibinfo {title} {{Continuous symmetry breaking in a two-dimensional Rydberg array}},\ }\href {https://doi.org/10.1038/s41586-023-05859-2} {\bibfield  {journal} {\bibinfo  {journal} {Nature}\ }\textbf {\bibinfo {volume} {616}},\ \bibinfo {pages} {691} (\bibinfo {year} {2023})}\BibitemShut {NoStop}%
\bibitem [{\citenamefont {Shaw}\ \emph {et~al.}(2024)\citenamefont {Shaw}, \citenamefont {Chen}, \citenamefont {Choi}, \citenamefont {Mark}, \citenamefont {Scholl}, \citenamefont {Finkelstein}, \citenamefont {Elben}, \citenamefont {Choi},\ and\ \citenamefont {Endres}}]{Shaw2024analogue}%
  \BibitemOpen
  \bibfield  {author} {\bibinfo {author} {\bibfnamefont {A.~L.}\ \bibnamefont {Shaw}}, \bibinfo {author} {\bibfnamefont {Z.}~\bibnamefont {Chen}}, \bibinfo {author} {\bibfnamefont {J.}~\bibnamefont {Choi}}, \bibinfo {author} {\bibfnamefont {D.~K.}\ \bibnamefont {Mark}}, \bibinfo {author} {\bibfnamefont {P.}~\bibnamefont {Scholl}}, \bibinfo {author} {\bibfnamefont {R.}~\bibnamefont {Finkelstein}}, \bibinfo {author} {\bibfnamefont {A.}~\bibnamefont {Elben}}, \bibinfo {author} {\bibfnamefont {S.}~\bibnamefont {Choi}},\ and\ \bibinfo {author} {\bibfnamefont {M.}~\bibnamefont {Endres}},\ }\bibfield  {title} {\bibinfo {title} {Benchmarking highly entangled states on a 60-atom analogue quantum simulator},\ }\href {https://doi.org/10.1038/s41586-024-07173-x} {\bibfield  {journal} {\bibinfo  {journal} {Nature}\ }\textbf {\bibinfo {volume} {628}},\ \bibinfo {pages} {71} (\bibinfo {year} {2024})}\BibitemShut {NoStop}%
\bibitem [{\citenamefont {Gonz{\'a}lez-Cuadra}\ \emph {et~al.}(2025)\citenamefont {Gonz{\'a}lez-Cuadra}, \citenamefont {Hamdan}, \citenamefont {Zache}, \citenamefont {Braverman}, \citenamefont {Kornja{\v c}a}, \citenamefont {Lukin}, \citenamefont {Cant{\'u}}, \citenamefont {Liu}, \citenamefont {Wang}, \citenamefont {Keesling}, \citenamefont {Lukin}, \citenamefont {Zoller},\ and\ \citenamefont {Bylinskii}}]{GonzalezCuadra2024String}%
  \BibitemOpen
  \bibfield  {author} {\bibinfo {author} {\bibfnamefont {D.}~\bibnamefont {Gonz{\'a}lez-Cuadra}}, \bibinfo {author} {\bibfnamefont {M.}~\bibnamefont {Hamdan}}, \bibinfo {author} {\bibfnamefont {T.~V.}\ \bibnamefont {Zache}}, \bibinfo {author} {\bibfnamefont {B.}~\bibnamefont {Braverman}}, \bibinfo {author} {\bibfnamefont {M.}~\bibnamefont {Kornja{\v c}a}}, \bibinfo {author} {\bibfnamefont {A.}~\bibnamefont {Lukin}}, \bibinfo {author} {\bibfnamefont {S.~H.}\ \bibnamefont {Cant{\'u}}}, \bibinfo {author} {\bibfnamefont {F.}~\bibnamefont {Liu}}, \bibinfo {author} {\bibfnamefont {S.-T.}\ \bibnamefont {Wang}}, \bibinfo {author} {\bibfnamefont {A.}~\bibnamefont {Keesling}}, \bibinfo {author} {\bibfnamefont {M.~D.}\ \bibnamefont {Lukin}}, \bibinfo {author} {\bibfnamefont {P.}~\bibnamefont {Zoller}},\ and\ \bibinfo {author} {\bibfnamefont {A.}~\bibnamefont {Bylinskii}},\ }\bibfield  {title} {\bibinfo {title} {{Observation of string breaking on a (2 + 1)D Rydberg quantum simulator}},\ }\href
  {https://doi.org/10.1038/s41586-025-09051-6} {\bibfield  {journal} {\bibinfo  {journal} {Nature}\ }\textbf {\bibinfo {volume} {642}},\ \bibinfo {pages} {321} (\bibinfo {year} {2025})}\BibitemShut {NoStop}%
\bibitem [{\citenamefont {Manovitz}\ \emph {et~al.}(2025)\citenamefont {Manovitz}, \citenamefont {Li}, \citenamefont {Ebadi}, \citenamefont {Samajdar}, \citenamefont {Geim}, \citenamefont {Evered}, \citenamefont {Bluvstein}, \citenamefont {Zhou}, \citenamefont {Koyluoglu}, \citenamefont {Feldmeier} \emph {et~al.}}]{manovitz2025quantum}%
  \BibitemOpen
  \bibfield  {author} {\bibinfo {author} {\bibfnamefont {T.}~\bibnamefont {Manovitz}}, \bibinfo {author} {\bibfnamefont {S.~H.}\ \bibnamefont {Li}}, \bibinfo {author} {\bibfnamefont {S.}~\bibnamefont {Ebadi}}, \bibinfo {author} {\bibfnamefont {R.}~\bibnamefont {Samajdar}}, \bibinfo {author} {\bibfnamefont {A.~A.}\ \bibnamefont {Geim}}, \bibinfo {author} {\bibfnamefont {S.~J.}\ \bibnamefont {Evered}}, \bibinfo {author} {\bibfnamefont {D.}~\bibnamefont {Bluvstein}}, \bibinfo {author} {\bibfnamefont {H.}~\bibnamefont {Zhou}}, \bibinfo {author} {\bibfnamefont {N.~U.}\ \bibnamefont {Koyluoglu}}, \bibinfo {author} {\bibfnamefont {J.}~\bibnamefont {Feldmeier}}, \emph {et~al.},\ }\bibfield  {title} {\bibinfo {title} {Quantum coarsening and collective dynamics on a programmable simulator},\ }\href {https://doi.org/10.1038/s41586-024-08353-5} {\bibfield  {journal} {\bibinfo  {journal} {Nature}\ }\textbf {\bibinfo {volume} {638}},\ \bibinfo {pages} {86} (\bibinfo {year} {2025})}\BibitemShut {NoStop}%
\bibitem [{\citenamefont {Fang}\ \emph {et~al.}(2025)\citenamefont {Fang}, \citenamefont {Wang}, \citenamefont {Liu}, \citenamefont {Wang}, \citenamefont {Cimmino}, \citenamefont {Wei}, \citenamefont {Bintz}, \citenamefont {Parr}, \citenamefont {Kemp}, \citenamefont {Ni},\ and\ \citenamefont {Yao}}]{Fang2025critical}%
  \BibitemOpen
  \bibfield  {author} {\bibinfo {author} {\bibfnamefont {F.}~\bibnamefont {Fang}}, \bibinfo {author} {\bibfnamefont {K.}~\bibnamefont {Wang}}, \bibinfo {author} {\bibfnamefont {V.~S.}\ \bibnamefont {Liu}}, \bibinfo {author} {\bibfnamefont {Y.}~\bibnamefont {Wang}}, \bibinfo {author} {\bibfnamefont {R.}~\bibnamefont {Cimmino}}, \bibinfo {author} {\bibfnamefont {J.}~\bibnamefont {Wei}}, \bibinfo {author} {\bibfnamefont {M.}~\bibnamefont {Bintz}}, \bibinfo {author} {\bibfnamefont {A.}~\bibnamefont {Parr}}, \bibinfo {author} {\bibfnamefont {J.}~\bibnamefont {Kemp}}, \bibinfo {author} {\bibfnamefont {K.-K.}\ \bibnamefont {Ni}},\ and\ \bibinfo {author} {\bibfnamefont {N.~Y.}\ \bibnamefont {Yao}},\ }\bibfield  {title} {\bibinfo {title} {Probing critical phenomena in open quantum systems using atom arrays},\ }\href {https://doi.org/10.1126/science.adq0278} {\bibfield  {journal} {\bibinfo  {journal} {Science}\ }\textbf {\bibinfo {volume} {390}},\ \bibinfo {pages} {601} (\bibinfo {year} {2025})}\BibitemShut {NoStop}%
\bibitem [{\citenamefont {Senoo}\ \emph {et~al.}(2025)\citenamefont {Senoo}, \citenamefont {Baumg{\"a}rtner}, \citenamefont {Lis}, \citenamefont {Vaidya}, \citenamefont {Zeng}, \citenamefont {Giudici}, \citenamefont {Pichler},\ and\ \citenamefont {Kaufman}}]{senoo2025high}%
  \BibitemOpen
  \bibfield  {author} {\bibinfo {author} {\bibfnamefont {A.}~\bibnamefont {Senoo}}, \bibinfo {author} {\bibfnamefont {A.}~\bibnamefont {Baumg{\"a}rtner}}, \bibinfo {author} {\bibfnamefont {J.~W.}\ \bibnamefont {Lis}}, \bibinfo {author} {\bibfnamefont {G.~M.}\ \bibnamefont {Vaidya}}, \bibinfo {author} {\bibfnamefont {Z.}~\bibnamefont {Zeng}}, \bibinfo {author} {\bibfnamefont {G.}~\bibnamefont {Giudici}}, \bibinfo {author} {\bibfnamefont {H.}~\bibnamefont {Pichler}},\ and\ \bibinfo {author} {\bibfnamefont {A.~M.}\ \bibnamefont {Kaufman}},\ }\bibfield  {title} {\bibinfo {title} {High-fidelity entanglement and coherent multi-qubit mapping in an atom array},\ }\href {https://arxiv.org/abs/2506.13632} {\bibfield  {journal} {\bibinfo  {journal} {arXiv:2506.13632}\ } (\bibinfo {year} {2025})}\BibitemShut {NoStop}%
\bibitem [{\citenamefont {Tsai}\ \emph {et~al.}(2025)\citenamefont {Tsai}, \citenamefont {Sun}, \citenamefont {Shaw}, \citenamefont {Finkelstein},\ and\ \citenamefont {Endres}}]{Tsai2025benchmarking}%
  \BibitemOpen
  \bibfield  {author} {\bibinfo {author} {\bibfnamefont {R.~B.-S.}\ \bibnamefont {Tsai}}, \bibinfo {author} {\bibfnamefont {X.}~\bibnamefont {Sun}}, \bibinfo {author} {\bibfnamefont {A.~L.}\ \bibnamefont {Shaw}}, \bibinfo {author} {\bibfnamefont {R.}~\bibnamefont {Finkelstein}},\ and\ \bibinfo {author} {\bibfnamefont {M.}~\bibnamefont {Endres}},\ }\bibfield  {title} {\bibinfo {title} {Benchmarking and fidelity response theory of high-fidelity rydberg entangling gates},\ }\href {https://doi.org/10.1103/PRXQuantum.6.010331} {\bibfield  {journal} {\bibinfo  {journal} {PRX Quantum}\ }\textbf {\bibinfo {volume} {6}},\ \bibinfo {pages} {010331} (\bibinfo {year} {2025})}\BibitemShut {NoStop}%
\bibitem [{\citenamefont {Emperauger}\ \emph {et~al.}(2025)\citenamefont {Emperauger}, \citenamefont {Qiao}, \citenamefont {Chen}, \citenamefont {Caleca}, \citenamefont {Bocini}, \citenamefont {Bintz}, \citenamefont {Bornet}, \citenamefont {Martin}, \citenamefont {G\'ely}, \citenamefont {Klein}, \citenamefont {Barredo}, \citenamefont {Chatterjee}, \citenamefont {Yao}, \citenamefont {Mezzacapo}, \citenamefont {Lahaye}, \citenamefont {Roscilde},\ and\ \citenamefont {Browaeys}}]{Emperauger2025Luttinger}%
  \BibitemOpen
  \bibfield  {author} {\bibinfo {author} {\bibfnamefont {G.}~\bibnamefont {Emperauger}}, \bibinfo {author} {\bibfnamefont {M.}~\bibnamefont {Qiao}}, \bibinfo {author} {\bibfnamefont {C.}~\bibnamefont {Chen}}, \bibinfo {author} {\bibfnamefont {F.}~\bibnamefont {Caleca}}, \bibinfo {author} {\bibfnamefont {S.}~\bibnamefont {Bocini}}, \bibinfo {author} {\bibfnamefont {M.}~\bibnamefont {Bintz}}, \bibinfo {author} {\bibfnamefont {G.}~\bibnamefont {Bornet}}, \bibinfo {author} {\bibfnamefont {R.}~\bibnamefont {Martin}}, \bibinfo {author} {\bibfnamefont {B.}~\bibnamefont {G\'ely}}, \bibinfo {author} {\bibfnamefont {L.}~\bibnamefont {Klein}}, \bibinfo {author} {\bibfnamefont {D.}~\bibnamefont {Barredo}}, \bibinfo {author} {\bibfnamefont {S.}~\bibnamefont {Chatterjee}}, \bibinfo {author} {\bibfnamefont {N.~Y.}\ \bibnamefont {Yao}}, \bibinfo {author} {\bibfnamefont {F.}~\bibnamefont {Mezzacapo}}, \bibinfo {author} {\bibfnamefont {T.}~\bibnamefont {Lahaye}}, \bibinfo {author} {\bibfnamefont {T.}~\bibnamefont {Roscilde}},\
  and\ \bibinfo {author} {\bibfnamefont {A.}~\bibnamefont {Browaeys}},\ }\bibfield  {title} {\bibinfo {title} {Tomonaga-luttinger liquid behavior in a rydberg-encoded spin chain},\ }\href {https://doi.org/10.1103/qfnp-6dpz} {\bibfield  {journal} {\bibinfo  {journal} {Phys. Rev. X}\ }\textbf {\bibinfo {volume} {15}},\ \bibinfo {pages} {031021} (\bibinfo {year} {2025})}\BibitemShut {NoStop}%
\bibitem [{\citenamefont {Bluvstein}\ \emph {et~al.}(2022)\citenamefont {Bluvstein}, \citenamefont {Levine}, \citenamefont {Semeghini}, \citenamefont {Wang}, \citenamefont {Ebadi}, \citenamefont {Kalinowski}, \citenamefont {Keesling}, \citenamefont {Maskara}, \citenamefont {Pichler}, \citenamefont {Greiner} \emph {et~al.}}]{bluvstein2022quantum}%
  \BibitemOpen
  \bibfield  {author} {\bibinfo {author} {\bibfnamefont {D.}~\bibnamefont {Bluvstein}}, \bibinfo {author} {\bibfnamefont {H.}~\bibnamefont {Levine}}, \bibinfo {author} {\bibfnamefont {G.}~\bibnamefont {Semeghini}}, \bibinfo {author} {\bibfnamefont {T.~T.}\ \bibnamefont {Wang}}, \bibinfo {author} {\bibfnamefont {S.}~\bibnamefont {Ebadi}}, \bibinfo {author} {\bibfnamefont {M.}~\bibnamefont {Kalinowski}}, \bibinfo {author} {\bibfnamefont {A.}~\bibnamefont {Keesling}}, \bibinfo {author} {\bibfnamefont {N.}~\bibnamefont {Maskara}}, \bibinfo {author} {\bibfnamefont {H.}~\bibnamefont {Pichler}}, \bibinfo {author} {\bibfnamefont {M.}~\bibnamefont {Greiner}}, \emph {et~al.},\ }\bibfield  {title} {\bibinfo {title} {A quantum processor based on coherent transport of entangled atom arrays},\ }\href {https://doi.org/10.1038/s41586-022-04592-6} {\bibfield  {journal} {\bibinfo  {journal} {Nature}\ }\textbf {\bibinfo {volume} {604}},\ \bibinfo {pages} {451} (\bibinfo {year} {2022})}\BibitemShut {NoStop}%
\bibitem [{\citenamefont {Evered}\ \emph {et~al.}(2025{\natexlab{b}})\citenamefont {Evered}, \citenamefont {Kalinowski}, \citenamefont {Geim}, \citenamefont {Manovitz}, \citenamefont {Bluvstein}, \citenamefont {Li}, \citenamefont {Maskara}, \citenamefont {Zhou}, \citenamefont {Ebadi}, \citenamefont {Xu} \emph {et~al.}}]{evered2025probing}%
  \BibitemOpen
  \bibfield  {author} {\bibinfo {author} {\bibfnamefont {S.~J.}\ \bibnamefont {Evered}}, \bibinfo {author} {\bibfnamefont {M.}~\bibnamefont {Kalinowski}}, \bibinfo {author} {\bibfnamefont {A.~A.}\ \bibnamefont {Geim}}, \bibinfo {author} {\bibfnamefont {T.}~\bibnamefont {Manovitz}}, \bibinfo {author} {\bibfnamefont {D.}~\bibnamefont {Bluvstein}}, \bibinfo {author} {\bibfnamefont {S.~H.}\ \bibnamefont {Li}}, \bibinfo {author} {\bibfnamefont {N.}~\bibnamefont {Maskara}}, \bibinfo {author} {\bibfnamefont {H.}~\bibnamefont {Zhou}}, \bibinfo {author} {\bibfnamefont {S.}~\bibnamefont {Ebadi}}, \bibinfo {author} {\bibfnamefont {M.}~\bibnamefont {Xu}}, \emph {et~al.},\ }\bibfield  {title} {\bibinfo {title} {Probing the kitaev honeycomb model on a neutral-atom quantum computer},\ }\href {https://doi.org/10.1038/s41586-025-09475-0} {\bibfield  {journal} {\bibinfo  {journal} {Nature}\ }\textbf {\bibinfo {volume} {645}},\ \bibinfo {pages} {341} (\bibinfo {year} {2025}{\natexlab{b}})}\BibitemShut {NoStop}%
\bibitem [{\citenamefont {Zeng}\ \emph {et~al.}(2017)\citenamefont {Zeng}, \citenamefont {Xu}, \citenamefont {He}, \citenamefont {Liu}, \citenamefont {Liu}, \citenamefont {Wang}, \citenamefont {Papoular}, \citenamefont {Shlyapnikov},\ and\ \citenamefont {Zhan}}]{Zeng2017dual}%
  \BibitemOpen
  \bibfield  {author} {\bibinfo {author} {\bibfnamefont {Y.}~\bibnamefont {Zeng}}, \bibinfo {author} {\bibfnamefont {P.}~\bibnamefont {Xu}}, \bibinfo {author} {\bibfnamefont {X.}~\bibnamefont {He}}, \bibinfo {author} {\bibfnamefont {Y.}~\bibnamefont {Liu}}, \bibinfo {author} {\bibfnamefont {M.}~\bibnamefont {Liu}}, \bibinfo {author} {\bibfnamefont {J.}~\bibnamefont {Wang}}, \bibinfo {author} {\bibfnamefont {D.~J.}\ \bibnamefont {Papoular}}, \bibinfo {author} {\bibfnamefont {G.~V.}\ \bibnamefont {Shlyapnikov}},\ and\ \bibinfo {author} {\bibfnamefont {M.}~\bibnamefont {Zhan}},\ }\bibfield  {title} {\bibinfo {title} {Entangling two individual atoms of different isotopes via rydberg blockade},\ }\href {https://doi.org/10.1103/PhysRevLett.119.160502} {\bibfield  {journal} {\bibinfo  {journal} {Phys. Rev. Lett.}\ }\textbf {\bibinfo {volume} {119}},\ \bibinfo {pages} {160502} (\bibinfo {year} {2017})}\BibitemShut {NoStop}%
\bibitem [{\citenamefont {Sheng}\ \emph {et~al.}(2022)\citenamefont {Sheng}, \citenamefont {Hou}, \citenamefont {He}, \citenamefont {Wang}, \citenamefont {Guo}, \citenamefont {Zhuang}, \citenamefont {Mamat}, \citenamefont {Xu}, \citenamefont {Liu}, \citenamefont {Wang},\ and\ \citenamefont {Zhan}}]{Sheng2022dual}%
  \BibitemOpen
  \bibfield  {author} {\bibinfo {author} {\bibfnamefont {C.}~\bibnamefont {Sheng}}, \bibinfo {author} {\bibfnamefont {J.}~\bibnamefont {Hou}}, \bibinfo {author} {\bibfnamefont {X.}~\bibnamefont {He}}, \bibinfo {author} {\bibfnamefont {K.}~\bibnamefont {Wang}}, \bibinfo {author} {\bibfnamefont {R.}~\bibnamefont {Guo}}, \bibinfo {author} {\bibfnamefont {J.}~\bibnamefont {Zhuang}}, \bibinfo {author} {\bibfnamefont {B.}~\bibnamefont {Mamat}}, \bibinfo {author} {\bibfnamefont {P.}~\bibnamefont {Xu}}, \bibinfo {author} {\bibfnamefont {M.}~\bibnamefont {Liu}}, \bibinfo {author} {\bibfnamefont {J.}~\bibnamefont {Wang}},\ and\ \bibinfo {author} {\bibfnamefont {M.}~\bibnamefont {Zhan}},\ }\bibfield  {title} {\bibinfo {title} {Defect-free arbitrary-geometry assembly of mixed-species atom arrays},\ }\href {https://doi.org/10.1103/PhysRevLett.128.083202} {\bibfield  {journal} {\bibinfo  {journal} {Phys. Rev. Lett.}\ }\textbf {\bibinfo {volume} {128}},\ \bibinfo {pages} {083202} (\bibinfo {year} {2022})}\BibitemShut
  {NoStop}%
\bibitem [{\citenamefont {Singh}\ \emph {et~al.}(2022)\citenamefont {Singh}, \citenamefont {Anand}, \citenamefont {Pocklington}, \citenamefont {Kemp},\ and\ \citenamefont {Bernien}}]{singh2022dual}%
  \BibitemOpen
  \bibfield  {author} {\bibinfo {author} {\bibfnamefont {K.}~\bibnamefont {Singh}}, \bibinfo {author} {\bibfnamefont {S.}~\bibnamefont {Anand}}, \bibinfo {author} {\bibfnamefont {A.}~\bibnamefont {Pocklington}}, \bibinfo {author} {\bibfnamefont {J.~T.}\ \bibnamefont {Kemp}},\ and\ \bibinfo {author} {\bibfnamefont {H.}~\bibnamefont {Bernien}},\ }\bibfield  {title} {\bibinfo {title} {Dual-element, two-dimensional atom array with continuous-mode operation},\ }\href {https://doi.org/10.1103/PhysRevX.12.011040} {\bibfield  {journal} {\bibinfo  {journal} {Phys. Rev. X}\ }\textbf {\bibinfo {volume} {12}},\ \bibinfo {pages} {011040} (\bibinfo {year} {2022})}\BibitemShut {NoStop}%
\bibitem [{\citenamefont {Singh}\ \emph {et~al.}(2023)\citenamefont {Singh}, \citenamefont {Bradley}, \citenamefont {Anand}, \citenamefont {Ramesh}, \citenamefont {White},\ and\ \citenamefont {Bernien}}]{singh2023mid}%
  \BibitemOpen
  \bibfield  {author} {\bibinfo {author} {\bibfnamefont {K.}~\bibnamefont {Singh}}, \bibinfo {author} {\bibfnamefont {C.~E.}\ \bibnamefont {Bradley}}, \bibinfo {author} {\bibfnamefont {S.}~\bibnamefont {Anand}}, \bibinfo {author} {\bibfnamefont {V.}~\bibnamefont {Ramesh}}, \bibinfo {author} {\bibfnamefont {R.}~\bibnamefont {White}},\ and\ \bibinfo {author} {\bibfnamefont {H.}~\bibnamefont {Bernien}},\ }\bibfield  {title} {\bibinfo {title} {Mid-circuit correction of correlated phase errors using an array of spectator qubits},\ }\href {https://doi.org/10.1126/science.ade5337} {\bibfield  {journal} {\bibinfo  {journal} {Science}\ }\textbf {\bibinfo {volume} {380}},\ \bibinfo {pages} {1265} (\bibinfo {year} {2023})}\BibitemShut {NoStop}%
\bibitem [{\citenamefont {Anand}\ \emph {et~al.}(2024)\citenamefont {Anand}, \citenamefont {Bradley}, \citenamefont {White}, \citenamefont {Ramesh}, \citenamefont {Singh},\ and\ \citenamefont {Bernien}}]{anand2024dual}%
  \BibitemOpen
  \bibfield  {author} {\bibinfo {author} {\bibfnamefont {S.}~\bibnamefont {Anand}}, \bibinfo {author} {\bibfnamefont {C.~E.}\ \bibnamefont {Bradley}}, \bibinfo {author} {\bibfnamefont {R.}~\bibnamefont {White}}, \bibinfo {author} {\bibfnamefont {V.}~\bibnamefont {Ramesh}}, \bibinfo {author} {\bibfnamefont {K.}~\bibnamefont {Singh}},\ and\ \bibinfo {author} {\bibfnamefont {H.}~\bibnamefont {Bernien}},\ }\bibfield  {title} {\bibinfo {title} {A dual-species rydberg array},\ }\href {https://doi.org/10.1038/s41567-024-02638-2} {\bibfield  {journal} {\bibinfo  {journal} {Nature Phys.}\ }\textbf {\bibinfo {volume} {20}},\ \bibinfo {pages} {1744} (\bibinfo {year} {2024})}\BibitemShut {NoStop}%
\bibitem [{\citenamefont {Prosen}(2007)}]{prosen2007chaos}%
  \BibitemOpen
  \bibfield  {author} {\bibinfo {author} {\bibfnamefont {T.}~\bibnamefont {Prosen}},\ }\bibfield  {title} {\bibinfo {title} {Chaos and complexity of quantum motion},\ }\href {https://doi.org/10.1088/1751-8113/40/28/S02} {\bibfield  {journal} {\bibinfo  {journal} {J. Phys. A: Math. Theor.}\ }\textbf {\bibinfo {volume} {40}},\ \bibinfo {pages} {7881} (\bibinfo {year} {2007})}\BibitemShut {NoStop}%
\bibitem [{\citenamefont {Prosen}(2002)}]{prosen2002general}%
  \BibitemOpen
  \bibfield  {author} {\bibinfo {author} {\bibfnamefont {T.}~\bibnamefont {Prosen}},\ }\bibfield  {title} {\bibinfo {title} {General relation between quantum ergodicity and fidelity of quantum dynamics},\ }\href {https://doi.org/10.1103/PhysRevE.65.036208} {\bibfield  {journal} {\bibinfo  {journal} {Phys. Rev. E}\ }\textbf {\bibinfo {volume} {65}},\ \bibinfo {pages} {036208} (\bibinfo {year} {2002})}\BibitemShut {NoStop}%
\bibitem [{\citenamefont {Kitaev}(2006)}]{kitaev2006anyons}%
  \BibitemOpen
  \bibfield  {author} {\bibinfo {author} {\bibfnamefont {A.}~\bibnamefont {Kitaev}},\ }\bibfield  {title} {\bibinfo {title} {Anyons in an exactly solved model and beyond},\ }\href {https://doi.org/10.1016/j.aop.2005.10.005} {\bibfield  {journal} {\bibinfo  {journal} {Ann. Phys.}\ }\textbf {\bibinfo {volume} {321}},\ \bibinfo {pages} {2} (\bibinfo {year} {2006})}\BibitemShut {NoStop}%
\bibitem [{\citenamefont {Lloyd}(1996)}]{lloyd1996universal}%
  \BibitemOpen
  \bibfield  {author} {\bibinfo {author} {\bibfnamefont {S.}~\bibnamefont {Lloyd}},\ }\bibfield  {title} {\bibinfo {title} {Universal quantum simulators},\ }\href {https://doi.org/10.1126/science.273.5278.1073} {\bibfield  {journal} {\bibinfo  {journal} {Science}\ }\textbf {\bibinfo {volume} {273}},\ \bibinfo {pages} {1073} (\bibinfo {year} {1996})}\BibitemShut {NoStop}%
\bibitem [{\citenamefont {Trotter}(1959)}]{trotter1959product}%
  \BibitemOpen
  \bibfield  {author} {\bibinfo {author} {\bibfnamefont {H.~F.}\ \bibnamefont {Trotter}},\ }\bibfield  {title} {\bibinfo {title} {On the product of semi-groups of operators},\ }\href {https://www.ams.org/journals/proc/1959-010-04/S0002-9939-1959-0108732-6/} {\bibfield  {journal} {\bibinfo  {journal} {Proc. American Math. Soc.}\ }\textbf {\bibinfo {volume} {10}},\ \bibinfo {pages} {545} (\bibinfo {year} {1959})}\BibitemShut {NoStop}%
\bibitem [{\citenamefont {Suzuki}(1991)}]{suzuki1991general}%
  \BibitemOpen
  \bibfield  {author} {\bibinfo {author} {\bibfnamefont {M.}~\bibnamefont {Suzuki}},\ }\bibfield  {title} {\bibinfo {title} {General theory of fractal path integrals with applications to many-body theories and statistical physics},\ }\href {https://doi.org/10.1063/1.529425} {\bibfield  {journal} {\bibinfo  {journal} {J. Math. Phys.}\ }\textbf {\bibinfo {volume} {32}},\ \bibinfo {pages} {400} (\bibinfo {year} {1991})}\BibitemShut {NoStop}%
\bibitem [{\citenamefont {Bohigas}\ \emph {et~al.}(1984)\citenamefont {Bohigas}, \citenamefont {Giannoni},\ and\ \citenamefont {Schmit}}]{bohigas1984characterization}%
  \BibitemOpen
  \bibfield  {author} {\bibinfo {author} {\bibfnamefont {O.}~\bibnamefont {Bohigas}}, \bibinfo {author} {\bibfnamefont {M.~J.}\ \bibnamefont {Giannoni}},\ and\ \bibinfo {author} {\bibfnamefont {C.}~\bibnamefont {Schmit}},\ }\bibfield  {title} {\bibinfo {title} {Characterization of chaotic quantum spectra and universality of level fluctuation laws},\ }\href {https://doi.org/10.1103/PhysRevLett.52.1} {\bibfield  {journal} {\bibinfo  {journal} {Phys. Rev. Lett.}\ }\textbf {\bibinfo {volume} {52}},\ \bibinfo {pages} {1} (\bibinfo {year} {1984})}\BibitemShut {NoStop}%
\bibitem [{\citenamefont {Haake}(1991)}]{haake_quantum_1991}%
  \BibitemOpen
  \bibfield  {author} {\bibinfo {author} {\bibfnamefont {F.}~\bibnamefont {Haake}},\ }\bibfield  {title} {\bibinfo {title} {Quantum {Signatures} of {Chaos}}\ }(\bibinfo  {publisher} {Springer US},\ \bibinfo {address} {Boston, MA},\ \bibinfo {year} {1991})\ pp.\ \bibinfo {pages} {583--595}\BibitemShut {NoStop}%
\bibitem [{\citenamefont {Gutzwiller}(2013)}]{gutzwiller2013chaos}%
  \BibitemOpen
  \bibfield  {author} {\bibinfo {author} {\bibfnamefont {M.~C.}\ \bibnamefont {Gutzwiller}},\ }\href@noop {} {\emph {\bibinfo {title} {Chaos in classical and quantum mechanics}}},\ Vol.~\bibinfo {volume} {1}\ (\bibinfo  {publisher} {Springer Science \& Business Media},\ \bibinfo {year} {2013})\BibitemShut {NoStop}%
\bibitem [{\citenamefont {Shenker}\ and\ \citenamefont {Stanford}(2014{\natexlab{a}})}]{shenker2014multiple}%
  \BibitemOpen
  \bibfield  {author} {\bibinfo {author} {\bibfnamefont {S.~H.}\ \bibnamefont {Shenker}}\ and\ \bibinfo {author} {\bibfnamefont {D.}~\bibnamefont {Stanford}},\ }\bibfield  {title} {\bibinfo {title} {Multiple shocks},\ }\href {https://doi.org/10.1007/JHEP12(2014)046} {\bibfield  {journal} {\bibinfo  {journal} {JHEP}\ }\textbf {\bibinfo {volume} {2014}}\bibinfo  {number} { (12)},\ \bibinfo {pages} {1}}\BibitemShut {NoStop}%
\bibitem [{\citenamefont {Roberts}\ and\ \citenamefont {Stanford}(2015)}]{roberts2015diagnosing}%
  \BibitemOpen
\bibfield  {number} {  }\bibfield  {author} {\bibinfo {author} {\bibfnamefont {D.~A.}\ \bibnamefont {Roberts}}\ and\ \bibinfo {author} {\bibfnamefont {D.}~\bibnamefont {Stanford}},\ }\bibfield  {title} {\bibinfo {title} {Diagnosing chaos using four-point functions in two-dimensional conformal field theory},\ }\href {https://doi.org/10.1103/PhysRevLett.115.131603} {\bibfield  {journal} {\bibinfo  {journal} {Phys. Rev. Lett.}\ }\textbf {\bibinfo {volume} {115}},\ \bibinfo {pages} {131603} (\bibinfo {year} {2015})}\BibitemShut {NoStop}%
\bibitem [{\citenamefont {Roberts}\ \emph {et~al.}(2015)\citenamefont {Roberts}, \citenamefont {Stanford},\ and\ \citenamefont {Susskind}}]{roberts2015localized}%
  \BibitemOpen
  \bibfield  {author} {\bibinfo {author} {\bibfnamefont {D.~A.}\ \bibnamefont {Roberts}}, \bibinfo {author} {\bibfnamefont {D.}~\bibnamefont {Stanford}},\ and\ \bibinfo {author} {\bibfnamefont {L.}~\bibnamefont {Susskind}},\ }\bibfield  {title} {\bibinfo {title} {Localized shocks},\ }\href {https://doi.org/10.1007/JHEP03(2015)051} {\bibfield  {journal} {\bibinfo  {journal} {JHEP}\ }\textbf {\bibinfo {volume} {2015}}\bibinfo  {number} { (3)},\ \bibinfo {pages} {1}}\BibitemShut {NoStop}%
\bibitem [{\citenamefont {Maldacena}\ \emph {et~al.}(2016)\citenamefont {Maldacena}, \citenamefont {Shenker},\ and\ \citenamefont {Stanford}}]{maldacena2016bound}%
  \BibitemOpen
\bibfield  {number} {  }\bibfield  {author} {\bibinfo {author} {\bibfnamefont {J.}~\bibnamefont {Maldacena}}, \bibinfo {author} {\bibfnamefont {S.~H.}\ \bibnamefont {Shenker}},\ and\ \bibinfo {author} {\bibfnamefont {D.}~\bibnamefont {Stanford}},\ }\bibfield  {title} {\bibinfo {title} {A bound on chaos},\ }\href {https://doi.org/10.1007/JHEP08(2016)106} {\bibfield  {journal} {\bibinfo  {journal} {JHEP}\ }\textbf {\bibinfo {volume} {2016}}\bibinfo  {number} { (8)},\ \bibinfo {pages} {1}}\BibitemShut {NoStop}%
\bibitem [{\citenamefont {von Keyserlingk}\ \emph {et~al.}(2018{\natexlab{b}})\citenamefont {von Keyserlingk}, \citenamefont {Rakovszky}, \citenamefont {Pollmann},\ and\ \citenamefont {Sondhi}}]{keyserlingk2018operator}%
  \BibitemOpen
\bibfield  {number} {  }\bibfield  {author} {\bibinfo {author} {\bibfnamefont {C.~W.}\ \bibnamefont {von Keyserlingk}}, \bibinfo {author} {\bibfnamefont {T.}~\bibnamefont {Rakovszky}}, \bibinfo {author} {\bibfnamefont {F.}~\bibnamefont {Pollmann}},\ and\ \bibinfo {author} {\bibfnamefont {S.~L.}\ \bibnamefont {Sondhi}},\ }\bibfield  {title} {\bibinfo {title} {Operator hydrodynamics, otocs, and entanglement growth in systems without conservation laws},\ }\href {https://doi.org/10.1103/PhysRevX.8.021013} {\bibfield  {journal} {\bibinfo  {journal} {Phys. Rev. X}\ }\textbf {\bibinfo {volume} {8}},\ \bibinfo {pages} {021013} (\bibinfo {year} {2018}{\natexlab{b}})}\BibitemShut {NoStop}%
\bibitem [{\citenamefont {Piroli}\ and\ \citenamefont {Cirac}(2020)}]{piroli2020quantumcellular}%
  \BibitemOpen
  \bibfield  {author} {\bibinfo {author} {\bibfnamefont {L.}~\bibnamefont {Piroli}}\ and\ \bibinfo {author} {\bibfnamefont {J.~I.}\ \bibnamefont {Cirac}},\ }\bibfield  {title} {\bibinfo {title} {Quantum cellular automata, tensor networks, and area laws},\ }\href {https://doi.org/10.1103/PhysRevLett.125.190402} {\bibfield  {journal} {\bibinfo  {journal} {Phys. Rev. Lett.}\ }\textbf {\bibinfo {volume} {125}},\ \bibinfo {pages} {190402} (\bibinfo {year} {2020})}\BibitemShut {NoStop}%
\bibitem [{\citenamefont {Abanin}\ \emph {et~al.}(2017)\citenamefont {Abanin}, \citenamefont {De~Roeck}, \citenamefont {Ho},\ and\ \citenamefont {Huveneers}}]{Abanin_2017}%
  \BibitemOpen
  \bibfield  {author} {\bibinfo {author} {\bibfnamefont {D.}~\bibnamefont {Abanin}}, \bibinfo {author} {\bibfnamefont {W.}~\bibnamefont {De~Roeck}}, \bibinfo {author} {\bibfnamefont {W.~W.}\ \bibnamefont {Ho}},\ and\ \bibinfo {author} {\bibfnamefont {F.}~\bibnamefont {Huveneers}},\ }\bibfield  {title} {\bibinfo {title} {A rigorous theory of many-body prethermalization for periodically driven and closed quantum systems},\ }\href {https://doi.org/10.1007/s00220-017-2930-x} {\bibfield  {journal} {\bibinfo  {journal} {Communications in Mathematical Physics}\ }\textbf {\bibinfo {volume} {354}},\ \bibinfo {pages} {809–827} (\bibinfo {year} {2017})}\BibitemShut {NoStop}%
\bibitem [{\citenamefont {Ishii}\ \emph {et~al.}(2018)\citenamefont {Ishii}, \citenamefont {Kuwahara}, \citenamefont {Mori},\ and\ \citenamefont {Hatano}}]{ishii2018heating}%
  \BibitemOpen
  \bibfield  {author} {\bibinfo {author} {\bibfnamefont {T.}~\bibnamefont {Ishii}}, \bibinfo {author} {\bibfnamefont {T.}~\bibnamefont {Kuwahara}}, \bibinfo {author} {\bibfnamefont {T.}~\bibnamefont {Mori}},\ and\ \bibinfo {author} {\bibfnamefont {N.}~\bibnamefont {Hatano}},\ }\bibfield  {title} {\bibinfo {title} {Heating in integrable time-periodic systems},\ }\href {https://doi.org/10.1103/PhysRevLett.120.220602} {\bibfield  {journal} {\bibinfo  {journal} {Phys. Rev. Lett.}\ }\textbf {\bibinfo {volume} {120}},\ \bibinfo {pages} {220602} (\bibinfo {year} {2018})}\BibitemShut {NoStop}%
\bibitem [{\citenamefont {Heyl}\ \emph {et~al.}(2019)\citenamefont {Heyl}, \citenamefont {Hauke},\ and\ \citenamefont {Zoller}}]{heyl2019quantum}%
  \BibitemOpen
  \bibfield  {author} {\bibinfo {author} {\bibfnamefont {M.}~\bibnamefont {Heyl}}, \bibinfo {author} {\bibfnamefont {P.}~\bibnamefont {Hauke}},\ and\ \bibinfo {author} {\bibfnamefont {P.}~\bibnamefont {Zoller}},\ }\bibfield  {title} {\bibinfo {title} {Quantum localization bounds trotter errors in digital quantum simulation},\ }\href {https://doi.org/10.1126/sciadv.aau8342} {\bibfield  {journal} {\bibinfo  {journal} {Science Adv.}\ }\textbf {\bibinfo {volume} {5}},\ \bibinfo {pages} {eaau8342} (\bibinfo {year} {2019})}\BibitemShut {NoStop}%
\bibitem [{\citenamefont {Chinni}\ \emph {et~al.}(2022)\citenamefont {Chinni}, \citenamefont {Mu\~noz Arias}, \citenamefont {Deutsch},\ and\ \citenamefont {Poggi}}]{chinni2022trotter}%
  \BibitemOpen
  \bibfield  {author} {\bibinfo {author} {\bibfnamefont {K.}~\bibnamefont {Chinni}}, \bibinfo {author} {\bibfnamefont {M.~H.}\ \bibnamefont {Mu\~noz Arias}}, \bibinfo {author} {\bibfnamefont {I.~H.}\ \bibnamefont {Deutsch}},\ and\ \bibinfo {author} {\bibfnamefont {P.~M.}\ \bibnamefont {Poggi}},\ }\bibfield  {title} {\bibinfo {title} {Trotter errors from dynamical structural instabilities of floquet maps in quantum simulation},\ }\href {https://doi.org/10.1103/PRXQuantum.3.010351} {\bibfield  {journal} {\bibinfo  {journal} {PRX Quantum}\ }\textbf {\bibinfo {volume} {3}},\ \bibinfo {pages} {010351} (\bibinfo {year} {2022})}\BibitemShut {NoStop}%
\bibitem [{\citenamefont {Vernier}\ \emph {et~al.}(2023)\citenamefont {Vernier}, \citenamefont {Bertini}, \citenamefont {Giudici},\ and\ \citenamefont {Piroli}}]{vernier2023integrable}%
  \BibitemOpen
  \bibfield  {author} {\bibinfo {author} {\bibfnamefont {E.}~\bibnamefont {Vernier}}, \bibinfo {author} {\bibfnamefont {B.}~\bibnamefont {Bertini}}, \bibinfo {author} {\bibfnamefont {G.}~\bibnamefont {Giudici}},\ and\ \bibinfo {author} {\bibfnamefont {L.}~\bibnamefont {Piroli}},\ }\bibfield  {title} {\bibinfo {title} {Integrable digital quantum simulation: Generalized gibbs ensembles and trotter transitions},\ }\href {https://doi.org/10.1103/PhysRevLett.130.260401} {\bibfield  {journal} {\bibinfo  {journal} {Phys. Rev. Lett.}\ }\textbf {\bibinfo {volume} {130}},\ \bibinfo {pages} {260401} (\bibinfo {year} {2023})}\BibitemShut {NoStop}%
\bibitem [{\citenamefont {Rudner}\ \emph {et~al.}(2013)\citenamefont {Rudner}, \citenamefont {Lindner}, \citenamefont {Berg},\ and\ \citenamefont {Levin}}]{PhysRevX.3.031005}%
  \BibitemOpen
  \bibfield  {author} {\bibinfo {author} {\bibfnamefont {M.~S.}\ \bibnamefont {Rudner}}, \bibinfo {author} {\bibfnamefont {N.~H.}\ \bibnamefont {Lindner}}, \bibinfo {author} {\bibfnamefont {E.}~\bibnamefont {Berg}},\ and\ \bibinfo {author} {\bibfnamefont {M.}~\bibnamefont {Levin}},\ }\bibfield  {title} {\bibinfo {title} {Anomalous edge states and the bulk-edge correspondence for periodically driven two-dimensional systems},\ }\href {https://doi.org/10.1103/PhysRevX.3.031005} {\bibfield  {journal} {\bibinfo  {journal} {Phys. Rev. X}\ }\textbf {\bibinfo {volume} {3}},\ \bibinfo {pages} {031005} (\bibinfo {year} {2013})}\BibitemShut {NoStop}%
\bibitem [{\citenamefont {Else}\ \emph {et~al.}(2017)\citenamefont {Else}, \citenamefont {Bauer},\ and\ \citenamefont {Nayak}}]{PhysRevX.7.011026}%
  \BibitemOpen
  \bibfield  {author} {\bibinfo {author} {\bibfnamefont {D.~V.}\ \bibnamefont {Else}}, \bibinfo {author} {\bibfnamefont {B.}~\bibnamefont {Bauer}},\ and\ \bibinfo {author} {\bibfnamefont {C.}~\bibnamefont {Nayak}},\ }\bibfield  {title} {\bibinfo {title} {Prethermal phases of matter protected by time-translation symmetry},\ }\href {https://doi.org/10.1103/PhysRevX.7.011026} {\bibfield  {journal} {\bibinfo  {journal} {Phys. Rev. X}\ }\textbf {\bibinfo {volume} {7}},\ \bibinfo {pages} {011026} (\bibinfo {year} {2017})}\BibitemShut {NoStop}%
\bibitem [{\citenamefont {Kyprianidis}\ \emph {et~al.}(2021)\citenamefont {Kyprianidis}, \citenamefont {Machado}, \citenamefont {Morong}, \citenamefont {Becker}, \citenamefont {Collins}, \citenamefont {Else}, \citenamefont {Feng}, \citenamefont {Hess}, \citenamefont {Nayak}, \citenamefont {Pagano}, \citenamefont {Yao},\ and\ \citenamefont {Monroe}}]{Kyprianidis_2021}%
  \BibitemOpen
  \bibfield  {author} {\bibinfo {author} {\bibfnamefont {A.}~\bibnamefont {Kyprianidis}}, \bibinfo {author} {\bibfnamefont {F.}~\bibnamefont {Machado}}, \bibinfo {author} {\bibfnamefont {W.}~\bibnamefont {Morong}}, \bibinfo {author} {\bibfnamefont {P.}~\bibnamefont {Becker}}, \bibinfo {author} {\bibfnamefont {K.~S.}\ \bibnamefont {Collins}}, \bibinfo {author} {\bibfnamefont {D.~V.}\ \bibnamefont {Else}}, \bibinfo {author} {\bibfnamefont {L.}~\bibnamefont {Feng}}, \bibinfo {author} {\bibfnamefont {P.~W.}\ \bibnamefont {Hess}}, \bibinfo {author} {\bibfnamefont {C.}~\bibnamefont {Nayak}}, \bibinfo {author} {\bibfnamefont {G.}~\bibnamefont {Pagano}}, \bibinfo {author} {\bibfnamefont {N.~Y.}\ \bibnamefont {Yao}},\ and\ \bibinfo {author} {\bibfnamefont {C.}~\bibnamefont {Monroe}},\ }\bibfield  {title} {\bibinfo {title} {Observation of a prethermal discrete time crystal},\ }\href {https://doi.org/10.1126/science.abg8102} {\bibfield  {journal} {\bibinfo  {journal} {Science}\ }\textbf {\bibinfo {volume} {372}},\
  \bibinfo {pages} {1192–1196} (\bibinfo {year} {2021})}\BibitemShut {NoStop}%
\bibitem [{\citenamefont {Kalinowski}\ \emph {et~al.}(2023)\citenamefont {Kalinowski}, \citenamefont {Maskara},\ and\ \citenamefont {Lukin}}]{PhysRevX.13.031008}%
  \BibitemOpen
  \bibfield  {author} {\bibinfo {author} {\bibfnamefont {M.}~\bibnamefont {Kalinowski}}, \bibinfo {author} {\bibfnamefont {N.}~\bibnamefont {Maskara}},\ and\ \bibinfo {author} {\bibfnamefont {M.~D.}\ \bibnamefont {Lukin}},\ }\bibfield  {title} {\bibinfo {title} {Non-abelian floquet spin liquids in a digital rydberg simulator},\ }\href {https://doi.org/10.1103/PhysRevX.13.031008} {\bibfield  {journal} {\bibinfo  {journal} {Phys. Rev. X}\ }\textbf {\bibinfo {volume} {13}},\ \bibinfo {pages} {031008} (\bibinfo {year} {2023})}\BibitemShut {NoStop}%
\bibitem [{\citenamefont {Lieb}\ and\ \citenamefont {Robinson}(1972)}]{lieb1972finite}%
  \BibitemOpen
  \bibfield  {author} {\bibinfo {author} {\bibfnamefont {E.~H.}\ \bibnamefont {Lieb}}\ and\ \bibinfo {author} {\bibfnamefont {D.~W.}\ \bibnamefont {Robinson}},\ }\bibfield  {title} {\bibinfo {title} {The finite group velocity of quantum spin systems},\ }\href {https://doi.org/10.1007/BF01645779} {\bibfield  {journal} {\bibinfo  {journal} {Comm. Math. Phys.}\ }\textbf {\bibinfo {volume} {28}},\ \bibinfo {pages} {251} (\bibinfo {year} {1972})}\BibitemShut {NoStop}%
\bibitem [{\citenamefont {Hastings}\ and\ \citenamefont {Koma}(2006)}]{hastings2006spectral}%
  \BibitemOpen
  \bibfield  {author} {\bibinfo {author} {\bibfnamefont {M.~B.}\ \bibnamefont {Hastings}}\ and\ \bibinfo {author} {\bibfnamefont {T.}~\bibnamefont {Koma}},\ }\bibfield  {title} {\bibinfo {title} {Spectral gap and exponential decay of correlations},\ }\href {https://doi.org/10.1007/s00220-006-0030-4} {\bibfield  {journal} {\bibinfo  {journal} {Comm. Math. Phys.}\ }\textbf {\bibinfo {volume} {265}},\ \bibinfo {pages} {781} (\bibinfo {year} {2006})}\BibitemShut {NoStop}%
\bibitem [{\citenamefont {Hastings}(2012)}]{hastings2010locality}%
  \BibitemOpen
  \bibfield  {author} {\bibinfo {author} {\bibfnamefont {M.~B.}\ \bibnamefont {Hastings}},\ }\bibfield  {title} {\bibinfo {title} {1713 {Locality} in quantum systems},\ }in\ \href {https://doi.org/10.1093/acprof:oso/9780199652495.003.0003} {\emph {\bibinfo {booktitle} {Quantum {Theory} from {Small} to {Large} {Scales}: {Lecture} {Notes} of the {Les} {Houches} {Summer} {School}: {Volume} 95, {August} 2010}}}\ (\bibinfo  {publisher} {Oxford University Press},\ \bibinfo {year} {2012})\BibitemShut {NoStop}%
\bibitem [{\citenamefont {Hosur}\ \emph {et~al.}(2016)\citenamefont {Hosur}, \citenamefont {Qi}, \citenamefont {Roberts},\ and\ \citenamefont {Yoshida}}]{hosur2016chaos}%
  \BibitemOpen
  \bibfield  {author} {\bibinfo {author} {\bibfnamefont {P.}~\bibnamefont {Hosur}}, \bibinfo {author} {\bibfnamefont {X.-L.}\ \bibnamefont {Qi}}, \bibinfo {author} {\bibfnamefont {D.~A.}\ \bibnamefont {Roberts}},\ and\ \bibinfo {author} {\bibfnamefont {B.}~\bibnamefont {Yoshida}},\ }\bibfield  {title} {\bibinfo {title} {Chaos in quantum channels},\ }\href {https://doi.org/10.1007/jhep02(2016)004} {\bibfield  {journal} {\bibinfo  {journal} {JHEP}\ }\textbf {\bibinfo {volume} {2016}}\bibinfo  {number} { (2)},\ \bibinfo {pages} {1}}\BibitemShut {NoStop}%
\bibitem [{\citenamefont {Gong}\ \emph {et~al.}(2022)\citenamefont {Gong}, \citenamefont {Nahum},\ and\ \citenamefont {Piroli}}]{gong2022coarse}%
  \BibitemOpen
\bibfield  {number} {  }\bibfield  {author} {\bibinfo {author} {\bibfnamefont {Z.}~\bibnamefont {Gong}}, \bibinfo {author} {\bibfnamefont {A.}~\bibnamefont {Nahum}},\ and\ \bibinfo {author} {\bibfnamefont {L.}~\bibnamefont {Piroli}},\ }\bibfield  {title} {\bibinfo {title} {Coarse-grained entanglement and operator growth in anomalous dynamics},\ }\href {https://doi.org/10.1103/PhysRevLett.128.080602} {\bibfield  {journal} {\bibinfo  {journal} {Phys. Rev. Lett.}\ }\textbf {\bibinfo {volume} {128}},\ \bibinfo {pages} {080602} (\bibinfo {year} {2022})}\BibitemShut {NoStop}%
\bibitem [{\citenamefont {Potter}\ and\ \citenamefont {Vasseur}(2022)}]{potter_entanglement_2022}%
  \BibitemOpen
  \bibfield  {author} {\bibinfo {author} {\bibfnamefont {A.~C.}\ \bibnamefont {Potter}}\ and\ \bibinfo {author} {\bibfnamefont {R.}~\bibnamefont {Vasseur}},\ }\bibfield  {title} {\bibinfo {title} {Entanglement {Dynamics} in {Hybrid} {Quantum} {Circuits}},\ }in\ \href {https://doi.org/10.1007/978-3-031-03998-0_9} {\emph {\bibinfo {booktitle} {Entanglement in {Spin} {Chains}: {From} {Theory} to {Quantum} {Technology} {Applications}}}},\ \bibinfo {editor} {edited by\ \bibinfo {editor} {\bibfnamefont {A.}~\bibnamefont {Bayat}}, \bibinfo {editor} {\bibfnamefont {S.}~\bibnamefont {Bose}},\ and\ \bibinfo {editor} {\bibfnamefont {H.}~\bibnamefont {Johannesson}}}\ (\bibinfo  {publisher} {Springer International Publishing},\ \bibinfo {address} {Cham},\ \bibinfo {year} {2022})\ pp.\ \bibinfo {pages} {211--249}\BibitemShut {NoStop}%
\bibitem [{\citenamefont {Gillman}\ \emph {et~al.}(2020)\citenamefont {Gillman}, \citenamefont {Carollo},\ and\ \citenamefont {Lesanovsky}}]{gillman2020nonequilibrium}%
  \BibitemOpen
  \bibfield  {author} {\bibinfo {author} {\bibfnamefont {E.}~\bibnamefont {Gillman}}, \bibinfo {author} {\bibfnamefont {F.}~\bibnamefont {Carollo}},\ and\ \bibinfo {author} {\bibfnamefont {I.}~\bibnamefont {Lesanovsky}},\ }\bibfield  {title} {\bibinfo {title} {Nonequilibrium phase transitions in ($1+1$)-dimensional quantum cellular automata with controllable quantum correlations},\ }\href {https://doi.org/10.1103/PhysRevLett.125.100403} {\bibfield  {journal} {\bibinfo  {journal} {Phys. Rev. Lett.}\ }\textbf {\bibinfo {volume} {125}},\ \bibinfo {pages} {100403} (\bibinfo {year} {2020})}\BibitemShut {NoStop}%
\bibitem [{\citenamefont {Noel}\ \emph {et~al.}(2022)\citenamefont {Noel}, \citenamefont {Niroula}, \citenamefont {Zhu}, \citenamefont {Risinger}, \citenamefont {Egan}, \citenamefont {Biswas}, \citenamefont {Cetina}, \citenamefont {Gorshkov}, \citenamefont {Gullans}, \citenamefont {Huse} \emph {et~al.}}]{noel2022measurement}%
  \BibitemOpen
  \bibfield  {author} {\bibinfo {author} {\bibfnamefont {C.}~\bibnamefont {Noel}}, \bibinfo {author} {\bibfnamefont {P.}~\bibnamefont {Niroula}}, \bibinfo {author} {\bibfnamefont {D.}~\bibnamefont {Zhu}}, \bibinfo {author} {\bibfnamefont {A.}~\bibnamefont {Risinger}}, \bibinfo {author} {\bibfnamefont {L.}~\bibnamefont {Egan}}, \bibinfo {author} {\bibfnamefont {D.}~\bibnamefont {Biswas}}, \bibinfo {author} {\bibfnamefont {M.}~\bibnamefont {Cetina}}, \bibinfo {author} {\bibfnamefont {A.~V.}\ \bibnamefont {Gorshkov}}, \bibinfo {author} {\bibfnamefont {M.~J.}\ \bibnamefont {Gullans}}, \bibinfo {author} {\bibfnamefont {D.~A.}\ \bibnamefont {Huse}}, \emph {et~al.},\ }\bibfield  {title} {\bibinfo {title} {Measurement-induced quantum phases realized in a trapped-ion quantum computer},\ }\href {https://doi.org/s41567-022-01619-7} {\bibfield  {journal} {\bibinfo  {journal} {Nature Phys.}\ }\textbf {\bibinfo {volume} {18}},\ \bibinfo {pages} {760} (\bibinfo {year} {2022})}\BibitemShut {NoStop}%
\bibitem [{\citenamefont {Koh}\ \emph {et~al.}(2023)\citenamefont {Koh}, \citenamefont {Sun}, \citenamefont {Motta},\ and\ \citenamefont {Minnich}}]{koh2023measurement}%
  \BibitemOpen
  \bibfield  {author} {\bibinfo {author} {\bibfnamefont {J.~M.}\ \bibnamefont {Koh}}, \bibinfo {author} {\bibfnamefont {S.-N.}\ \bibnamefont {Sun}}, \bibinfo {author} {\bibfnamefont {M.}~\bibnamefont {Motta}},\ and\ \bibinfo {author} {\bibfnamefont {A.~J.}\ \bibnamefont {Minnich}},\ }\bibfield  {title} {\bibinfo {title} {Measurement-induced entanglement phase transition on a superconducting quantum processor with mid-circuit readout},\ }\href {https://doi.org/s41567-023-02076-6#citeas} {\bibfield  {journal} {\bibinfo  {journal} {Nature Phys.}\ }\textbf {\bibinfo {volume} {19}},\ \bibinfo {pages} {1314} (\bibinfo {year} {2023})}\BibitemShut {NoStop}%
\bibitem [{\citenamefont {Serbyn}\ \emph {et~al.}(2021)\citenamefont {Serbyn}, \citenamefont {Abanin},\ and\ \citenamefont {Papi{\'c}}}]{serbyn2021quantum}%
  \BibitemOpen
  \bibfield  {author} {\bibinfo {author} {\bibfnamefont {M.}~\bibnamefont {Serbyn}}, \bibinfo {author} {\bibfnamefont {D.~A.}\ \bibnamefont {Abanin}},\ and\ \bibinfo {author} {\bibfnamefont {Z.}~\bibnamefont {Papi{\'c}}},\ }\bibfield  {title} {\bibinfo {title} {Quantum many-body scars and weak breaking of ergodicity},\ }\href {https://doi.org/10.1038/s41567-021-01230-2} {\bibfield  {journal} {\bibinfo  {journal} {Nature Phys.}\ }\textbf {\bibinfo {volume} {17}},\ \bibinfo {pages} {675} (\bibinfo {year} {2021})}\BibitemShut {NoStop}%
\bibitem [{\citenamefont {Beterov}\ and\ \citenamefont {Saffman}(2015)}]{beterov2015rydberg}%
  \BibitemOpen
  \bibfield  {author} {\bibinfo {author} {\bibfnamefont {I.~I.}\ \bibnamefont {Beterov}}\ and\ \bibinfo {author} {\bibfnamefont {M.}~\bibnamefont {Saffman}},\ }\bibfield  {title} {\bibinfo {title} {Rydberg blockade, f\"orster resonances, and quantum state measurements with different atomic species},\ }\href {https://doi.org/10.1103/PhysRevA.92.042710} {\bibfield  {journal} {\bibinfo  {journal} {Phys. Rev. A}\ }\textbf {\bibinfo {volume} {92}},\ \bibinfo {pages} {042710} (\bibinfo {year} {2015})}\BibitemShut {NoStop}%
\bibitem [{\citenamefont {Aharonov}\ and\ \citenamefont {Anandan}(1987)}]{aharonov1987phase}%
  \BibitemOpen
  \bibfield  {author} {\bibinfo {author} {\bibfnamefont {Y.}~\bibnamefont {Aharonov}}\ and\ \bibinfo {author} {\bibfnamefont {J.}~\bibnamefont {Anandan}},\ }\bibfield  {title} {\bibinfo {title} {Phase change during a cyclic quantum evolution},\ }\href {https://doi.org/10.1103/PhysRevLett.58.1593} {\bibfield  {journal} {\bibinfo  {journal} {Phys. Rev. Lett.}\ }\textbf {\bibinfo {volume} {58}},\ \bibinfo {pages} {1593} (\bibinfo {year} {1987})}\BibitemShut {NoStop}%
\bibitem [{\citenamefont {Jaksch}\ \emph {et~al.}(2000)\citenamefont {Jaksch}, \citenamefont {Cirac}, \citenamefont {Zoller}, \citenamefont {Rolston}, \citenamefont {C\^ot\'e},\ and\ \citenamefont {Lukin}}]{jaksch2000fast}%
  \BibitemOpen
  \bibfield  {author} {\bibinfo {author} {\bibfnamefont {D.}~\bibnamefont {Jaksch}}, \bibinfo {author} {\bibfnamefont {J.~I.}\ \bibnamefont {Cirac}}, \bibinfo {author} {\bibfnamefont {P.}~\bibnamefont {Zoller}}, \bibinfo {author} {\bibfnamefont {S.~L.}\ \bibnamefont {Rolston}}, \bibinfo {author} {\bibfnamefont {R.}~\bibnamefont {C\^ot\'e}},\ and\ \bibinfo {author} {\bibfnamefont {M.~D.}\ \bibnamefont {Lukin}},\ }\bibfield  {title} {\bibinfo {title} {Fast quantum gates for neutral atoms},\ }\href {https://doi.org/10.1103/PhysRevLett.85.2208} {\bibfield  {journal} {\bibinfo  {journal} {Phys. Rev. Lett.}\ }\textbf {\bibinfo {volume} {85}},\ \bibinfo {pages} {2208} (\bibinfo {year} {2000})}\BibitemShut {NoStop}%
\bibitem [{\citenamefont {Cesa}\ and\ \citenamefont {Pichler}(2023)}]{cesa2023universal}%
  \BibitemOpen
  \bibfield  {author} {\bibinfo {author} {\bibfnamefont {F.}~\bibnamefont {Cesa}}\ and\ \bibinfo {author} {\bibfnamefont {H.}~\bibnamefont {Pichler}},\ }\bibfield  {title} {\bibinfo {title} {Universal quantum computation in globally driven rydberg atom arrays},\ }\href {https://doi.org/10.1103/PhysRevLett.131.170601} {\bibfield  {journal} {\bibinfo  {journal} {Phys. Rev. Lett.}\ }\textbf {\bibinfo {volume} {131}},\ \bibinfo {pages} {170601} (\bibinfo {year} {2023})}\BibitemShut {NoStop}%
\bibitem [{\citenamefont {Pagano}\ \emph {et~al.}(2022)\citenamefont {Pagano}, \citenamefont {Weber}, \citenamefont {Jaschke}, \citenamefont {Pfau}, \citenamefont {Meinert}, \citenamefont {Montangero},\ and\ \citenamefont {B\"uchler}}]{pagano2022error}%
  \BibitemOpen
  \bibfield  {author} {\bibinfo {author} {\bibfnamefont {A.}~\bibnamefont {Pagano}}, \bibinfo {author} {\bibfnamefont {S.}~\bibnamefont {Weber}}, \bibinfo {author} {\bibfnamefont {D.}~\bibnamefont {Jaschke}}, \bibinfo {author} {\bibfnamefont {T.}~\bibnamefont {Pfau}}, \bibinfo {author} {\bibfnamefont {F.}~\bibnamefont {Meinert}}, \bibinfo {author} {\bibfnamefont {S.}~\bibnamefont {Montangero}},\ and\ \bibinfo {author} {\bibfnamefont {H.~P.}\ \bibnamefont {B\"uchler}},\ }\bibfield  {title} {\bibinfo {title} {Error budgeting for a controlled-phase gate with strontium-88 rydberg atoms},\ }\href {https://doi.org/10.1103/PhysRevResearch.4.033019} {\bibfield  {journal} {\bibinfo  {journal} {Phys. Rev. Res.}\ }\textbf {\bibinfo {volume} {4}},\ \bibinfo {pages} {033019} (\bibinfo {year} {2022})}\BibitemShut {NoStop}%
\bibitem [{\citenamefont {Fromonteil}\ \emph {et~al.}(2024)\citenamefont {Fromonteil}, \citenamefont {Tricarico}, \citenamefont {Cesa},\ and\ \citenamefont {Pichler}}]{fromonteil2024HJB}%
  \BibitemOpen
  \bibfield  {author} {\bibinfo {author} {\bibfnamefont {C.}~\bibnamefont {Fromonteil}}, \bibinfo {author} {\bibfnamefont {R.}~\bibnamefont {Tricarico}}, \bibinfo {author} {\bibfnamefont {F.}~\bibnamefont {Cesa}},\ and\ \bibinfo {author} {\bibfnamefont {H.}~\bibnamefont {Pichler}},\ }\bibfield  {title} {\bibinfo {title} {Hamilton-jacobi-bellman equations for rydberg-blockade processes},\ }\href {https://doi.org/10.1103/PhysRevResearch.6.033333} {\bibfield  {journal} {\bibinfo  {journal} {Phys. Rev. Res.}\ }\textbf {\bibinfo {volume} {6}},\ \bibinfo {pages} {033333} (\bibinfo {year} {2024})}\BibitemShut {NoStop}%
\bibitem [{\citenamefont {Astolfi}\ \emph {et~al.}(2025)\citenamefont {Astolfi}, \citenamefont {Jandura},\ and\ \citenamefont {Pupillo}}]{Astolfi2025}%
  \BibitemOpen
  \bibfield  {author} {\bibinfo {author} {\bibfnamefont {F.~A.}\ \bibnamefont {Astolfi}}, \bibinfo {author} {\bibfnamefont {S.}~\bibnamefont {Jandura}},\ and\ \bibinfo {author} {\bibfnamefont {G.}~\bibnamefont {Pupillo}},\ }\bibfield  {title} {\bibinfo {title} {Pontryagin maximum principle for rydberg-blockaded state-to-state transfers: A semi-analytic approach},\ }\href {https://arxiv.org/abs/2512.13549} {\bibfield  {journal} {\bibinfo  {journal} {arXiv:2512.13549}\ } (\bibinfo {year} {2025})}\BibitemShut {NoStop}%
\bibitem [{\citenamefont {Jandura}\ and\ \citenamefont {Pupillo}(2022)}]{jandura2022control}%
  \BibitemOpen
  \bibfield  {author} {\bibinfo {author} {\bibfnamefont {S.}~\bibnamefont {Jandura}}\ and\ \bibinfo {author} {\bibfnamefont {G.}~\bibnamefont {Pupillo}},\ }\bibfield  {title} {\bibinfo {title} {Time-optimal two- and three-qubit gates for rydberg atoms},\ }\href {https://doi.org/10.22331/q-2022-05-13-712} {\bibfield  {journal} {\bibinfo  {journal} {Quantum}\ }\textbf {\bibinfo {volume} {6}},\ \bibinfo {pages} {712} (\bibinfo {year} {2022})}\BibitemShut {NoStop}%
\bibitem [{\citenamefont {Bertini}\ \emph {et~al.}(2019{\natexlab{b}})\citenamefont {Bertini}, \citenamefont {Kos},\ and\ \citenamefont {Prosen}}]{bertini2019exact}%
  \BibitemOpen
  \bibfield  {author} {\bibinfo {author} {\bibfnamefont {B.}~\bibnamefont {Bertini}}, \bibinfo {author} {\bibfnamefont {P.}~\bibnamefont {Kos}},\ and\ \bibinfo {author} {\bibfnamefont {T.}~\bibnamefont {Prosen}},\ }\bibfield  {title} {\bibinfo {title} {Exact correlation functions for dual-unitary lattice models in $1+1$ dimensions},\ }\href {https://doi.org/10.1103/PhysRevLett.123.210601} {\bibfield  {journal} {\bibinfo  {journal} {Phys. Rev. Lett.}\ }\textbf {\bibinfo {volume} {123}},\ \bibinfo {pages} {210601} (\bibinfo {year} {2019}{\natexlab{b}})}\BibitemShut {NoStop}%
\bibitem [{\citenamefont {Daley}\ \emph {et~al.}(2022)\citenamefont {Daley}, \citenamefont {Bloch}, \citenamefont {Kokail}, \citenamefont {Flannigan}, \citenamefont {Pearson}, \citenamefont {Troyer},\ and\ \citenamefont {Zoller}}]{daley2022practical}%
  \BibitemOpen
  \bibfield  {author} {\bibinfo {author} {\bibfnamefont {A.~J.}\ \bibnamefont {Daley}}, \bibinfo {author} {\bibfnamefont {I.}~\bibnamefont {Bloch}}, \bibinfo {author} {\bibfnamefont {C.}~\bibnamefont {Kokail}}, \bibinfo {author} {\bibfnamefont {S.}~\bibnamefont {Flannigan}}, \bibinfo {author} {\bibfnamefont {N.}~\bibnamefont {Pearson}}, \bibinfo {author} {\bibfnamefont {M.}~\bibnamefont {Troyer}},\ and\ \bibinfo {author} {\bibfnamefont {P.}~\bibnamefont {Zoller}},\ }\bibfield  {title} {\bibinfo {title} {Practical quantum advantage in quantum simulation},\ }\href {https://doi.org/10.1038/s41586-022-04940-6} {\bibfield  {journal} {\bibinfo  {journal} {Nature}\ }\textbf {\bibinfo {volume} {607}},\ \bibinfo {pages} {667} (\bibinfo {year} {2022})}\BibitemShut {NoStop}%
\bibitem [{\citenamefont {Kim}\ \emph {et~al.}(2023)\citenamefont {Kim}, \citenamefont {Eddins}, \citenamefont {Anand}, \citenamefont {Wei}, \citenamefont {Van Den~Berg}, \citenamefont {Rosenblatt}, \citenamefont {Nayfeh}, \citenamefont {Wu}, \citenamefont {Zaletel}, \citenamefont {Temme} \emph {et~al.}}]{kim2023evidence}%
  \BibitemOpen
  \bibfield  {author} {\bibinfo {author} {\bibfnamefont {Y.}~\bibnamefont {Kim}}, \bibinfo {author} {\bibfnamefont {A.}~\bibnamefont {Eddins}}, \bibinfo {author} {\bibfnamefont {S.}~\bibnamefont {Anand}}, \bibinfo {author} {\bibfnamefont {K.~X.}\ \bibnamefont {Wei}}, \bibinfo {author} {\bibfnamefont {E.}~\bibnamefont {Van Den~Berg}}, \bibinfo {author} {\bibfnamefont {S.}~\bibnamefont {Rosenblatt}}, \bibinfo {author} {\bibfnamefont {H.}~\bibnamefont {Nayfeh}}, \bibinfo {author} {\bibfnamefont {Y.}~\bibnamefont {Wu}}, \bibinfo {author} {\bibfnamefont {M.}~\bibnamefont {Zaletel}}, \bibinfo {author} {\bibfnamefont {K.}~\bibnamefont {Temme}}, \emph {et~al.},\ }\bibfield  {title} {\bibinfo {title} {Evidence for the utility of quantum computing before fault tolerance},\ }\href {https://doi.org/10.1038/s41586-023-06096-3} {\bibfield  {journal} {\bibinfo  {journal} {Nature}\ }\textbf {\bibinfo {volume} {618}},\ \bibinfo {pages} {500} (\bibinfo {year} {2023})}\BibitemShut {NoStop}%
\bibitem [{\citenamefont {Tindall}\ \emph {et~al.}(2024)\citenamefont {Tindall}, \citenamefont {Fishman}, \citenamefont {Stoudenmire},\ and\ \citenamefont {Sels}}]{tindall2024efficient}%
  \BibitemOpen
  \bibfield  {author} {\bibinfo {author} {\bibfnamefont {J.}~\bibnamefont {Tindall}}, \bibinfo {author} {\bibfnamefont {M.}~\bibnamefont {Fishman}}, \bibinfo {author} {\bibfnamefont {E.~M.}\ \bibnamefont {Stoudenmire}},\ and\ \bibinfo {author} {\bibfnamefont {D.}~\bibnamefont {Sels}},\ }\bibfield  {title} {\bibinfo {title} {Efficient tensor network simulation of ibm's eagle kicked ising experiment},\ }\href {https://doi.org/10.1103/PRXQuantum.5.010308} {\bibfield  {journal} {\bibinfo  {journal} {PRX Quantum}\ }\textbf {\bibinfo {volume} {5}},\ \bibinfo {pages} {010308} (\bibinfo {year} {2024})}\BibitemShut {NoStop}%
\bibitem [{\citenamefont {Liao}\ \emph {et~al.}(2023)\citenamefont {Liao}, \citenamefont {Wang}, \citenamefont {Zhou}, \citenamefont {Zhang},\ and\ \citenamefont {Xiang}}]{liao2023simulation}%
  \BibitemOpen
  \bibfield  {author} {\bibinfo {author} {\bibfnamefont {H.-J.}\ \bibnamefont {Liao}}, \bibinfo {author} {\bibfnamefont {K.}~\bibnamefont {Wang}}, \bibinfo {author} {\bibfnamefont {Z.-S.}\ \bibnamefont {Zhou}}, \bibinfo {author} {\bibfnamefont {P.}~\bibnamefont {Zhang}},\ and\ \bibinfo {author} {\bibfnamefont {T.}~\bibnamefont {Xiang}},\ }\bibfield  {title} {\bibinfo {title} {Simulation of ibm's kicked ising experiment with projected entangled pair operator},\ }\href {https://arxiv.org/abs/2308.03082} {\bibfield  {journal} {\bibinfo  {journal} {arXiv:2308.03082}\ } (\bibinfo {year} {2023})}\BibitemShut {NoStop}%
\bibitem [{\citenamefont {Patra}\ \emph {et~al.}(2024)\citenamefont {Patra}, \citenamefont {Jahromi}, \citenamefont {Singh},\ and\ \citenamefont {Or\'us}}]{patra2024efficient}%
  \BibitemOpen
  \bibfield  {author} {\bibinfo {author} {\bibfnamefont {S.}~\bibnamefont {Patra}}, \bibinfo {author} {\bibfnamefont {S.~S.}\ \bibnamefont {Jahromi}}, \bibinfo {author} {\bibfnamefont {S.}~\bibnamefont {Singh}},\ and\ \bibinfo {author} {\bibfnamefont {R.}~\bibnamefont {Or\'us}},\ }\bibfield  {title} {\bibinfo {title} {Efficient tensor network simulation of ibm's largest quantum processors},\ }\href {https://doi.org/10.1103/PhysRevResearch.6.013326} {\bibfield  {journal} {\bibinfo  {journal} {Phys. Rev. Res.}\ }\textbf {\bibinfo {volume} {6}},\ \bibinfo {pages} {013326} (\bibinfo {year} {2024})}\BibitemShut {NoStop}%
\bibitem [{\citenamefont {Childs}\ \emph {et~al.}(2021)\citenamefont {Childs}, \citenamefont {Su}, \citenamefont {Tran}, \citenamefont {Wiebe},\ and\ \citenamefont {Zhu}}]{childs2021theory}%
  \BibitemOpen
  \bibfield  {author} {\bibinfo {author} {\bibfnamefont {A.~M.}\ \bibnamefont {Childs}}, \bibinfo {author} {\bibfnamefont {Y.}~\bibnamefont {Su}}, \bibinfo {author} {\bibfnamefont {M.~C.}\ \bibnamefont {Tran}}, \bibinfo {author} {\bibfnamefont {N.}~\bibnamefont {Wiebe}},\ and\ \bibinfo {author} {\bibfnamefont {S.}~\bibnamefont {Zhu}},\ }\bibfield  {title} {\bibinfo {title} {Theory of trotter error with commutator scaling},\ }\href {https://doi.org/10.1103/PhysRevX.11.011020} {\bibfield  {journal} {\bibinfo  {journal} {Phys. Rev. X}\ }\textbf {\bibinfo {volume} {11}},\ \bibinfo {pages} {011020} (\bibinfo {year} {2021})}\BibitemShut {NoStop}%
\bibitem [{\citenamefont {Xu}\ and\ \citenamefont {Swingle}(2024)}]{xu2024scrambling}%
  \BibitemOpen
  \bibfield  {author} {\bibinfo {author} {\bibfnamefont {S.}~\bibnamefont {Xu}}\ and\ \bibinfo {author} {\bibfnamefont {B.}~\bibnamefont {Swingle}},\ }\bibfield  {title} {\bibinfo {title} {Scrambling dynamics and out-of-time-ordered correlators in quantum many-body systems},\ }\href {https://doi.org/10.1103/PRXQuantum.5.010201} {\bibfield  {journal} {\bibinfo  {journal} {PRX Quantum}\ }\textbf {\bibinfo {volume} {5}},\ \bibinfo {pages} {010201} (\bibinfo {year} {2024})}\BibitemShut {NoStop}%
\bibitem [{\citenamefont {D'Alessio}\ \emph {et~al.}(2016)\citenamefont {D'Alessio}, \citenamefont {Kafri}, \citenamefont {Polkovnikov},\ and\ \citenamefont {Rigol}}]{d2016quantum}%
  \BibitemOpen
  \bibfield  {author} {\bibinfo {author} {\bibfnamefont {L.}~\bibnamefont {D'Alessio}}, \bibinfo {author} {\bibfnamefont {Y.}~\bibnamefont {Kafri}}, \bibinfo {author} {\bibfnamefont {A.}~\bibnamefont {Polkovnikov}},\ and\ \bibinfo {author} {\bibfnamefont {M.}~\bibnamefont {Rigol}},\ }\bibfield  {title} {\bibinfo {title} {From quantum chaos and eigenstate thermalization to statistical mechanics and thermodynamics},\ }\href {https://doi.org/10.1080/00018732.2016.1198134} {\bibfield  {journal} {\bibinfo  {journal} {Adv. Phys.}\ }\textbf {\bibinfo {volume} {65}},\ \bibinfo {pages} {239} (\bibinfo {year} {2016})}\BibitemShut {NoStop}%
\bibitem [{\citenamefont {Kos}\ \emph {et~al.}(2018)\citenamefont {Kos}, \citenamefont {Ljubotina},\ and\ \citenamefont {Prosen}}]{kos2018many}%
  \BibitemOpen
  \bibfield  {author} {\bibinfo {author} {\bibfnamefont {P.}~\bibnamefont {Kos}}, \bibinfo {author} {\bibfnamefont {M.}~\bibnamefont {Ljubotina}},\ and\ \bibinfo {author} {\bibfnamefont {T.}~\bibnamefont {Prosen}},\ }\bibfield  {title} {\bibinfo {title} {Many-body quantum chaos: Analytic connection to random matrix theory},\ }\href {https://doi.org/10.1103/PhysRevX.8.021062} {\bibfield  {journal} {\bibinfo  {journal} {Phys. Rev. X}\ }\textbf {\bibinfo {volume} {8}},\ \bibinfo {pages} {021062} (\bibinfo {year} {2018})}\BibitemShut {NoStop}%
\bibitem [{\citenamefont {Shenker}\ and\ \citenamefont {Stanford}(2014{\natexlab{b}})}]{shenker2014black}%
  \BibitemOpen
  \bibfield  {author} {\bibinfo {author} {\bibfnamefont {S.~H.}\ \bibnamefont {Shenker}}\ and\ \bibinfo {author} {\bibfnamefont {D.}~\bibnamefont {Stanford}},\ }\bibfield  {title} {\bibinfo {title} {Black holes and the butterfly effect},\ }\href {https://doi.org/10.1007/JHEP03(2014)067} {\bibfield  {journal} {\bibinfo  {journal} {JHEP}\ }\textbf {\bibinfo {volume} {2014}}\bibinfo  {number} { (3)},\ \bibinfo {pages} {1}}\BibitemShut {NoStop}%
\bibitem [{\citenamefont {Joshi}\ \emph {et~al.}(2020)\citenamefont {Joshi}, \citenamefont {Elben}, \citenamefont {Vermersch}, \citenamefont {Brydges}, \citenamefont {Maier}, \citenamefont {Zoller}, \citenamefont {Blatt},\ and\ \citenamefont {Roos}}]{joshi2020quantum}%
  \BibitemOpen
\bibfield  {number} {  }\bibfield  {author} {\bibinfo {author} {\bibfnamefont {M.~K.}\ \bibnamefont {Joshi}}, \bibinfo {author} {\bibfnamefont {A.}~\bibnamefont {Elben}}, \bibinfo {author} {\bibfnamefont {B.}~\bibnamefont {Vermersch}}, \bibinfo {author} {\bibfnamefont {T.}~\bibnamefont {Brydges}}, \bibinfo {author} {\bibfnamefont {C.}~\bibnamefont {Maier}}, \bibinfo {author} {\bibfnamefont {P.}~\bibnamefont {Zoller}}, \bibinfo {author} {\bibfnamefont {R.}~\bibnamefont {Blatt}},\ and\ \bibinfo {author} {\bibfnamefont {C.~F.}\ \bibnamefont {Roos}},\ }\bibfield  {title} {\bibinfo {title} {Quantum information scrambling in a trapped-ion quantum simulator with tunable range interactions},\ }\href {https://doi.org/10.1103/PhysRevLett.124.240505} {\bibfield  {journal} {\bibinfo  {journal} {Phys. Rev. Lett.}\ }\textbf {\bibinfo {volume} {124}},\ \bibinfo {pages} {240505} (\bibinfo {year} {2020})}\BibitemShut {NoStop}%
\bibitem [{\citenamefont {Green}\ \emph {et~al.}(2022)\citenamefont {Green}, \citenamefont {Elben}, \citenamefont {Alderete}, \citenamefont {Joshi}, \citenamefont {Nguyen}, \citenamefont {Zache}, \citenamefont {Zhu}, \citenamefont {Sundar},\ and\ \citenamefont {Linke}}]{green2022experimental}%
  \BibitemOpen
  \bibfield  {author} {\bibinfo {author} {\bibfnamefont {A.~M.}\ \bibnamefont {Green}}, \bibinfo {author} {\bibfnamefont {A.}~\bibnamefont {Elben}}, \bibinfo {author} {\bibfnamefont {C.~H.}\ \bibnamefont {Alderete}}, \bibinfo {author} {\bibfnamefont {L.~K.}\ \bibnamefont {Joshi}}, \bibinfo {author} {\bibfnamefont {N.~H.}\ \bibnamefont {Nguyen}}, \bibinfo {author} {\bibfnamefont {T.~V.}\ \bibnamefont {Zache}}, \bibinfo {author} {\bibfnamefont {Y.}~\bibnamefont {Zhu}}, \bibinfo {author} {\bibfnamefont {B.}~\bibnamefont {Sundar}},\ and\ \bibinfo {author} {\bibfnamefont {N.~M.}\ \bibnamefont {Linke}},\ }\bibfield  {title} {\bibinfo {title} {Experimental measurement of out-of-time-ordered correlators at finite temperature},\ }\href {https://doi.org/10.1103/PhysRevLett.128.140601} {\bibfield  {journal} {\bibinfo  {journal} {Phys. Rev. Lett.}\ }\textbf {\bibinfo {volume} {128}},\ \bibinfo {pages} {140601} (\bibinfo {year} {2022})}\BibitemShut {NoStop}%
\bibitem [{\citenamefont {Braum{\"u}ller}\ \emph {et~al.}(2022)\citenamefont {Braum{\"u}ller}, \citenamefont {Karamlou}, \citenamefont {Yanay}, \citenamefont {Kannan}, \citenamefont {Kim}, \citenamefont {Kjaergaard}, \citenamefont {Melville}, \citenamefont {Niedzielski}, \citenamefont {Sung}, \citenamefont {Veps{\"a}l{\"a}inen} \emph {et~al.}}]{braumuller2022probing}%
  \BibitemOpen
  \bibfield  {author} {\bibinfo {author} {\bibfnamefont {J.}~\bibnamefont {Braum{\"u}ller}}, \bibinfo {author} {\bibfnamefont {A.~H.}\ \bibnamefont {Karamlou}}, \bibinfo {author} {\bibfnamefont {Y.}~\bibnamefont {Yanay}}, \bibinfo {author} {\bibfnamefont {B.}~\bibnamefont {Kannan}}, \bibinfo {author} {\bibfnamefont {D.}~\bibnamefont {Kim}}, \bibinfo {author} {\bibfnamefont {M.}~\bibnamefont {Kjaergaard}}, \bibinfo {author} {\bibfnamefont {A.}~\bibnamefont {Melville}}, \bibinfo {author} {\bibfnamefont {B.~M.}\ \bibnamefont {Niedzielski}}, \bibinfo {author} {\bibfnamefont {Y.}~\bibnamefont {Sung}}, \bibinfo {author} {\bibfnamefont {A.}~\bibnamefont {Veps{\"a}l{\"a}inen}}, \emph {et~al.},\ }\bibfield  {title} {\bibinfo {title} {Probing quantum information propagation with out-of-time-ordered correlators},\ }\href {https://doi.org/10.1038/s41567-021-01430-w} {\bibfield  {journal} {\bibinfo  {journal} {Nature Phys.}\ }\textbf {\bibinfo {volume} {18}},\ \bibinfo {pages} {172} (\bibinfo {year} {2022})}\BibitemShut
  {NoStop}%
\bibitem [{\citenamefont {G{\"a}rttner}\ \emph {et~al.}(2017)\citenamefont {G{\"a}rttner}, \citenamefont {Bohnet}, \citenamefont {Safavi-Naini}, \citenamefont {Wall}, \citenamefont {Bollinger},\ and\ \citenamefont {Rey}}]{garttner2017measuring}%
  \BibitemOpen
  \bibfield  {author} {\bibinfo {author} {\bibfnamefont {M.}~\bibnamefont {G{\"a}rttner}}, \bibinfo {author} {\bibfnamefont {J.~G.}\ \bibnamefont {Bohnet}}, \bibinfo {author} {\bibfnamefont {A.}~\bibnamefont {Safavi-Naini}}, \bibinfo {author} {\bibfnamefont {M.~L.}\ \bibnamefont {Wall}}, \bibinfo {author} {\bibfnamefont {J.~J.}\ \bibnamefont {Bollinger}},\ and\ \bibinfo {author} {\bibfnamefont {A.~M.}\ \bibnamefont {Rey}},\ }\bibfield  {title} {\bibinfo {title} {Measuring out-of-time-order correlations and multiple quantum spectra in a trapped-ion quantum magnet},\ }\href {https://doi.org/10.1038/nphys4119} {\bibfield  {journal} {\bibinfo  {journal} {Nature Phys.}\ }\textbf {\bibinfo {volume} {13}},\ \bibinfo {pages} {781} (\bibinfo {year} {2017})}\BibitemShut {NoStop}%
\bibitem [{\citenamefont {Qi}\ \emph {et~al.}(2019)\citenamefont {Qi}, \citenamefont {Davis}, \citenamefont {Periwal},\ and\ \citenamefont {Schleier-Smith}}]{qi2019measuring}%
  \BibitemOpen
  \bibfield  {author} {\bibinfo {author} {\bibfnamefont {X.-L.}\ \bibnamefont {Qi}}, \bibinfo {author} {\bibfnamefont {E.~J.}\ \bibnamefont {Davis}}, \bibinfo {author} {\bibfnamefont {A.}~\bibnamefont {Periwal}},\ and\ \bibinfo {author} {\bibfnamefont {M.}~\bibnamefont {Schleier-Smith}},\ }\bibfield  {title} {\bibinfo {title} {Measuring operator size growth in quantum quench experiments},\ }\href {http://arxiv.org/abs/1906.00524} {\bibfield  {journal} {\bibinfo  {journal} {arXiv:1906.00524}\ } (\bibinfo {year} {2019})}\BibitemShut {NoStop}%
\bibitem [{\citenamefont {Mele}(2024)}]{mele2024introduction}%
  \BibitemOpen
  \bibfield  {author} {\bibinfo {author} {\bibfnamefont {A.~A.}\ \bibnamefont {Mele}},\ }\bibfield  {title} {\bibinfo {title} {Introduction to haar measure tools in quantum information: A beginner's tutorial},\ }\href {https://doi.org/10.22331/q-2024-05-08-1340} {\bibfield  {journal} {\bibinfo  {journal} {Quantum}\ }\textbf {\bibinfo {volume} {8}},\ \bibinfo {pages} {1340} (\bibinfo {year} {2024})}\BibitemShut {NoStop}%
\bibitem [{\citenamefont {Abanin}\ \emph {et~al.}(2019)\citenamefont {Abanin}, \citenamefont {Altman}, \citenamefont {Bloch},\ and\ \citenamefont {Serbyn}}]{abanin2019colloquium}%
  \BibitemOpen
  \bibfield  {author} {\bibinfo {author} {\bibfnamefont {D.~A.}\ \bibnamefont {Abanin}}, \bibinfo {author} {\bibfnamefont {E.}~\bibnamefont {Altman}}, \bibinfo {author} {\bibfnamefont {I.}~\bibnamefont {Bloch}},\ and\ \bibinfo {author} {\bibfnamefont {M.}~\bibnamefont {Serbyn}},\ }\bibfield  {title} {\bibinfo {title} {Colloquium: Many-body localization, thermalization, and entanglement},\ }\href {https://doi.org/10.1103/RevModPhys.91.021001} {\bibfield  {journal} {\bibinfo  {journal} {Rev. Mod. Phys.}\ }\textbf {\bibinfo {volume} {91}},\ \bibinfo {pages} {021001} (\bibinfo {year} {2019})}\BibitemShut {NoStop}%
\bibitem [{\citenamefont {Huang}\ \emph {et~al.}(2017)\citenamefont {Huang}, \citenamefont {Zhang},\ and\ \citenamefont {Chen}}]{huang2017out}%
  \BibitemOpen
  \bibfield  {author} {\bibinfo {author} {\bibfnamefont {Y.}~\bibnamefont {Huang}}, \bibinfo {author} {\bibfnamefont {Y.-L.}\ \bibnamefont {Zhang}},\ and\ \bibinfo {author} {\bibfnamefont {X.}~\bibnamefont {Chen}},\ }\bibfield  {title} {\bibinfo {title} {Out-of-time-ordered correlators in many-body localized systems},\ }\href {https://doi.org/10.1002/andp.201600318} {\bibfield  {journal} {\bibinfo  {journal} {Ann. der Phys.}\ }\textbf {\bibinfo {volume} {529}},\ \bibinfo {pages} {1600318} (\bibinfo {year} {2017})}\BibitemShut {NoStop}%
\bibitem [{\citenamefont {Fan}\ \emph {et~al.}(2017)\citenamefont {Fan}, \citenamefont {Zhang}, \citenamefont {Shen},\ and\ \citenamefont {Zhai}}]{fan2017out}%
  \BibitemOpen
  \bibfield  {author} {\bibinfo {author} {\bibfnamefont {R.}~\bibnamefont {Fan}}, \bibinfo {author} {\bibfnamefont {P.}~\bibnamefont {Zhang}}, \bibinfo {author} {\bibfnamefont {H.}~\bibnamefont {Shen}},\ and\ \bibinfo {author} {\bibfnamefont {H.}~\bibnamefont {Zhai}},\ }\bibfield  {title} {\bibinfo {title} {Out-of-time-order correlation for many-body localization},\ }\href {https://doi.org/10.1016/j.scib.2017.04.011} {\bibfield  {journal} {\bibinfo  {journal} {Science bull.}\ }\textbf {\bibinfo {volume} {62}},\ \bibinfo {pages} {707} (\bibinfo {year} {2017})}\BibitemShut {NoStop}%
\bibitem [{\citenamefont {Zhou}\ and\ \citenamefont {Luitz}(2017)}]{zhou2017operator}%
  \BibitemOpen
  \bibfield  {author} {\bibinfo {author} {\bibfnamefont {T.}~\bibnamefont {Zhou}}\ and\ \bibinfo {author} {\bibfnamefont {D.~J.}\ \bibnamefont {Luitz}},\ }\bibfield  {title} {\bibinfo {title} {Operator entanglement entropy of the time evolution operator in chaotic systems},\ }\href {https://doi.org/10.1103/PhysRevB.95.094206} {\bibfield  {journal} {\bibinfo  {journal} {Phys. Rev. B}\ }\textbf {\bibinfo {volume} {95}},\ \bibinfo {pages} {094206} (\bibinfo {year} {2017})}\BibitemShut {NoStop}%
\bibitem [{\citenamefont {Renes}\ \emph {et~al.}(2004)\citenamefont {Renes}, \citenamefont {Blume-Kohout}, \citenamefont {Scott},\ and\ \citenamefont {Caves}}]{renes2004symmetric}%
  \BibitemOpen
  \bibfield  {author} {\bibinfo {author} {\bibfnamefont {J.~M.}\ \bibnamefont {Renes}}, \bibinfo {author} {\bibfnamefont {R.}~\bibnamefont {Blume-Kohout}}, \bibinfo {author} {\bibfnamefont {A.~J.}\ \bibnamefont {Scott}},\ and\ \bibinfo {author} {\bibfnamefont {C.~M.}\ \bibnamefont {Caves}},\ }\bibfield  {title} {\bibinfo {title} {Symmetric informationally complete quantum measurements},\ }\href {https://doi.org/10.1063/1.1737053} {\bibfield  {journal} {\bibinfo  {journal} {J. Math. Phys.}\ }\textbf {\bibinfo {volume} {45}},\ \bibinfo {pages} {2171} (\bibinfo {year} {2004})}\BibitemShut {NoStop}%
\bibitem [{\citenamefont {Jordan}\ and\ \citenamefont {Wigner}(1993)}]{jordan1993}%
  \BibitemOpen
  \bibfield  {author} {\bibinfo {author} {\bibfnamefont {P.}~\bibnamefont {Jordan}}\ and\ \bibinfo {author} {\bibfnamefont {E.~P.}\ \bibnamefont {Wigner}},\ }\bibinfo {title} {{\"U}ber das paulische {\"a}quivalenzverbot},\ in\ \href {https://doi.org/10.1007/978-3-662-02781-3_9} {\emph {\bibinfo {booktitle} {The Collected Works of Eugene Paul Wigner: Part A: The Scientific Papers}}},\ \bibinfo {editor} {edited by\ \bibinfo {editor} {\bibfnamefont {A.~S.}\ \bibnamefont {Wightman}}}\ (\bibinfo  {publisher} {Springer Berlin Heidelberg},\ \bibinfo {address} {Berlin, Heidelberg},\ \bibinfo {year} {1993})\ pp.\ \bibinfo {pages} {109--129}\BibitemShut {NoStop}%
\bibitem [{\citenamefont {Nielsen}\ and\ \citenamefont {Chuang}(2010)}]{nielsen2010quantum}%
  \BibitemOpen
  \bibfield  {author} {\bibinfo {author} {\bibfnamefont {M.~A.}\ \bibnamefont {Nielsen}}\ and\ \bibinfo {author} {\bibfnamefont {I.~L.}\ \bibnamefont {Chuang}},\ }\href@noop {} {\emph {\bibinfo {title} {Quantum computation and quantum information}}}\ (\bibinfo  {publisher} {Cambridge university press, Cambridge, UK},\ \bibinfo {year} {2010})\BibitemShut {NoStop}%
\bibitem [{\citenamefont {Gottesman}(1998)}]{gottesman1998theory}%
  \BibitemOpen
  \bibfield  {author} {\bibinfo {author} {\bibfnamefont {D.}~\bibnamefont {Gottesman}},\ }\bibfield  {title} {\bibinfo {title} {Theory of fault-tolerant quantum computation},\ }\href {https://doi.org/10.1103/PhysRevA.57.127} {\bibfield  {journal} {\bibinfo  {journal} {Phys. Rev. A}\ }\textbf {\bibinfo {volume} {57}},\ \bibinfo {pages} {127} (\bibinfo {year} {1998})}\BibitemShut {NoStop}%
\bibitem [{\citenamefont {Moeckel}\ and\ \citenamefont {Kehrein}(2008)}]{moeckel2008interaction}%
  \BibitemOpen
  \bibfield  {author} {\bibinfo {author} {\bibfnamefont {M.}~\bibnamefont {Moeckel}}\ and\ \bibinfo {author} {\bibfnamefont {S.}~\bibnamefont {Kehrein}},\ }\bibfield  {title} {\bibinfo {title} {Interaction quench in the hubbard model},\ }\href {https://doi.org/10.1103/PhysRevLett.100.175702} {\bibfield  {journal} {\bibinfo  {journal} {Phys. Rev. Lett.}\ }\textbf {\bibinfo {volume} {100}},\ \bibinfo {pages} {175702} (\bibinfo {year} {2008})}\BibitemShut {NoStop}%
\bibitem [{\citenamefont {Rosch}\ \emph {et~al.}(2008)\citenamefont {Rosch}, \citenamefont {Rasch}, \citenamefont {Binz},\ and\ \citenamefont {Vojta}}]{rosch2008metastable}%
  \BibitemOpen
  \bibfield  {author} {\bibinfo {author} {\bibfnamefont {A.}~\bibnamefont {Rosch}}, \bibinfo {author} {\bibfnamefont {D.}~\bibnamefont {Rasch}}, \bibinfo {author} {\bibfnamefont {B.}~\bibnamefont {Binz}},\ and\ \bibinfo {author} {\bibfnamefont {M.}~\bibnamefont {Vojta}},\ }\bibfield  {title} {\bibinfo {title} {Metastable superfluidity of repulsive fermionic atoms in optical lattices},\ }\href {https://doi.org/10.1103/PhysRevLett.101.265301} {\bibfield  {journal} {\bibinfo  {journal} {Phys. Rev. Lett.}\ }\textbf {\bibinfo {volume} {101}},\ \bibinfo {pages} {265301} (\bibinfo {year} {2008})}\BibitemShut {NoStop}%
\bibitem [{\citenamefont {Marcuzzi}\ \emph {et~al.}(2013)\citenamefont {Marcuzzi}, \citenamefont {Marino}, \citenamefont {Gambassi},\ and\ \citenamefont {Silva}}]{marcuzzi2013prethermalization}%
  \BibitemOpen
  \bibfield  {author} {\bibinfo {author} {\bibfnamefont {M.}~\bibnamefont {Marcuzzi}}, \bibinfo {author} {\bibfnamefont {J.}~\bibnamefont {Marino}}, \bibinfo {author} {\bibfnamefont {A.}~\bibnamefont {Gambassi}},\ and\ \bibinfo {author} {\bibfnamefont {A.}~\bibnamefont {Silva}},\ }\bibfield  {title} {\bibinfo {title} {Prethermalization in a nonintegrable quantum spin chain after a quench},\ }\href {https://doi.org/10.1103/PhysRevLett.111.197203} {\bibfield  {journal} {\bibinfo  {journal} {Phys. Rev. Lett.}\ }\textbf {\bibinfo {volume} {111}},\ \bibinfo {pages} {197203} (\bibinfo {year} {2013})}\BibitemShut {NoStop}%
\bibitem [{\citenamefont {Bertini}\ \emph {et~al.}(2015)\citenamefont {Bertini}, \citenamefont {Essler}, \citenamefont {Groha},\ and\ \citenamefont {Robinson}}]{bertini2015prethermalization}%
  \BibitemOpen
  \bibfield  {author} {\bibinfo {author} {\bibfnamefont {B.}~\bibnamefont {Bertini}}, \bibinfo {author} {\bibfnamefont {F.~H.~L.}\ \bibnamefont {Essler}}, \bibinfo {author} {\bibfnamefont {S.}~\bibnamefont {Groha}},\ and\ \bibinfo {author} {\bibfnamefont {N.~J.}\ \bibnamefont {Robinson}},\ }\bibfield  {title} {\bibinfo {title} {Prethermalization and thermalization in models with weak integrability breaking},\ }\href {https://doi.org/10.1103/PhysRevLett.115.180601} {\bibfield  {journal} {\bibinfo  {journal} {Phys. Rev. Lett.}\ }\textbf {\bibinfo {volume} {115}},\ \bibinfo {pages} {180601} (\bibinfo {year} {2015})}\BibitemShut {NoStop}%
\bibitem [{\citenamefont {Brandino}\ \emph {et~al.}(2015)\citenamefont {Brandino}, \citenamefont {Caux},\ and\ \citenamefont {Konik}}]{brandino2015glimmers}%
  \BibitemOpen
  \bibfield  {author} {\bibinfo {author} {\bibfnamefont {G.~P.}\ \bibnamefont {Brandino}}, \bibinfo {author} {\bibfnamefont {J.-S.}\ \bibnamefont {Caux}},\ and\ \bibinfo {author} {\bibfnamefont {R.~M.}\ \bibnamefont {Konik}},\ }\bibfield  {title} {\bibinfo {title} {Glimmers of a quantum kam theorem: Insights from quantum quenches in one-dimensional bose gases},\ }\href {https://doi.org/10.1103/PhysRevX.5.041043} {\bibfield  {journal} {\bibinfo  {journal} {Phys. Rev. X}\ }\textbf {\bibinfo {volume} {5}},\ \bibinfo {pages} {041043} (\bibinfo {year} {2015})}\BibitemShut {NoStop}%
\bibitem [{\citenamefont {Evered}\ \emph {et~al.}(2023)\citenamefont {Evered}, \citenamefont {Bluvstein}, \citenamefont {Kalinowski}, \citenamefont {Ebadi}, \citenamefont {Manovitz}, \citenamefont {Zhou}, \citenamefont {Li}, \citenamefont {Geim}, \citenamefont {Wang}, \citenamefont {Maskara} \emph {et~al.}}]{evered2023high}%
  \BibitemOpen
  \bibfield  {author} {\bibinfo {author} {\bibfnamefont {S.~J.}\ \bibnamefont {Evered}}, \bibinfo {author} {\bibfnamefont {D.}~\bibnamefont {Bluvstein}}, \bibinfo {author} {\bibfnamefont {M.}~\bibnamefont {Kalinowski}}, \bibinfo {author} {\bibfnamefont {S.}~\bibnamefont {Ebadi}}, \bibinfo {author} {\bibfnamefont {T.}~\bibnamefont {Manovitz}}, \bibinfo {author} {\bibfnamefont {H.}~\bibnamefont {Zhou}}, \bibinfo {author} {\bibfnamefont {S.~H.}\ \bibnamefont {Li}}, \bibinfo {author} {\bibfnamefont {A.~A.}\ \bibnamefont {Geim}}, \bibinfo {author} {\bibfnamefont {T.~T.}\ \bibnamefont {Wang}}, \bibinfo {author} {\bibfnamefont {N.}~\bibnamefont {Maskara}}, \emph {et~al.},\ }\bibfield  {title} {\bibinfo {title} {High-fidelity parallel entangling gates on a neutral-atom quantum computer},\ }\href {https://doi.org/10.1038/s41586-023-06481-y} {\bibfield  {journal} {\bibinfo  {journal} {Nature}\ }\textbf {\bibinfo {volume} {622}},\ \bibinfo {pages} {268} (\bibinfo {year} {2023})}\BibitemShut {NoStop}%
\bibitem [{\citenamefont {Skinner}\ \emph {et~al.}(2019)\citenamefont {Skinner}, \citenamefont {Ruhman},\ and\ \citenamefont {Nahum}}]{skinner2019measurement}%
  \BibitemOpen
  \bibfield  {author} {\bibinfo {author} {\bibfnamefont {B.}~\bibnamefont {Skinner}}, \bibinfo {author} {\bibfnamefont {J.}~\bibnamefont {Ruhman}},\ and\ \bibinfo {author} {\bibfnamefont {A.}~\bibnamefont {Nahum}},\ }\bibfield  {title} {\bibinfo {title} {Measurement-induced phase transitions in the dynamics of entanglement},\ }\href {https://doi.org/10.1103/PhysRevX.9.031009} {\bibfield  {journal} {\bibinfo  {journal} {Phys. Rev. X}\ }\textbf {\bibinfo {volume} {9}},\ \bibinfo {pages} {031009} (\bibinfo {year} {2019})}\BibitemShut {NoStop}%
\bibitem [{\citenamefont {Li}\ \emph {et~al.}(2018)\citenamefont {Li}, \citenamefont {Chen},\ and\ \citenamefont {Fisher}}]{li2018quantum}%
  \BibitemOpen
  \bibfield  {author} {\bibinfo {author} {\bibfnamefont {Y.}~\bibnamefont {Li}}, \bibinfo {author} {\bibfnamefont {X.}~\bibnamefont {Chen}},\ and\ \bibinfo {author} {\bibfnamefont {M.~P.~A.}\ \bibnamefont {Fisher}},\ }\bibfield  {title} {\bibinfo {title} {Quantum zeno effect and the many-body entanglement transition},\ }\href {https://doi.org/10.1103/PhysRevB.98.205136} {\bibfield  {journal} {\bibinfo  {journal} {Phys. Rev. B}\ }\textbf {\bibinfo {volume} {98}},\ \bibinfo {pages} {205136} (\bibinfo {year} {2018})}\BibitemShut {NoStop}%
\bibitem [{\citenamefont {Kogut}(1979)}]{RevModPhys.51.659}%
  \BibitemOpen
  \bibfield  {author} {\bibinfo {author} {\bibfnamefont {J.~B.}\ \bibnamefont {Kogut}},\ }\bibfield  {title} {\bibinfo {title} {An introduction to lattice gauge theory and spin systems},\ }\href {https://doi.org/10.1103/RevModPhys.51.659} {\bibfield  {journal} {\bibinfo  {journal} {Rev. Mod. Phys.}\ }\textbf {\bibinfo {volume} {51}},\ \bibinfo {pages} {659} (\bibinfo {year} {1979})}\BibitemShut {NoStop}%
\bibitem [{\citenamefont {Raussendorf}(2005)}]{raussendorf2005quantum}%
  \BibitemOpen
  \bibfield  {author} {\bibinfo {author} {\bibfnamefont {R.}~\bibnamefont {Raussendorf}},\ }\bibfield  {title} {\bibinfo {title} {Quantum cellular automaton for universal quantum computation},\ }\href {https://doi.org/10.1103/PhysRevA.72.022301} {\bibfield  {journal} {\bibinfo  {journal} {Phys. Rev. A}\ }\textbf {\bibinfo {volume} {72}},\ \bibinfo {pages} {022301} (\bibinfo {year} {2005})}\BibitemShut {NoStop}%
\bibitem [{\citenamefont {D'Alessandro}(2021)}]{Dalessandro2021}%
  \BibitemOpen
  \bibfield  {author} {\bibinfo {author} {\bibfnamefont {D.}~\bibnamefont {D'Alessandro}},\ }\href {https://api.pageplace.de/preview/DT0400.9781000394993_A40919156/preview-9781000394993_A40919156.pdf} {\emph {\bibinfo {title} {Introduction to Quantum Control and Dynamics}}},\ \bibinfo {edition} {2nd}\ ed.,\ Advances in Applied Mathematics\ (\bibinfo  {publisher} {Taylor \& Francis / CRC Press},\ \bibinfo {address} {Abingdon, UK},\ \bibinfo {year} {2021})\BibitemShut {NoStop}%
\bibitem [{\citenamefont {Levine}\ \emph {et~al.}(2019)\citenamefont {Levine}, \citenamefont {Keesling}, \citenamefont {Semeghini}, \citenamefont {Omran}, \citenamefont {Wang}, \citenamefont {Ebadi}, \citenamefont {Bernien}, \citenamefont {Greiner}, \citenamefont {Vuleti\ifmmode~\acute{c}\else \'{c}\fi{}}, \citenamefont {Pichler},\ and\ \citenamefont {Lukin}}]{levine2019high}%
  \BibitemOpen
  \bibfield  {author} {\bibinfo {author} {\bibfnamefont {H.}~\bibnamefont {Levine}}, \bibinfo {author} {\bibfnamefont {A.}~\bibnamefont {Keesling}}, \bibinfo {author} {\bibfnamefont {G.}~\bibnamefont {Semeghini}}, \bibinfo {author} {\bibfnamefont {A.}~\bibnamefont {Omran}}, \bibinfo {author} {\bibfnamefont {T.~T.}\ \bibnamefont {Wang}}, \bibinfo {author} {\bibfnamefont {S.}~\bibnamefont {Ebadi}}, \bibinfo {author} {\bibfnamefont {H.}~\bibnamefont {Bernien}}, \bibinfo {author} {\bibfnamefont {M.}~\bibnamefont {Greiner}}, \bibinfo {author} {\bibfnamefont {V.}~\bibnamefont {Vuleti\ifmmode~\acute{c}\else \'{c}\fi{}}}, \bibinfo {author} {\bibfnamefont {H.}~\bibnamefont {Pichler}},\ and\ \bibinfo {author} {\bibfnamefont {M.~D.}\ \bibnamefont {Lukin}},\ }\bibfield  {title} {\bibinfo {title} {Parallel implementation of high-fidelity multiqubit gates with neutral atoms},\ }\href {https://doi.org/10.1103/PhysRevLett.123.170503} {\bibfield  {journal} {\bibinfo  {journal} {Phys. Rev. Lett.}\ }\textbf {\bibinfo {volume}
  {123}},\ \bibinfo {pages} {170503} (\bibinfo {year} {2019})}\BibitemShut {NoStop}%
\bibitem [{\citenamefont {Khaneja}\ \emph {et~al.}(2005)\citenamefont {Khaneja}, \citenamefont {Reiss}, \citenamefont {Kehlet}, \citenamefont {Schulte-Herbrüggen},\ and\ \citenamefont {Glaser}}]{khaneja2005GRAPE}%
  \BibitemOpen
  \bibfield  {author} {\bibinfo {author} {\bibfnamefont {N.}~\bibnamefont {Khaneja}}, \bibinfo {author} {\bibfnamefont {T.}~\bibnamefont {Reiss}}, \bibinfo {author} {\bibfnamefont {C.}~\bibnamefont {Kehlet}}, \bibinfo {author} {\bibfnamefont {T.}~\bibnamefont {Schulte-Herbrüggen}},\ and\ \bibinfo {author} {\bibfnamefont {S.~J.}\ \bibnamefont {Glaser}},\ }\bibfield  {title} {\bibinfo {title} {Optimal control of coupled spin dynamics: design of nmr pulse sequences by gradient ascent algorithms},\ }\href {https://doi.org/https://doi.org/10.1016/j.jmr.2004.11.004} {\bibfield  {journal} {\bibinfo  {journal} {J. Magn. Res.}\ }\textbf {\bibinfo {volume} {172}},\ \bibinfo {pages} {296} (\bibinfo {year} {2005})}\BibitemShut {NoStop}%
\bibitem [{\citenamefont {Ma}\ \emph {et~al.}(2023)\citenamefont {Ma}, \citenamefont {Liu}, \citenamefont {Peng}, \citenamefont {Zhang}, \citenamefont {Jandura}, \citenamefont {Claes}, \citenamefont {Burgers}, \citenamefont {Pupillo}, \citenamefont {Puri},\ and\ \citenamefont {Thompson}}]{ma2023high}%
  \BibitemOpen
  \bibfield  {author} {\bibinfo {author} {\bibfnamefont {S.}~\bibnamefont {Ma}}, \bibinfo {author} {\bibfnamefont {G.}~\bibnamefont {Liu}}, \bibinfo {author} {\bibfnamefont {P.}~\bibnamefont {Peng}}, \bibinfo {author} {\bibfnamefont {B.}~\bibnamefont {Zhang}}, \bibinfo {author} {\bibfnamefont {S.}~\bibnamefont {Jandura}}, \bibinfo {author} {\bibfnamefont {J.}~\bibnamefont {Claes}}, \bibinfo {author} {\bibfnamefont {A.~P.}\ \bibnamefont {Burgers}}, \bibinfo {author} {\bibfnamefont {G.}~\bibnamefont {Pupillo}}, \bibinfo {author} {\bibfnamefont {S.}~\bibnamefont {Puri}},\ and\ \bibinfo {author} {\bibfnamefont {J.~D.}\ \bibnamefont {Thompson}},\ }\bibfield  {title} {\bibinfo {title} {High-fidelity gates and mid-circuit erasure conversion in an atomic qubit},\ }\href {https://doi.org/10.1038/s41586-023-06438-1} {\bibfield  {journal} {\bibinfo  {journal} {Nature}\ }\textbf {\bibinfo {volume} {622}},\ \bibinfo {pages} {279} (\bibinfo {year} {2023})}\BibitemShut {NoStop}%
\bibitem [{\citenamefont {Giudici}\ \emph {et~al.}(2025)\citenamefont {Giudici}, \citenamefont {Veroni}, \citenamefont {Giudice}, \citenamefont {Pichler},\ and\ \citenamefont {Zeiher}}]{giudici2025dipole}%
  \BibitemOpen
  \bibfield  {author} {\bibinfo {author} {\bibfnamefont {G.}~\bibnamefont {Giudici}}, \bibinfo {author} {\bibfnamefont {S.}~\bibnamefont {Veroni}}, \bibinfo {author} {\bibfnamefont {G.}~\bibnamefont {Giudice}}, \bibinfo {author} {\bibfnamefont {H.}~\bibnamefont {Pichler}},\ and\ \bibinfo {author} {\bibfnamefont {J.}~\bibnamefont {Zeiher}},\ }\bibfield  {title} {\bibinfo {title} {Fast entangling gates for rydberg atoms via resonant dipole-dipole interaction},\ }\href {https://doi.org/10.1103/5d8p-3hm1} {\bibfield  {journal} {\bibinfo  {journal} {PRX Quantum}\ }\textbf {\bibinfo {volume} {6}},\ \bibinfo {pages} {030308} (\bibinfo {year} {2025})}\BibitemShut {NoStop}%
\bibitem [{\citenamefont {Maskara}\ \emph {et~al.}(2025)\citenamefont {Maskara}, \citenamefont {Ostermann}, \citenamefont {Shee}, \citenamefont {Kalinowski}, \citenamefont {McClain~Gomez}, \citenamefont {Araiza~Bravo}, \citenamefont {Wang}, \citenamefont {Krylov}, \citenamefont {Yao}, \citenamefont {Head-Gordon}, \citenamefont {Lukin},\ and\ \citenamefont {Yelin}}]{maskara2025programmable}%
  \BibitemOpen
  \bibfield  {author} {\bibinfo {author} {\bibfnamefont {N.}~\bibnamefont {Maskara}}, \bibinfo {author} {\bibfnamefont {S.}~\bibnamefont {Ostermann}}, \bibinfo {author} {\bibfnamefont {J.}~\bibnamefont {Shee}}, \bibinfo {author} {\bibfnamefont {M.}~\bibnamefont {Kalinowski}}, \bibinfo {author} {\bibfnamefont {A.}~\bibnamefont {McClain~Gomez}}, \bibinfo {author} {\bibfnamefont {R.}~\bibnamefont {Araiza~Bravo}}, \bibinfo {author} {\bibfnamefont {D.~S.}\ \bibnamefont {Wang}}, \bibinfo {author} {\bibfnamefont {A.~I.}\ \bibnamefont {Krylov}}, \bibinfo {author} {\bibfnamefont {N.~Y.}\ \bibnamefont {Yao}}, \bibinfo {author} {\bibfnamefont {M.}~\bibnamefont {Head-Gordon}}, \bibinfo {author} {\bibfnamefont {M.~D.}\ \bibnamefont {Lukin}},\ and\ \bibinfo {author} {\bibfnamefont {S.~F.}\ \bibnamefont {Yelin}},\ }\bibfield  {title} {\bibinfo {title} {Programmable simulations of molecules and materials with reconfigurable quantum processors},\ }\href {https://doi.org/10.1038/s41567-024-02738-z} {\bibfield  {journal}
  {\bibinfo  {journal} {Nature Physics}\ }\textbf {\bibinfo {volume} {21}},\ \bibinfo {pages} {289} (\bibinfo {year} {2025})}\BibitemShut {NoStop}%
\bibitem [{\citenamefont {Zeng}\ \emph {et~al.}(2025)\citenamefont {Zeng}, \citenamefont {Giudici}, \citenamefont {Senoo}, \citenamefont {Baumg{\"a}rtner}, \citenamefont {Kaufman},\ and\ \citenamefont {Pichler}}]{Zeng2025AdiabaticEcho}%
  \BibitemOpen
  \bibfield  {author} {\bibinfo {author} {\bibfnamefont {Z.}~\bibnamefont {Zeng}}, \bibinfo {author} {\bibfnamefont {G.}~\bibnamefont {Giudici}}, \bibinfo {author} {\bibfnamefont {A.}~\bibnamefont {Senoo}}, \bibinfo {author} {\bibfnamefont {A.}~\bibnamefont {Baumg{\"a}rtner}}, \bibinfo {author} {\bibfnamefont {A.~M.}\ \bibnamefont {Kaufman}},\ and\ \bibinfo {author} {\bibfnamefont {H.}~\bibnamefont {Pichler}},\ }\bibfield  {title} {\bibinfo {title} {Adiabatic echo protocols for robust quantum many-body state preparation},\ }\href {https://arxiv.org/abs/2506.12138} {\bibfield  {journal} {\bibinfo  {journal} {arXiv:2506.12138}\ } (\bibinfo {year} {2025})}\BibitemShut {NoStop}%
\bibitem [{\citenamefont {White}(1992)}]{white1992density}%
  \BibitemOpen
  \bibfield  {author} {\bibinfo {author} {\bibfnamefont {S.~R.}\ \bibnamefont {White}},\ }\bibfield  {title} {\bibinfo {title} {Density matrix formulation for quantum renormalization groups},\ }\href {https://doi.org/10.1103/PhysRevLett.69.2863} {\bibfield  {journal} {\bibinfo  {journal} {Phys. Rev. Lett.}\ }\textbf {\bibinfo {volume} {69}},\ \bibinfo {pages} {2863} (\bibinfo {year} {1992})}\BibitemShut {NoStop}%
\bibitem [{\citenamefont {Vidal}(2003)}]{vidal2003efficient}%
  \BibitemOpen
  \bibfield  {author} {\bibinfo {author} {\bibfnamefont {G.}~\bibnamefont {Vidal}},\ }\bibfield  {title} {\bibinfo {title} {Efficient classical simulation of slightly entangled quantum computations},\ }\href {https://doi.org/10.1103/PhysRevLett.91.147902} {\bibfield  {journal} {\bibinfo  {journal} {Phys. Rev. Lett.}\ }\textbf {\bibinfo {volume} {91}},\ \bibinfo {pages} {147902} (\bibinfo {year} {2003})}\BibitemShut {NoStop}%
\bibitem [{\citenamefont {Cirac}\ and\ \citenamefont {Verstraete}(2004)}]{verstraete2004renormalization}%
  \BibitemOpen
  \bibfield  {author} {\bibinfo {author} {\bibfnamefont {J.~I.}\ \bibnamefont {Cirac}}\ and\ \bibinfo {author} {\bibfnamefont {F.}~\bibnamefont {Verstraete}},\ }\bibfield  {title} {\bibinfo {title} {Renormalization algorithms for quantum-many body systems in two and higher dimensions},\ }\href {https://arxiv.org/abs/cond-mat/0407066} {\bibfield  {journal} {\bibinfo  {journal} {arXiv cond-mat/0407066}\ } (\bibinfo {year} {2004})}\BibitemShut {NoStop}%
\bibitem [{\citenamefont {Fishman}\ \emph {et~al.}(2022)\citenamefont {Fishman}, \citenamefont {White},\ and\ \citenamefont {Stoudenmire}}]{ITensor}%
  \BibitemOpen
  \bibfield  {author} {\bibinfo {author} {\bibfnamefont {M.}~\bibnamefont {Fishman}}, \bibinfo {author} {\bibfnamefont {S.~R.}\ \bibnamefont {White}},\ and\ \bibinfo {author} {\bibfnamefont {E.~M.}\ \bibnamefont {Stoudenmire}},\ }\bibfield  {title} {\bibinfo {title} {{The ITensor Software Library for Tensor Network Calculations}},\ }\href {https://doi.org/10.21468/SciPostPhysCodeb.4} {\bibfield  {journal} {\bibinfo  {journal} {SciPost Phys. Codebases}\ ,\ \bibinfo {pages} {4}} (\bibinfo {year} {2022})}\BibitemShut {NoStop}%
\bibitem [{\citenamefont {Phien}\ \emph {et~al.}(2015)\citenamefont {Phien}, \citenamefont {Bengua}, \citenamefont {Tuan}, \citenamefont {Corboz},\ and\ \citenamefont {Or\'us}}]{phien2015infinite}%
  \BibitemOpen
  \bibfield  {author} {\bibinfo {author} {\bibfnamefont {H.~N.}\ \bibnamefont {Phien}}, \bibinfo {author} {\bibfnamefont {J.~A.}\ \bibnamefont {Bengua}}, \bibinfo {author} {\bibfnamefont {H.~D.}\ \bibnamefont {Tuan}}, \bibinfo {author} {\bibfnamefont {P.}~\bibnamefont {Corboz}},\ and\ \bibinfo {author} {\bibfnamefont {R.}~\bibnamefont {Or\'us}},\ }\bibfield  {title} {\bibinfo {title} {Infinite projected entangled pair states algorithm improved: Fast full update and gauge fixing},\ }\href {https://doi.org/10.1103/PhysRevB.92.035142} {\bibfield  {journal} {\bibinfo  {journal} {Phys. Rev. B}\ }\textbf {\bibinfo {volume} {92}},\ \bibinfo {pages} {035142} (\bibinfo {year} {2015})}\BibitemShut {NoStop}%
\bibitem [{\citenamefont {Jiang}\ \emph {et~al.}(2008)\citenamefont {Jiang}, \citenamefont {Weng},\ and\ \citenamefont {Xiang}}]{jiang2008accurate}%
  \BibitemOpen
  \bibfield  {author} {\bibinfo {author} {\bibfnamefont {H.~C.}\ \bibnamefont {Jiang}}, \bibinfo {author} {\bibfnamefont {Z.~Y.}\ \bibnamefont {Weng}},\ and\ \bibinfo {author} {\bibfnamefont {T.}~\bibnamefont {Xiang}},\ }\bibfield  {title} {\bibinfo {title} {Accurate determination of tensor network state of quantum lattice models in two dimensions},\ }\href {https://doi.org/10.1103/PhysRevLett.101.090603} {\bibfield  {journal} {\bibinfo  {journal} {Phys. Rev. Lett.}\ }\textbf {\bibinfo {volume} {101}},\ \bibinfo {pages} {090603} (\bibinfo {year} {2008})}\BibitemShut {NoStop}%
\bibitem [{\citenamefont {Corboz}(2020)}]{corboz2020ipeps}%
  \BibitemOpen
  \bibfield  {author} {\bibinfo {author} {\bibfnamefont {P.}~\bibnamefont {Corboz}},\ }\href@noop {} {\bibinfo {title} {Benasque lecture on tensor network algorithms}},\ \bibinfo {howpublished} {\url{https://www.benasque.org/2020scs/talks_contr/113_tensornetworks_lecture2.pdf}} (\bibinfo {year} {2020})\BibitemShut {NoStop}%
\bibitem [{\citenamefont {Lubasch}\ \emph {et~al.}(2014{\natexlab{a}})\citenamefont {Lubasch}, \citenamefont {Cirac},\ and\ \citenamefont {Banuls}}]{lubasch2014unifying}%
  \BibitemOpen
  \bibfield  {author} {\bibinfo {author} {\bibfnamefont {M.}~\bibnamefont {Lubasch}}, \bibinfo {author} {\bibfnamefont {J.~I.}\ \bibnamefont {Cirac}},\ and\ \bibinfo {author} {\bibfnamefont {M.-C.}\ \bibnamefont {Banuls}},\ }\bibfield  {title} {\bibinfo {title} {Unifying projected entangled pair state contractions},\ }\href {https://doi.org/10.1088/1367-2630/16/3/033014} {\bibfield  {journal} {\bibinfo  {journal} {New J. Phys.}\ }\textbf {\bibinfo {volume} {16}},\ \bibinfo {pages} {033014} (\bibinfo {year} {2014}{\natexlab{a}})}\BibitemShut {NoStop}%
\bibitem [{\citenamefont {Lubasch}\ \emph {et~al.}(2014{\natexlab{b}})\citenamefont {Lubasch}, \citenamefont {Cirac},\ and\ \citenamefont {Ba\~nuls}}]{lubasch2014algorithms}%
  \BibitemOpen
  \bibfield  {author} {\bibinfo {author} {\bibfnamefont {M.}~\bibnamefont {Lubasch}}, \bibinfo {author} {\bibfnamefont {J.~I.}\ \bibnamefont {Cirac}},\ and\ \bibinfo {author} {\bibfnamefont {M.-C.}\ \bibnamefont {Ba\~nuls}},\ }\bibfield  {title} {\bibinfo {title} {Algorithms for finite projected entangled pair states},\ }\href {https://doi.org/10.1103/PhysRevB.90.064425} {\bibfield  {journal} {\bibinfo  {journal} {Phys. Rev. B}\ }\textbf {\bibinfo {volume} {90}},\ \bibinfo {pages} {064425} (\bibinfo {year} {2014}{\natexlab{b}})}\BibitemShut {NoStop}%
\end{thebibliography}%

\end{document}